\documentclass[aps,prb,floatfix,superscriptaddress,reprint,
longbibliography,noshowpacs,floatfix]{revtex4-2}
\usepackage{graphics,graphicx,epsfig,amsmath,amssymb,epstopdf,wasysym,comment}
\usepackage[T1]{fontenc}
\usepackage{frcursive}
\usepackage{color}
\usepackage[table]{xcolor}
\usepackage{hyperref}
\newenvironment{frcseries}{\fontfamily{frc}\selectfont}{}

\usepackage{hyperref}%
\hypersetup{colorlinks=true,allcolors=blue}
\usepackage{physics}
\usepackage{amsmath}
\usepackage{longtable}
\usepackage{array,multirow}
\usepackage{rotating}

\begin{document}

\title{Twistronics of Janus transition metal dichalcogenide bilayers}

\author{Mattia Angeli}
\affiliation{ John A. Paulson School of Engineering and Applied Sciences,
Harvard University, Cambridge, Massachusetts 02138, USA} 

\author{Gabriel R. Schleder}
\affiliation{ John A. Paulson School of Engineering and Applied Sciences,
Harvard University, Cambridge, Massachusetts 02138, USA} 

\author{Efthimios Kaxiras}
\affiliation{ John A. Paulson School of Engineering and Applied Sciences,
Harvard University, Cambridge, Massachusetts 02138, USA} 
\affiliation{Department of Physics, Harvard University, Cambridge, Massachusetts 02138, USA} 

\begin{abstract}
Twisted multilayers of two-dimensional (2D) materials are an increasingly important platform for investigating quantum phases of matter, and in particular, strongly correlated electrons. The moiré pattern introduced by the relative twist between layers creates effective potentials of long-wavelength, leading to electron localization.
However, in contrast to the abundance of 2D materials, few twisted heterostructures have been studied until now.
Here we develop a first-principle continuum theory to study the electronic bands introduced by moire patterns of twisted Janus transition metal dichalcogenides (TMD) homo- and hetero-bilayers.
The model includes lattice relaxation, stacking-dependent effective mass, and Rashba spin-orbit coupling.
We then perform a high-throughput generation and characterization of DFT-extracted continuum models for more than a hundred possible combinations of materials and stackings.
Our model predicts that the moiré physics and emergent symmetries depend on chemical composition, vertical layer orientation, and twist angle,
so that the minibands wavefunctions can form triangular, honeycomb, and Kagome networks.
Rashba spin-orbit effects, peculiar of these systems, can dominate the moir\'e bandwidth at small angles.
Our work enables the detailed investigation of Janus twisted heterostructures, allowing the discovery and control of novel electronic phenomena.
\end{abstract}

\date{\today}          
                         
\maketitle

\section{Introduction}

Strong interactions between particles are a central pillar of quantum physical phenomena, leading to many different phases of matter, such as superconductors and fractional quantum Hall states \cite{Ginzburg2004,Stormer1999}. These phenomena are investigated in condensed matter systems in the so-called quantum materials \cite{Keimer2017,Basov2017,Giustino2020}, where quantum effects dominate their properties. These properties can in turn be tuned through materials-engineering "knobs" that pertain to each type of material, such as chemical composition, structural changes, and external perturbations.

Recently, two-dimensional few-layer systems, such as graphene and transition metal dichalcogenides (TMDs), have proven to be one of the leading platforms to study many of these effects, owing to their low dimensionality.
The possibility to manipulate interlayer interactions of van der Waals (vdW) materials by changing the twist angle between two layers has been triggered by the breakthroughs in magic-angle graphene multilayers \cite{Herrero-1, Herrero-2, Lu2019, He2021, Yanko_science, Zhou2021, Park2021, Li_2018, Gadelha2021, Saito2020, Wong2020}.
In these systems, the relative twist angle results in a moiré superlattice, which acts as a long-wavelength modulating potential changing the electronic, vibrational, and structural properties \cite{ Andrei2021, Carr2020_review,Lau2022}.
Flat bands, non-trivial topology, emergent symmetries, enhanced correlations, and strong electron-phonon coupling can arise as a result of the charge localization induced by the moiré superlattice \cite{Bistritzer_PNAS,Carr2017_twistronics, Song_PRL, Fabrizio_PRB, Po_PRX, Fabrizio_PRX, Yoo2019, Bernevig_PRL, Blason2021}.\\
Indeed, recent experiments have moved beyond graphene. As an example, work on twisted TMDs quickly uncovered a plethora of novel phenomena such as correlated insulating states, quantum anomalous Hall states, superconductivity and moiré excitons \cite{Ghiotto_2021,Grzeszczyk_2021,Jie_2022,Tang_2020,Wang_2020, Tran_2019}.
Given the huge number of possible combinations to create these moiré materials \cite{Kennes2021, Tritsaris2021} and the experimental challenges in achieving them,  only a small subset of systems have been studied to date. %
Nonetheless, by creating moiré structures with different properties in their individual constituents, new physical properties are possible in the interacting system.

Transition-metal dichalcogenide layers with different chalcogen atoms on either side of the metal plane \cite{Rodrick_PRB}, referred to as "Janus" layers \cite{Zhang_2017}, have attracted intense interest due to the special properties introduced by the broken mirror symmetry of the corresponding pure layers \cite{Rodrick_PRB}.
Compared to $\text{MX}_2$ monolayers with $\text{D}_{3v}$ symmetry such as $\text{MoS}_2$, Janus
group-VI chalcogenides $\text{MXY}$ ($\text{X}, \text{Y} = \text{S, Se, Te};\;\; \text{X} \neq \text{Y}$) are non-centrosymmetric compounds with $\text{C}_{3v}$ symmetry \cite{Li_2018}. 
The difference in electronegativity between the two chalcogen layers leads to a built-in electric dipole oriented in the out-of-plane direction, generating strong Rashba spin–orbit couplings (SOC). 
Furthermore, the intrinsic electric field provides an additional degree of freedom with which to tune the van der Waals interaction between adjacent layers.
As such, these systems present new ingredients and properties to the already known tuning knobs available for probing moiré-driven phenomena.

In this paper, we present results from an $ab \; initio$ continuum model designed for twisted Janus TMDs. 
The model extends the one-band continuum Hamiltonian of Refs.~\cite{Wu_PRL, Angeli_PNAS} to better capture the rich physics of polar TMDs.
A set of inexpensive bilayer calculations is used to compute the electronic and structural potentials which are then used to include lattice relaxation and compute the bandstructure.
The effective mass approximation of the band edges is relaxed by the inclusion of a moiré-dependent term.
The flat bands of these systems are thoroughly analyzed and characterized in terms of the symmetries and centers of the corresponding Wannier orbitals.
Finally, we create a database of relevant physical parameters that can be used as a reference, to reproduce the continuum model bands of more than a hundred twisted bilayers, and also to guide experimental and theoretical efforts toward those systems which exhibit the most interesting physics.
Additionally, the models presented in this work can serve as a starting point to study the quantum phases of twisted Janus materials with Hartree-Fock \cite{HF1, HF2, Blason2021, HF3, HF4, Liu_PRR, HF5, HF6, HF7} and exact diagonalization \cite{Duran_PRB, Duran_PRL} techniques, or to derive a small set of Wannier orbitals \cite{Carr_wannier} used in DMRG calculations that treat strong correlations \cite{Kennes_PRB, Xian2021, Kennes2020, Claassen2022, Xian2019, Zaletel_PRB}.

The work is organized as follows: in Sec. \ref{sec:res} we start by briefly introducing the geometrical features of twisted bilayers. 
In Sec.~\ref{sec:res:ham} we focus on the continuum model Hamiltonian and the methodology employed to obtain its potentials from $ab \;initio$ calculations.  
In Sec. \ref{sec:res:homo} we discuss a set of twisted bilayers in which Rashba SOC is negligible and the moir\'e bands develop emergent symmetries.
In Sec. \ref{sec:res:hetero} we focus on those bilayers in which Rashba effects are important. The last section (Sec. \ref{sec:discuss}) is devoted to discussing some of the exotic physics that could be observed in these systems.
The parameters that characterize the continuum models of more than a hundred of twisted bilayers are listed in Tables~\ref{tab:homo} and \ref{tab:hetero}.

\section{Results} \label{sec:res}

\begin{figure}[h!]
\begin{center}
\includegraphics[width=\linewidth]{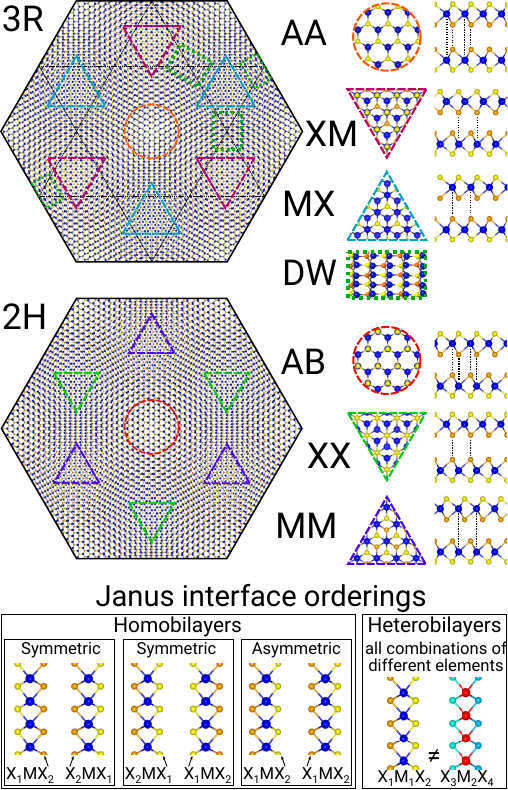}
\caption{Moir\'e superlattices created by twisting TMDs bilayers and possible interface orderings of Janus TMDs. Two different configurations are possible, 3R or 2H. From an aligned reference, these two configurations differ by a 180° rotation of one layer. 
The high-symmetry stacking regions of each superlattice are highlighted with dashed circles and triangles, together with the corresponding top and side views.\\
In 3R twisted bilayers, perfect layer alignment is found in the AA regions.
MX/XM are Bernal-stacked areas in which a metal atom is on top/below a pair of chalcogens. These regions form a hexagonal pattern.
At the interface between Bernal domains lie the domain wall (DW) regions. Three of these areas are highlighted in dotted green. As pointed out by black lines, the DWs form a Kagome network.
In 2H twisted bilayers, the horizontal displacement of two layers overlaps in the AB regions where metal and chalcogens lie on top of each other. In the XX/MM Bernal domains, only the chalcogen/metal layers align.}
\label{fig:superlattice}
\end{center}
\end{figure}
Twisted heterostructures between TMD monolayers,
occur in two distinct configurations \cite{Weston_Nature}, known as 2H and 3R. 
The two configurations differ by a $180^\circ$ rotation of the top layer around a vertical axis passing through the position of the transition metal atom, and are not related by any shift of one layer with respect to the other.
In Fig.~\ref{fig:superlattice} we show 3R and 2H twisted bilayers.
In the 3R configuration, the perfectly aligned areas called AA are 
surrounded by six Bernal-stacked regions (MX and XM) that form a honeycomb network. 
In the MX/XM areas, a metal atom (M) on one layer is directly on top/below a chalcogen atom (X) on the other layer.
The two regions are related by a reflection that exchanges the two layers. 
In 2H bilayers, metal/chalcogen atoms are on top of each other in the MM/XX Bernal stacked regions.  Vertical alignment between the layers is obtained in the AB regions, where different atoms are on top of each other.
In both configurations, the interface regions between Bernal domains are the so-called domain walls (DWs) which form a Kagome network.\\
In general, the geometry of the moir\'e pattern is characterized by the presence of high-symmetry regions orderly distributed to form hexagonal Bravais lattices \cite{nowick_1995}.
The AA/AB regions form a triangular network while two adjacent MX/XM or XX/MM regions are the basis of a honeycomb lattice. Finally, the DWs centers correspond to the three-site basis of a Kagome lattice.\\

\begin{figure*}
\includegraphics[width=0.9\textwidth]{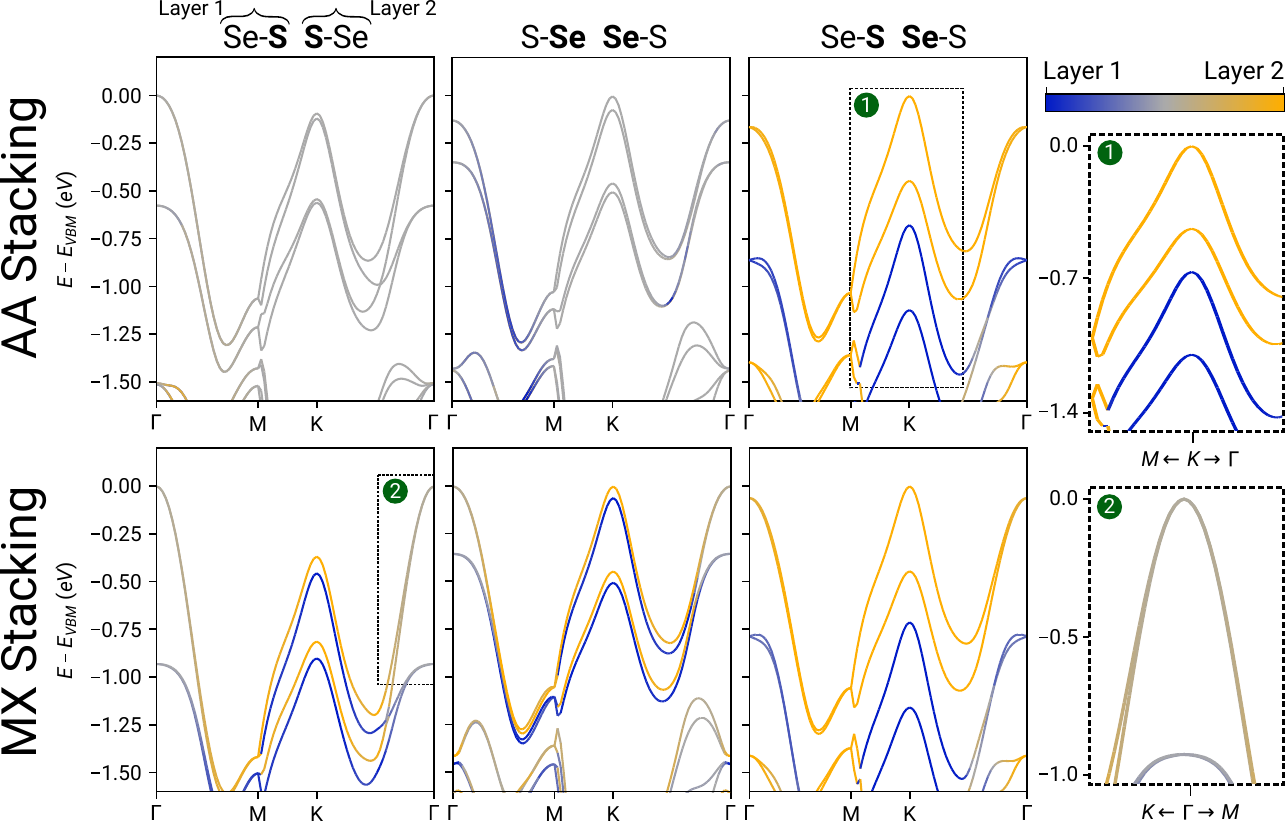}
\caption{Homobilayers band structures of WSSe at different stackings, projected on each monolayer (blue to orange). The energy is shifted to the valence band maximum of each configuration.
Inset 1 highlights the K-valley of an asymmetrically ordered bilayer (Se-S-Se-S), presenting both maximal layer polarization and strong spin-orbit spin-splitting. 
Inset 2 highlights the $\Gamma$-valley of a symmetrically ordered bilayer (Se-S-S-Se), where strong interlayer interaction especially between $p_z$ and $d_{z^2}$ orbitals leads to a large band splitting, isolating the valence band maximum (antibonding state) from other states.}
\label{fig:bilayer_bands}
\end{figure*}

Unlike in non-polar TMDs, the electronic properties of Janus bilayers depend on how the chalcogen layers are vertically ordered \cite{Zhou_PRB}. This ordering will determine the interface and interlayer interactions between the layers, dictating where the valence band maximum and conduction band minimum (VBM/CBM) are located in momentum space. \\
As an example, in Fig.~\ref{fig:bilayer_bands} the bandstructure of AA and MX stacked WSSe is shown for the three possible orderings.
In WSSe bilayers with the Se-S-Se-S interface configuration, named only by the chalcogen planes, there is an obvious layer polarization caused by the intrinsic dipole moment of the layers and the asymmetry of the order. At the interface, the potential energy on the side of the S atomic plane is larger, so the energy of the top layer is larger than in the other layer. 
Since the layers are weakly coupled, this results in the bottom (top) layer contributing to the CBM (VBM).\\
On the contrary, in bilayers with symmetric orders (S-Se-Se-S and Se-S-S-Se), the effective dipole in each layer points in opposite directions and the VBM/CBM combine states from both layers.
Therefore, the electronic structure of twisted Janus TMDs bilayers and their consequent physical properties will sensitively depend not only on the chemical compositions, but also on the interface orderings.

\subsection{Twisted bilayers moir\'e Hamiltonian} \label{sec:res:ham}

We derive the moiré Hamiltonian from first principles following the approach outlined in Ref.~\cite{Jung_PRB}. The procedure requires
a relatively inexpensive set of bilayer untwisted calculations.
These calculations are intended to extract how the Bloch states near a band edge are affected by the misalignment between the layers. This information is then used to derive a continuum model Hamiltonian which targets the moir\'e bands that form in twisted bilayers.
This model is further augmented to include lattice relaxation \cite{Carr_PRB}, an effective mass correction, and Rashba spin-orbit coupling terms.\\\\

For twisted semiconductors, the moiré bands closest to the band gap can be derived from the untwisted parabolic band extrema (or valleys) in the original Brillouin zone. 
In particular, in Janus bilayers, we can identify two distinct situations.\\
In the first one, when the two layers are dissimilar or weakly coupled, the moir\'e physics derives mostly from one layer only. This condition is not relevant only for heterobilayers, but also for asymmetrically ordered homobilayers (X-Y-X-Y). In this case, the bilayer's effective dipole shifts in energy each layer's band edge, so that the VBM/CBM are mostly layer polarized states.\\
A different second situation, is that of symmetrically ordered homobilayers (X-Y-Y-X or Y-X-X-Y) where the system recovers the mirror symmetry that relates the chalcogen planes, so that the layer's effective dipoles cancel out.  This further implies they are $\text{D}_3$ centrosymmetric structures, in which Rashba spin splittings are forbidden by symmetry.
Therefore, the main differences between asymmetrically ordered homobilayers or heterobilayers and symmetrically ordered homobilayers are the presence or not of Rashba effects and the degree of layer polarization.\\

In momentum space, the valence band maximum is located either at the Brillouin zone corners (K/K') or at $\Gamma$.
In the former case with the VBM at K/K', the strong Ising spin-orbit splitting isolates in energy one state per valley, realizing a situation similar to that of non-polar TMD heterobilayers \cite{Wu_PRL}.\\
In the latter case with the VBM at $\Gamma$, the interlayer interaction is particularly important as out of plane $d_{z^2} / p_z$ orbitals are involved at the band edge. This leads to a large band splitting where the VBM is the layer antibonding state that, depending on the degree of layer polarization, is separated by its bonding counterpart by several tens or hundreds of meV.
When Rashba SOC vanishes due to symmetry, the physics is then similar to that of non-polar TMD homobilayers \cite{Angeli_PNAS} where the $\Gamma$ band edge is spin-degenerate. \\
The conduction band minimum is, either at K/K' or the Q points \cite{TMDs_falko}. The physics of the moir\'e bands forming around the Q points will be discussed in future work.\\

\begin{figure*}
\begin{center}
\includegraphics[width=1.0\textwidth]{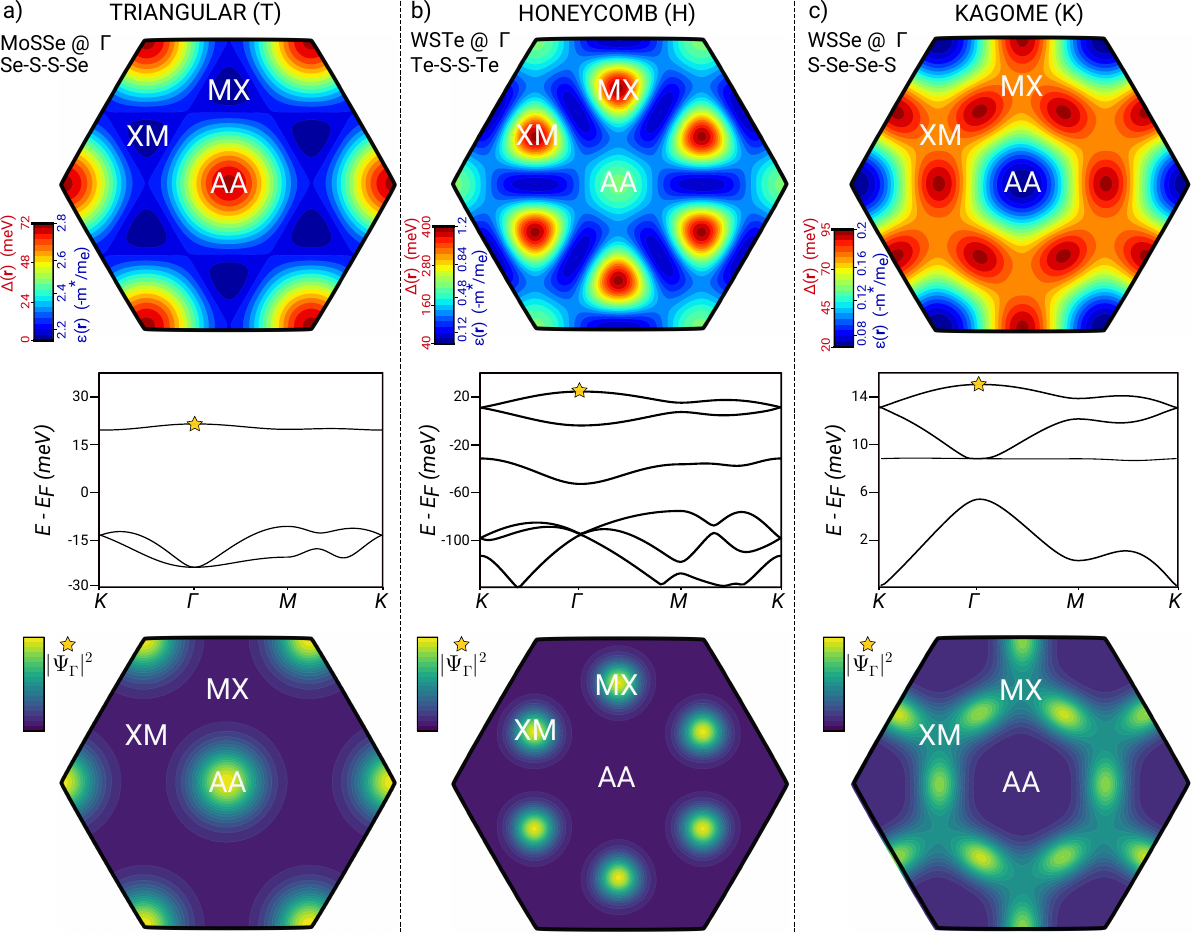}
\caption{Continuum model moiré and effective mass potentials, bandstructure, and charge density distribution are shown for three representative examples, each displaying the different possible physics of twisted Janus TMDs. 
The examples are (a) 3R MoSSe, (b) 3R WSTe, and (c) 3R WSSe twisted homobilayers with twist angles $4.0^\circ$, $3.0^\circ$ and $1.5^\circ$, respectively. The moiré $\Delta (\textbf{r})$ and effective mass potential $\varepsilon(\textbf{r})$ follow the same real space distribution (encoded in colors).  The charge distribution of the topmost Bloch state (denoted by a star) is shown. The charge is localized around the highly-symmetric moiré potential maxima that form, in each example, triangular, honeycomb, and Kagome networks.
In the last column of Tables~\ref{tab:homo} and \ref{tab:hetero}, we present the characterization of each system in terms of which moiré superlattice it forms, along with the corresponding continuum model parameters.
}
\label{fig:homo}
\end{center}
\end{figure*}

\vspace{0.5cm}
\subsubsection{Continuum model Hamiltonian}
The low energy physics of twisted hetero/homobilayers can be modeled by means of a low energy continuum model Hamiltonian.
In both symmetrically and asymmetrically ordered twisted bilayers the conduction and valence band edges are isolated in energy by at least 50-100 meV.\\
The corresponding moir\'e physics is well described by the one-band continuum model Hamiltonian \cite{Wu_PRL, Angeli_PNAS}:
\begin{equation}
  \mathcal{H}= -\frac{\hbar^2 \textbf{k}^2}{2m^*_0} + \Delta(\textbf{r}) + \varepsilon_{m^*}(\textbf{r}) - \alpha_R (\boldsymbol{\sigma} \cross \textbf{k}) \cdot \hat{z} ,
   \label{moire_H} 
\end{equation}
where $m^*_0$ and $\alpha_R$ are the average (over all the stacking configurations) effective mass and Rashba parameter, $\Delta(\textbf{r})$ is the moiré potential, $\varepsilon_{m^*}(\textbf{r})$ is the effective mass correction, $\textbf{r}$ tracks the local stacking configuration, and $\boldsymbol{\sigma}$ are Pauli matrices. \\
$\Delta(\textbf{r})$ and $\varepsilon_{m^*}(\textbf{r})$ are scalar functions that capture the energy and effective mass of holes/electrons in the VBM/CBM as a function of the relative in-plane displacement $\textbf{r}$ between the two aligned layers.
To capture the stacking-dependent electronic  details necessary for the determination of these potentials, we
perform density functional theory (DFT) calculations on non-twisted bilayers. Accordingly, we refer to this model as a first principle effective Hamiltonian.
The DFT calculations are performed with the aim of sampling how the physics locally changes with stacking: the top layer is shifted relative to the bottom layer over a 9$\cross$9 grid in the unit cell.  \\
These calculations are performed with Quantum Espresso \cite{QE1,QE2} using the generalized gradient approximation with the weak vdW
forces acting between the layers taken into account by the nonlocal vdW functional vdW-DF2-C09 \cite{Hamada_PRB,Berland_2015}. All calculations include spin-orbit coupling.
For each displacement, the in-plane position of the atoms is fixed while the interlayer distance is relaxed.
The total energy, band edge energy, and effective mass are extracted for each bilayer stacking, and then used in the continuum model.
To find the effective mass and Rashba coupling, we fit the values obtained from the aligned bilayers band edges to the $\varepsilon \textbf{k}^2 + \alpha |\textbf{k}|$ function.
The quadratic term tracks the effective mass $\varepsilon$  while the linear term takes into account the Rashba spin splitting $\alpha$. In general, we found that, unlike for the effective mass, $\alpha$ is only weakly dependent on stacking. We therefore consider in the continuum model only the average (over stacking configurations) Rashba parameter $\alpha_R$.
In the Supplemental Material Fig. S1~\cite{SM}, we validate the continuum model against explicit large-scale DFT calculations and corroborate the good agreement near the band extremum.
\\

\subsubsection{Lattice relaxation}

To evaluate in-plane relaxation effects, we first compute the monolayers' bulk and shear moduli $\mathcal{G}$ and $\mathcal{K}$ (see Table~\ref{tab:strain}).
For the interlayer energy, we extract from the bilayers' DFT calculations their total energies, which are then combined in the generalized stacking
fault energy  functional (GSFE) $\Omega(\textbf{r})$ \cite{Carr_PRB, GSFE_1, GSFE_2}. The GSFE tracks how the interlayer energy depends on the relative stacking
between the layers. \\
Then, we minimize the total mechanical energy of the moiré pattern obtaining  the top/bottom layer relaxation displacements $\textbf{u}_{t/b}$. 
In general, the relaxation distorts the layers to maximize the area of the lowest-energy stacking between layers. This is achieved via small in-plane deformations characterized by strain fields that locally change the layer alignment \cite{Fabrizio_PRB, Carr_PRB}. Out-of-plane relaxation
is implicitly included at the level of the DFT calculations,
with each stacking allowed to move to its ideal interlayer
separation.\\
Finally, we incorporate the effect of atomic relaxations on the
electronic Hamiltonian of Eq.~\eqref{moire_H} through a linear mapping 
that updates the unrelaxed configuration $\textbf{r} \to \textbf{r} + \textbf{u}_t + \textbf{u}_b$ \cite{Tang_PRB}.\\
The relaxed GSFE of several twisted bilayers is shown in the Supplemental Material~\cite{SM}.
\begin{table}[h]
\caption{Bulk ($\mathcal{K}$) and shear moduli ($\mathcal{G}$), in meV per unit cell, of the six Janus TMDs monolayers considered.}
\begin{tabular}{c  c c c c c c  }
  & MoSSe & WSSe & MoSTe & WSTe & MoSeTe & WSeTe \\ \hline
 $\mathcal{K}$ & 49.7 & 45.8 & 39.7 & 42.7 & 41.7 & 39.1 \\ 
 $\mathcal{G}$ & 34.2 & 30.7 & 25.4 &  29.6 & 27.7 & 27.7\\  \hline
\end{tabular}
\label{tab:strain}
\end{table}

\subsubsection{Fourier expansion}
The two-dimensional periodicity of the aligned bilayers implies that the potentials $\Delta(\textbf{r})$,  $\varepsilon_{m^*}(\textbf{r})$, and $\Omega (\textbf{r})$ are periodic functions that can be Fourier transformed in momentum space.
All the information relative to the moir\'e pattern, from its structural relaxation to its effect on the bilayer effective masses, is therefore conveniently stored in few Fourier coefficients.\\
The number of these coefficients can be reduced by taking into account symmetry.
Three-fold rotations with respect to the out-of-plane z-axis ($\text{C}_{3z}$) require that the potentials and their Fourier transforms are invariant under $120^\circ$ rotations.
This, and the fact that the potentials are real functions, reduces the number of independent coefficients to two: an amplitude and phase $\phi$  \cite{Wu_PRL, Angeli_PNAS}.\\ 
Therefore, the potentials can be expressed as follows:
\begin{subequations}
  \begin{align}
      \Omega(\textbf{r})=\sum_{l=1}^3 \sum_{j=1}^6 W_l \exp(i \textbf{g}_j^l\cdot\textbf{r} + \phi_W^l), \label{eq:omega}\\
    \Delta(\textbf{r})=\sum_{l=1}^3 \sum_{j=1}^6 V_l \exp(i \textbf{g}_j^l\cdot\textbf{r} + \phi_V^l), \label{eq:delta}\\
    \varepsilon(\textbf{r})=\sum_{j=1}^6 U_1 \exp(i \textbf{g}_j^1\cdot\textbf{r} + \phi_U^1), 
  \end{align}
 \label{moire_pot}
\end{subequations}
where $\textbf{g}_{j+1}^l=\text{C}_{6z} \textbf{g}_j^l$ is the $l$-th shell of six moir\'e $\textbf{g}$ vectors ordered with increasing $|\textbf{g}|$. Unlike for the effective mass correction, the moir\'e potential and GSFE have to be expanded up to the third momentum shell to accurately capture the moir\'e physics. \\

\subsection{Moiré bands without Rashba effects: symmetric homobilayers and emergent symmetries} \label{sec:res:homo}

We start by focusing on symmetrically ordered homobilayers (X-Y-Y-X or Y-X-X-Y). In these cases, the system recovers the mirror symmetry that relates the chalcogen planes. Being centrosymmetric structures, Rashba spin splittings are forbidden by symmetry.\\
We consider a simplified version of the Hamiltonian in Eq.~\ref{moire_H} and solve for the Bloch states by expanding in plane waves:

\begin{equation}
    \bra{\textbf{k}+\textbf{g}'}\mathcal{H}\ket{\textbf{k}+\textbf{g}}=-\delta_{\textbf{g},\textbf{g}'}\frac{\hbar^2|\textbf{k+g}|^2}{2m_0^*} + \Delta(\textbf{g}-\textbf{g}') + \varepsilon(\textbf{g}-\textbf{g}'),
    \label{moiré_H_bands}
\end{equation}
where $\textbf{k}$ is a wavevector in the moir\'e Brillouin-zone.
We then use the continuum model to calculate the bandstructure of twisted Janus TMDs at angles where full microscopic calculations are prohibitive.\\

\subsubsection{Emergent symmetries}
An interesting feature of moiré materials is that the potentials in Eq.~\eqref{moire_pot} can develop symmetries additional to those of the underlying twisted lattice, with profound consequences on the electronic and elastic properties of the system \cite{Senthil_PRB, Fabrizio_PRB, Angeli_PNAS, Fabrizio_PRX, Fabrizio_EPJ}.
In symmetrically ordered 3R bilayers with $\text{D}_3$ symmetry, two bilayers stacked by $\textbf{r}$ and $\text{C}_{2z}\textbf{r}=-\textbf{r}$
are mapped into each other by a $z\leftrightarrow-z$ mirror operation and hence have the same band and total energies. This property, which is peculiar to 3R homobilayers, further implies that $\Delta(\textbf{r})=\Delta(\text{C}_{2z}\textbf{r})$, $\varepsilon_{m^*}(\textbf{r})=\varepsilon_{m^*}(\text{C}_{2z}\textbf{r})$, and $\Omega(\textbf{r})=\Omega (\text{C}_{2z}(\textbf{r}))$, {\it i.e.}, that the potentials are inversion-symmetric functions. In terms of their expansion in momentum space, this constraints the phases $\phi$ in Eq.\eqref{moire_pot} to be either $0$ or $\pi$ \cite{Angeli_PNAS}. \\
The moir\'e Hamiltonian in Eq.~\eqref{moire_H} and the GSFE inherit this additional symmetry. This, combined with the three-fold rotations, promotes them to be six-fold ($\text{D}_6$) symmetric functions. 
This allows the moir\'e bandstructure to have distinct features, such as Dirac nodes, which would otherwise be forbidden by symmetry.\\
It is important to stress that the moir\'e Hamiltonian in Eq.~\eqref{moire_H} and also the GSFE develop the emergent $\text{C}_{2z}$ symmetry of the group $\text{D}_6$ despite the lattice having only three-fold $\text{D}_3$ symmetry.\\

\subsubsection{Bandstructure symmetry analysis}
We employ the Topological Quantum Chemistry theory \cite{Bernevig_Nature} to identify the symmetries and centers of the Wannier orbitals underlying the moir\'e bands.
This procedure involves computing the symmetry of the Bloch states, classifying them in terms of the irreducible representations (irreps) of the little groups at the corresponding high symmetry points, and then comparing the list of irreps with the Elementary Band Representations (EBR) listed on the Bilbao Crystallographic server \cite{Bilbao_server}. The centers of the Wannier orbitals for all the twisted bilayers' considered are listed in Tables~\ref{tab:homo} and ~\ref{tab:hetero}.\\

In Fig.~\ref{fig:homo}, the valence band structure of three different twisted bilayers is shown along with their moir\'e/effective mass potentials and charge distribution.
They are three representative examples of the different physics that can occur in twisted Janus TMDs.
As can be seen, the maxima of the potentials are located around high symmetry regions of the moiré pattern. As a consequence, the charge tends to localize around these maxima, as shown by the charge distribution of the topmost moir\'e bands.\\
Thus, the moir\'e pattern acts as a long-wavelength confining potential, localizing holes/electrons around triangular, honeycomb, and Kagome lattice sites.
This behavior can be understood ~\cite{Wu_PRL, Angeli_PNAS, Carr_PRB} by making a harmonic approximation to the moir\'e potential 
near its minima/maxima, thus identifying the harmonic oscillators wavefunctions as Wannier functions. 
The Bloch states symmetry analysis agrees with this picture. The Wannier orbitals generating the topmost set of moir\'e bands transform as a fully symmetric $s$ state while the set of bands below is spanned by $p_x \pm i p_y$.\\
Thus, Fig.~\ref{fig:homo} summarizes the richness of the physics found in TMDs Janus homobilayers, and in Table~\ref{tab:homo} we present  the characterization of each system in terms of which moiré superlattice it forms, along with the corresponding continuum model parameters.
Depending on valley, chalcogen ordering, 3R/2H twist configuration, and chemical composition, the system realizes a platform to simulate triangular, honeycomb, and Kagome flat bands, respectively shown in Fig.~\ref{fig:homo}(a-c).
The first case is usual, and commonly known for twisted hexagonal systems including TMDs.
The last two cases are a peculiarity of 3R symmetrically ordered twisted bilayers and the emergent six-fold symmetry of the continuum model Hamiltonian.
In particular, in Fig.~\ref{fig:homo}(b), the charge localizes around the honeycomb network formed by the MX/XM regions. The topmost set of bands are non-degenerate at $\Gamma$, form a Dirac node at K, and are topologically equivalent to the $\pi$ bands of graphene.
The bands in Fig.~\ref{fig:homo}(c) are even more intriguing as they exhibit an extremely narrow band degenerate at $\Gamma$ and Dirac nodes.
These bands are those of the geometrically frustrated Kagome lattice, an exotic situation that can host interesting correlated phenomena such as magnetic frustration and spin liquid physics \cite{SpinLiquid,Quantum_spin_liquids}.
The charge distribution, which is localized around the Kagome network formed by the domain walls centers, further confirms this picture.

\subsection{Moiré bands with Rashba effects: asymmetric homobilayers and heterobilayers} \label{sec:res:hetero}

\begin{figure}
\begin{center}
\includegraphics[width=0.9\linewidth]{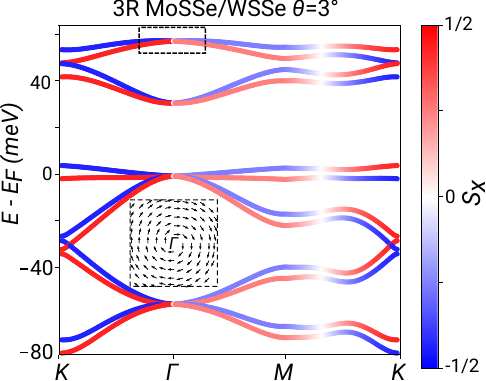}
\caption{$\Gamma$-valley moiré valence bands of 3R MoSSe/WSSe with Se-S-S-Se ordering, at $\theta=3^\circ$. The $\text{S}_x$ spin component is encoded in colors. The in-plane spin texture of the Bloch states around $\Gamma$ in the topmost moiré band is shown in the inset. The spin distribution in momentum space follows a spiral pattern winding around $\Gamma$.}
\label{fig:hetero}
\end{center}
\end{figure}

We now discuss the physics of heterobilayers and asymmetrically ordered homobilayers.
In these cases, due to the dissimilarity between the layers or chalcogen ordering, the system is non-centrosymmetric.
This further implies that Rashba SOC is not prevented to arise around the valence band edge at $\Gamma$.
By expanding in plane waves the full Hamiltonian in Eq.~\ref{moire_H}, we obtain:
\begin{align} \label{H_soc_bands}
    \bra{\textbf{k}+\textbf{g},s}\mathcal{H}\ket{\textbf{k}+\textbf{g}',s'}= &
    -\delta_{\textbf{g},\textbf{g}'} \delta_{\textbf{s},\textbf{s}'} \frac{\hbar^2|\textbf{k+g}|^2}{2m_0^*} \nonumber \\
    &+ \delta_{\textbf{s},\textbf{s}'} \Delta(\textbf{g}-\textbf{g}') \nonumber \\
    &+ \delta_{\textbf{s},\textbf{s}'} \varepsilon(\textbf{g}-\textbf{g}') \nonumber \\
    &+ \delta_{\textbf{g},\textbf{g}'} \alpha_R[ {\scriptstyle (k_y+g_y) \sigma_x - (k_x+g_x) \sigma_y }], \nonumber \\ %
\end{align}
where $s$ is the spin index, $\sigma_{x,y,z}$ are Pauli matrices, and otherwise the same notation as before.\\

In Fig.~\ref{fig:hetero} we show the spin-resolved bandstructure of a MoSSe/WSSe heterobilayer twisted at 3 degrees (more examples are shown in the Supplemental Material~\cite{SM}).
Given the 3R configuration and the symmetry of chalcogen layers' ordering, the bandstructure resembles that of the honeycomb network moir\'e materials (Fig.~\ref{fig:homo}b), with the first set of bands similar to that of graphene. The second set of 4 bands resembles that of $p_x \pm ip_y$ orbitals on a honeycomb lattice. 
However, in heterobilayers, the fourfold Dirac degeneracy at the K points is lifted.
First of all, the moir\'e potential is not six-fold symmetric because the MX regions, with a molybdenum atom on top of a pair of chalcogens, has a slightly different energy than the XM regions, where the chalcogens are on top of a tungsten atom. The imbalance between the MX/XM honeycomb lattice sites corresponds to a mass term in the Dirac Hamiltonian \cite{graphene_review}.
On top of this, the Rashba spin-orbit coupling further spoils the Dirac nodes by lifting the spin degeneracy.
The spin texture distribution in momentum space follows a spiral pattern winding around $\Gamma$.\\
The continuum model parameters for the Janus TMDs heterobilayers, all having triangular moiré superlattices, are listed in Table~\ref{tab:hetero}, while the parameters for the asymmetric homobilayers are part of Table~\ref{tab:homo}.
In particular, the Rashba parameters are of the order of tens or a few hundred meV/\AA{}, greatly exceeding the moir\'e bandwidth at small angles. 
Therefore, in the small angle regime, Rashba effects are likely to affect, if not dominate, the correlated physics of these systems.

\section{Discussion} \label{sec:discuss}
In conclusion, we have presented an extended continuum model able to accurately calculate the electronic structure of twisted Janus TMDs.
The model features terms that properly take into account lattice relaxation, the stacking dependence of the effective mass, and Rashba spin-orbit coupling.
The application of this methodology to more than a hundred twisted bilayers reveals the richness of their physics, and that it can be carefully tuned by choosing atomic composition, stacking structure, interface ordering, and twist angle.\\
In symmetrically ordered homobilayers, the layers' inversion symmetry prevents Rashba spin splittings and allows for the emergence of symmetries and Dirac nodes in the moir\'e bands.
Thus, they are able to realize not only common triangular moiré superlattices, but also honeycomb and Kagome networks \cite{Morell2019}, potentially displaying rich correlated-electronic phenomena and enabling their use as a platform of quantum simulators \cite{Kennes2021}.
The honeycomb bands that form in most 3R twisted homobilayers provide a convenient realization of artificial graphene, but with a fine-structure constant that can be tuned at will simply by varying twist angle \cite{Chiral}. 
The Kagome lattice models of 3R WSSe and MoSSe are expected to host different phases, including spin-liquids ~\cite{SpinLiquid, Quantum_spin_liquids}.  
Furthermore, by manipulating spin-orbit coupling with gate electric fields or by proximity effects, these bands can exhibit topologically non-trivial states \cite{Tang,Iimura,Kane,Wang}.
Since all systems have bandwidths that can be adjusted simply by varying twist angles, they provide an enticing toolbox to study the exotic properties of strongly correlated physics in hexagonal Bravais lattices \cite{WU2,WU3,Gregory,Natori,Kiesel}. \\
In heterobilayers and asymmetrically ordered homobilayers, Rashba spin-orbit coupling constants can greatly exceed the systems' bandwidth, therefore being expected to have an important impact on the correlated states. 
Indeed, this situation could provide opportunities for observing non-trivial phenomena such as the intrinsic spin Hall effect \cite{Sinova_PRL}, non-centrosymmetric superconductivity \cite{Bauer_PRL}, and Majorana fermions \cite{Sato_PRB}.
Furthermore, the presence of a magnetic substrate can strongly impact the non-coplanar magnetic texture of the moiré unit cell \cite{Soriano_2021}, providing an additional degree of freedom to tune the system's properties. \\
Additionally, light-matter interactions can have a significant impact on the bilayer physics \cite{PRB_Topp, PRR_Topp, Latini2019, PRR_Babak}. This is because the quenching of the kinetic energy of flatband electrons might allow one to reach the regime of strong light-matter interactions more easily than in dispersive materials.\\
Finally, with the data employed in the continuum model calculations and shown in Tables \ref{tab:homo} and \ref{tab:hetero}, we present a step toward the creation of a twistronics database, containing data from different systems and able to calculate properties for any, including arbitrarily small, twist angles. This information can serve as a basis for guiding both theoretical and experimental studies, relevant to different research strategies, such as direct and inverse design, and artificial intelligence investigations \cite{Tritsaris2021,ML2021}. 
We expect our results to have implications for the theoretical and experimental study of correlated physics in moir\'e materials, which could pave the way for the use of these systems for magnetism, spintronics, and other nanoscale devices.

\newcolumntype{H}{>{\setbox0=\hbox\bgroup}c<{\egroup}@{}} %
{\makeatletter
	\renewcommand\table@hook{\footnotesize}
	\makeatother
\begin{longtable*}{p{0.5cm}p{1.75cm}Hccccccccc} 
\caption{ Homobilayer continuum model parameters.
The first line for each material corresponds to the valence band edge (VBM), which occurs at $\Gamma$, and the second line corresponds to the conduction band edge (CBM), which occurs at $K$. The exceptions to this rule are marked by an asterisk, in which case both
the VBM and CBM occur at $K$.
The GSFE ($\Omega$) and moiré potential ($\Delta$) fourier coefficients (modulus and phase $\phi$) are listed up to the third moiré G-vectors shell.
The effective mass ($m_0^*$) and potential ($\varepsilon$) fourier coeffient are in bare electron mass units.
The Rashba coefficient ($\alpha_R$) is in meV/\AA{}.
The moiré superlattice ($MSL$) formed by the topmost/lowest valence/conduction band's localized charge distribution is listed in the last column, where T stands for triangular, H for honeycomb, and K for Kagome.}
\label{tab:homo} \\
\rotatebox{60}{Material} & Stack ordering & Band Edge & $W_{1,2,3}$& $\phi_W^{1,2,3}$ &$V_{1,2,3}$ & $\phi_V^{1,2,3}$ & $m_0^*$ &$U_1$ & $\phi_U^1$ & $\alpha_R$ & MSL \\ \hline

\multirow{9}{*}{\begin{sideways}\colorbox{blue!5}{\hspace*{5mm} 3R MoSSe \hspace*{5mm}}\end{sideways}} & \multirow{2}{*}{ Se-S-Se-S $\bigg\vert$} & $\Gamma$ @ VB  &(8.7,0.2,0.1)& (1.0,0.2, 2.6)& (3.6,0.1,0.1) & (74.6,62.5,10.6) & -0.1 & 0.01 & 163.5& 66.3 &T @ MX \\
 &  & $K$ @ CB  &(8.7,0.2,0.1)& (1.0,0.2, 2.6)& (3.7,0.14,0.16) & (92.3,81.0,87.3) & 0.92 & 0.01 & 2.6& 0.0 & T @ MX\\\\

 & \multirow{2}{*}{ S-Se-Se-S $\bigg\vert$} & $\Gamma$ @ VB  &(7.4,1.0,0.1)& (0.0,180.0,180.0)& (8.7,0.1,1.7) & (180.0,-180.0,0.0) & -0.37 & 0.01 & 0.0& 0.0 & H @ MX/XM\\

 &  & $K$ @ CB  &(7.4,1.0,0.1)& (0.0,180.0,180.0)& (5.9,0.6,0.2) & (0.0,180.0,0.0) & 0.37 & 0.08 & 0.0& 0.0  &H @ MX/XM\\\\

 & \multirow{2}{*}{ Se-S-S-Se $\bigg\vert$} & $\Gamma$ @ VB  &(6.1,0.2,0.2)& (0.0,0.0, 0.0)& (7.8,0.2,0.3) & (0.0,0.0,0.0) & -0.4 & 0.01 & 0.0& 0.0 &T @ AA\\
 
 &  & $K$ @ CB  &(6.1,0.2,0.2)& (0.0,0.0, 0.0)& (0.5,1.1,0.51) & (0.0,180.0,0.0) & 0.65 & 0.01 & 180.0& 0.0 &H @ MX/XM\\\\

\multirow{9}{*}{\begin{sideways}\colorbox{orange!10}{\hspace*{5mm} 2H MoSSe \hspace*{5mm}}\end{sideways}} & \multirow{3}{*}{ Se-S-Se-S $\bigg\vert$} & $\Gamma$ @ VB  &(7.9,0.1,0.1)& (131.9, 1.2, 85.4)& (4.9,0.1,0.1) & (125.9,156.9,140.5) & -0.36 & 0.0 & 0.0& 16.6 &T @ MM\\

&  & $K$ @ CB  &(7.9,0.1,0.1)& (131.9, 1.2, 85.4)& (0.7,0.3,0.1) & (122.2,175.5,73.0) & 0.83& 0.01 & 144.0& 0.0 &T @ XX\\\\

  & \multirow{2}{*}{ S-Se-Se-S $\bigg\vert$} & $\Gamma$ @ VB  &(7.5,1.3,0.4)& (143.7, 180.0, 87.5)& (11.9,0.96,1.3) & (58.5,175.5,116.0) & -0.42 & 0.02 & 180.0& 0.0 &T @ AB\\

  &  &  $K$ @ CB  &(7.5,1.3,0.4)& (143.7, 180.0, 87.5)& (5.9,0.6,0.2) & (115.0,8.6,38.8) & 0.29 & 0.06 & 150.0& 0.0 &T @ XX\\\\

  & \multirow{3}{*}{ Se-S-S-Se $\bigg\vert$} & $\Gamma$ @ VB  &(8.9,1.4,0.2)& (129.2, 180.0, 46.7)& (35.0,14.5,7.7) & (60.5,0.8,128.0) & -0.62 & 0.08 & 127.6& 0.0&T @ XX\\
  
  &  &  $K$ @ CB  &(8.9,1.4,0.2)& (129.2, 180.0, 46.7)& (0.7,0.0,0.2) & (24.0,7.3,64.8) & 0.5 & 0.03 & 122.9& 0.0 &T @ MM\\\\

\multirow{9}{*}{\begin{sideways}\colorbox{blue!5}{\hspace*{5mm} 3R WSSe \hspace*{5mm}}\end{sideways}} & \multirow{2}{*}{ Se-S-Se-S\textbf{*}$\bigg\vert$} &  $K$ @ VB  &(8.3,0.2,0.1)& (0.0,0.0, -2.4)& (4.9,0.6,0.9) & (92.9,-162.7,-56.7) & -2.64 & 0.0 & 0.0& 0.0&T @ XM\\

 &  & $K$ @ CB  &(8.3,0.2,0.1)& (0.0,0.0, -2.4)& (3.8,0.2,0.2) & (-79.1,39.7,93.9) & 2.07 & 0.01 & -2.5& 0.0&T @ XM\\\\

 & \multirow{2}{*}{ S-Se-Se-S $\bigg\vert$} & $\Gamma$ @ VB  &(7.6,1.3,0.6)& (0.0,-180.0, -180.0)& (8.9,2.8,0.5) & (180.0,180.0,0.0) & -0.7 & 0.02 & 0.0& 0.0& K @ DW \\

 &  & $K$ @ CB  &(7.6,1.3,0.6)& (0.0,-180.0, -180.0)& (2.6,0.8,1.5) & (0.0,0.0,0.0) & 1.19 & 0.02 & 0.0& 0.0 & H @ XM/XM\\\\

 & \multirow{2}{*}{ Se-S-S-Se $\bigg\vert$} & $\Gamma$ @ VB  &(6.0,0.1,0.1)& (0.0,0.0,0.0)& (7.9,0.2,0.06) & (0.0,0.0,-120.0) & -0.82 & 0.01 & 0.0& 0.0&T @ AA \\

 &  & $K$ @ CB   &(6.0,0.1,0.1)& (0.0,0.0,0.0)& (2.4,2.6,1.3) & (0.0,180.0,0.0) & 1.9 & 0.01 & 180.0& 0.0 &H @ XM/XM\\\\

\multirow{9}{*}{\begin{sideways}\colorbox{orange!10}{\hspace*{5mm} 2H WSSe \hspace*{5mm}}\end{sideways}} & \multirow{2}{*}{ S-Se-S-Se\textbf{*}$\bigg\vert$} &   $K$ @ VB  &(8.3,0.2,0.1)& (0.0, 0.0, -2.0)& (0.7,0.4,0.3) & (-168.0,-71.0,142.5) & -2.63 & 0.0 & 0.0& 0.0& T @ XX\\

 &  & $K$ @ CB  &(8.3,0.2,0.1)& (0.0, 0.0, -2.0)& (1.2,0.1,0.2)& (-160.5,-66.1,-134.9)  & 2.1 & 0.01 & 115.0& 0.0&T @ AB\\\\

 & \multirow{2}{*}{ S-Se-Se-S $\bigg\vert$} & $\Gamma$ @ VB  &(7.2,0.1,0.7)& (148.1, -109.9, 60.5)& (3.6,1.2,1.5) & (93.4,178.9,29.7) & -0.72 & 0.03 & 125.0& 0.0& T @ M\\

 &  &  $K$ @ CB  &(7.2,0.1,0.7)& (148.1, -109.9, 60.5)& (2.8,0.8,0.2)& (113.4,5.0,-45.1)  & 1.3 & 0.05 & 132.3& 0.0 & T @ MM\\\\

 & \multirow{2}{*}{ Se-S-S-Se $\bigg\vert$} & $\Gamma$ @ VB  &(7.8,1.9,0.5)& (127.7, 180.0, 75.5)& (20.2,20.6,9.5) & (-66.6,0.0,-125.7) & -0.99 & 0.07 & 125.0& 0.0 &T @ XX\\

 &  &  $K$ @ CB  &(7.8,1.9,0.5)& (127.7, 180.0, 75.5)& (0.8,0.1,0.1)& (-23.7,164.1,60.0)  & 2.0 & 0.06 & 55.0& 0.0&T @ MM\\\\

\multirow{9}{*}{\begin{sideways}\colorbox{blue!5}{\hspace*{5mm} 3R MoSTe \hspace*{5mm}}\end{sideways}} & \multirow{2}{*}{ Te-S-Te-S $\bigg\vert$} & $\Gamma$ @ VB  &(10.9,1.8,0.5)& (-7.4,180.0, 177.4)& (4.9,0.7,0.5) & (-29.0,26.4,-171.9) & -0.2 & 0.0 & 0.0& 0.0& T @ AA \\

 &  & $K$ @ CB  &(10.9,1.8,0.5)& (-7.4,180.0, 177.4)&(3.7,0.1,0.1) & (92.1,-86.9,81.9) & 0.96 & 0.01 & -2.5& 0.0 &T @ MX\\\\

 & \multirow{2}{*}{ Te-S-S-Te $\bigg\vert$} & $\Gamma$ @ VB  &(6.4,0.2,0.2)& (0.0,0.0, 0.0)& (12.6,0.1,0.4) & (0.0,-67.1,2.2) & -1.0 & 0.02 & 0.0& 0.0 &T @ AA \\

 &  & $K$ @ CB  &(6.4,0.2,0.2)& (0.0,0.0, 0.0)& (0.5,1.1,0.5) & (0.0,180.0,0.0) & 0.68 & 0.01 & 180.0& 0.0 &H @ MX/XM\\\\

 & \multirow{2}{*}{ S-Te-Te-S $\bigg\vert$}& $\Gamma$ @ VB  &(9.0,1.6,0.3)& (0.0,180.0, 180.0)& (1.5,0.4,0.2) & (180.0,0.0,-165.2) & -0.19 & 0.01 & 180.0& 0.0&H @ MX/XM \\

 &  & $K$ @ CB  &(9.0,1.6,0.3)& (0.0,180.0, 180.0)& (4.3,1.6,0.2) & (0.0,180.0,0.0) & 0.1 & 0.08 & 0.0& 0.0& H @ MX/XM\\\\

\multirow{9}{*}{\begin{sideways}\colorbox{orange!10}{\hspace*{5mm} 2H MoSTe \hspace*{5mm}}\end{sideways}} & \multirow{3}{*}{ S-Te-S-Te\textbf{*}$\bigg\vert$} & $K$ @ VB &(9.9,2.4,1.4)& (136.2, 180.0, 53.6)& (5.2,0.4,0.2)& (-128.5,153.4,-137.9)  & -2.3 & 0.0 & 0.0& 0.0 &T @ MM\\

 &  &  $K$ @ CB &(9.9,2.4,1.4)& (136.2, 180.0, 53.6)& (4.8,4.3,2.0)& (114.5,0.0,-128.4)  & 1.2 & 0.1 & -107.7& 0.0 &T @ AA\\\\

 & \multirow{3}{*}{ Te-S-S-Te $\bigg\vert$} & $\Gamma$ @ VB  &(10.9,2.0,0.7)& (112.2, 180.0, 20.8)& (43.5,13.5,12.4) & (90.5,0.0,-138.9) & -1.49 & 0.17 & -104.4.0& 0.0 &T @ AB\\

 & &  $K$ @ CB  &(10.9,2.0,0.7)& (112.2, 180.0, 20.8)& (1.36,0.98,0.68)& (49.8,180.0,27.5)  & 1.07 & 0.01 & 40.7& 0.0 &T @ XX\\\\

 & \multirow{3}{*}{ S-Te-Te-S $\bigg\vert$} & $\Gamma$ @ VB  &(9.4,1.4,0.56)& (155.3, 180.0, 129.0)& (1.21,0.22,0.28) & (7.6,3.0,14.6) & -0.3 & 0.01 & 155.0& 0.0 &T @ MM\\

 &  &  $K$ @ CB  &(9.4,1.4,0.56)& (155.3, 180.0, 129.0)& (3.64,0.33,0.36)& (122.8,172.0,55.5)  & 0.13 & 0.09 & 178.9& 0.0 &T @ MM\\\\

\multirow{9}{*}{\begin{sideways}\colorbox{blue!5}{\hspace*{5mm} 3R WSTe \hspace*{5mm}}\end{sideways}} & \multirow{3}{*}{ Te-S-Te-S\textbf{*}$\bigg\vert$}  & $K$ @ VB  &(9.5,1.86,0.8)& (142.4,180.0, 114.7)& (1.1,0.65,1.1) & (152.4,-21.3,-64.8) & -2.02 & 0.01 & 40.0& 0.0&T @ XM\\

 &  & $K$ @ CB  &(9.5,1.86,0.8)& (142.4,180.0, 114.7)& (1.64,0.53,0.75) & (-143.1,9.0,75.5) & 0.96 & 0.01 & -2.5& 0.0&T @ AA\\\\

 & \multirow{3}{*}{ Te-S-S-Te $\bigg\vert$} & $\Gamma$ @ VB  &(9.6,2.8,2.0)& (0.0,180.0, 0.0)& (14.7,33.6,13.4) & (180.0,0.0,180.0) & -1.3 & 0.09 & 0.0& 0.0 &H @ MX/XM \\

 &  & $K$ @ CB  &(9.6,2.8,2.0)& (0.0,180.0, 0.0)& (1.1,2.2,1.4) & (0.0,180.0,0.0) & 1.52 & 0.0 & 0.0& 0.0 & H @ MX/XM\\\\

 & \multirow{3}{*}{ S-Te-Te-S $\bigg\vert$} & $\Gamma$ @ VB  &(8.6,1.1,0.16)& (0.0,180.0, 0.0)& (2.5,0.5,0.3) & (130.0,0.0,-135.2) & -0.3 & 0.01 & 180.0& 0.0 &T @ MX \\

 &  & $K$ @ CB &(8.6,1.1,0.16)& (0.0,180.0, 0.0)& (6.7,2.6,0.8) & (0.0,180.0,180.0) & 0.42 & 0.07 & 0.0& 0.0 &H @ MX/XM\\\\

\multirow{9}{*}{\begin{sideways}\colorbox{orange!10}{\hspace*{5mm} 2H WSTe \hspace*{5mm}}\end{sideways}} & \multirow{3}{*}{ S-Te-S-Te\textbf{*}$\bigg\vert$}  &  $K$ @ VB &(9.5,1.9,0.8)& (142.5, 180.0, 114.7)& (1.1,0.6,1.1)& (152.5,-21.0,-64.4) & -2.0 & 0.01 & 40.2& 0.0& T @ MM\\

 &  &  $K$ @ CB  &(9.5,1.9,0.8)& (142.5, 180.0, 114.7)& (1.6,0.5,0.75)& (-141.1,9.0,75.4)  & 1.6 & 0.02 & 112.7& 0.0&T @ AA\\\\

 & \multirow{2}{*}{ Te-S-S-Te $\bigg\vert$} & $\Gamma$ @ VB  &(9.3,1.3,0.2)& (116.2, 180.0, 0.0)& (37.7,8.9,2.9) & (-82.5,0.0,-171.1) & -1.52 & 0.13 & 110.4 & 0.0 &T @ XX\\

 &  &  $K$ @ CB    &(9.3,1.3,0.2)& (116.2, 180.0, 0.0)& (2.85, 0.4,0.2)& (-62.8,0.0,-107.5)  & 1.54 & 0.02 & 45.7& 0.0 &T @ MM\\\\

 & \multirow{2}{*}{ S-Te-Te-S $\bigg\vert$} & $\Gamma$ @ VB  &(9.4,1.4,0.6)& (155.3, 180.0, 129.0)& (0.2,0.4,0.2) & (140.6,24.0,14.6) & -0.4 & 0.02 & 70.0& 0.0 &T @ XX\\

 &   &  $K$ @ CB  &(9.4,1.4,0.6)& (155.3, 180.0, 129.0)& (6.7,0.3,0.3)& (148.8,-169.0,20.5)  & 0.36 & 0.12 & 176.9& 0.0 &T @ AB\\\\

\multirow{9}{*}{\begin{sideways}\colorbox{blue!5}{\hspace*{5mm} 3R WSeTe \hspace*{5mm}}\end{sideways}} & \multirow{2}{*}{ Te-Se-Te-Se $\bigg\vert$} & $\Gamma$ @ VB  &(11.8,1.7,1.25)& (2.4,180.0, 8.6)& (30.7,1.9,5.5) & (-177.0,-168.4,35.7) & -0.4 & 0.0 & 0.0& 0.0 &T @ MX \\

 & & $K$ @ CB  &(11.8,1.7,1.25)& (2.4,180.0, 8.6)& (1.6,0.5,0.8) & (-143.1,9.0,75.5) & 0.96 & 0.01 & -2.5& 0.0 &T @ MX\\\\

 & \multirow{2}{*}{ Te-Se-Se-Te $\bigg\vert$} & $\Gamma$ @ VB  &(9.6,2.8,2.0)& (0.0,180.0, 0.0)& (14.7,33.6,13.4) & (180.0,0.0,180.0) & -1.3 & 0.09 & 0.0& 0.0 &H @ MX/XM \\

 &  & $K$ @ CB  &(9.6,2.8,2.0)& (0.0,180.0, 0.0)& (1.1,2.2,1.36) & (0.0,180.0,0.0) & 1.52 & 0.0 & 0.0& 0.0 &H @ MX/XM\\\\

 & \multirow{3}{*}{ Se-Te-Te-Se $\bigg\vert$} & $\Gamma$ @ VB  &(11.2,0.3,0.4)& (0.0,180.0, 0.0)& (3.9,5.2,5.7) & (180.0,0.0,0.0) & -0.9 & 0.07 & 0.0& 0.0 &T @ AA \\

 &  & $K$ @ CB   &(11.2,0.3,0.4)& (0.0,180.0, 0.0)& (1.4,1.8,1.6) & (0.0,180.0,0.0) & 1.6 & 0.01 & 180.0& 0.0 &H @ MX/XM\\\\

\multirow{9}{*}{\begin{sideways}\colorbox{orange!10}{\hspace*{5mm} 2H WSeTe \hspace*{5mm}}\end{sideways}} & \multirow{3}{*}{ Se-Te-Se-Te $\bigg\vert$} & $\Gamma$ @ VB  &(10.3,0.8,1.0) & (151.2, -172.0, -103.7)& (24.5,4.7,6.0) & (-55.5,-6.0,-155.4) & -0.3 & 0.13 & 123.4& 107.4 & T @ XX\\

 &  &  $K$ @ CB   &(10.3,0.8,1.0) & (151.2, -172.0, -103.7)& (3.6,0.5,0.6)& (-78.9,-165.9,20.4)  & 1.7 & 0.03 & 94.7& 0.0 &T @ XX\\\\

 & \multirow{3}{*}{ Te-Se-Se-Te $\bigg\vert$} & $\Gamma$ @ VB  &(12.0,0.7,0.5)& (160.0, -167.0, -48.0)& (9.9,2.7,1.6) & (-9.0,34.5,90.0) & -0.3 & 0.03 & 70.0 & 0.0 &T @ AB\\

 &  &  $K$ @ CB  &(12.0,0.7,0.5)& (160.0, -167.0, -48.0)& (5.2, 0.4,0.3)& (-153.9,174.0,-58.5)  & 0.9 & 0.12 & 166.7& 0.0 &T @ AB\\\\

 & \multirow{3}{*}{ Se-Te-Te-Se $\bigg\vert$} & $\Gamma$ @ VB  &(9.1,0.7,0.2)& (128.3, 180.0, -43.0)& (21.1,3.4,1.4) & (-65.3,-177.0,-142.1) & -0.9 & 0.07 & 125.0& 0.0 &T @ XX\\

 &   &  $K$ @ CB  &(9.1,0.7,0.2)& (128.3, 180.0, -43.0)&  (22.5,23.3,11.0)& (-62.5,78.0,-107.5)  & 1.7 & 0.01 & 76.9& 0.0 &T @ MM\\\\

\multirow{9}{*}{\begin{sideways}\colorbox{blue!5}{\hspace*{5mm} 3R MoSeTe \hspace*{5mm}}\end{sideways}} & \multirow{3}{*}{ Te-Se-Te-Se $\bigg\vert$} & $\Gamma$ @ VB  &(11.4,2.1,1.3)& (-3.8,180.0, 9.6)& (55.7,6.1,5.4) & (177.7,-173.9,36.9) & -0.17 & 0.15 & 0.0& 70.3 &T @  MX \\

 &  & $K$ @ CB  &(11.4,2.1,1.3)& (-3.8,180.0, 9.6)& (22.4,8.5,3.8) & (19.1,2.0,1.1) & 0.4 & 0.07 & 3.5& 0.0& T @ MX\\\\

 & \multirow{3}{*}{ Te-Se-Se-Te $\bigg\vert$} & $\Gamma$ @ VB  &(11.8,0.1,1.0)& (0.0,0.0, 0.0)& (6.3,9.3,14.4) & (180.0,180.0,180.0) & -0.8 & 0.06 & 0.0& 0.0 &T@MX \\

 &  & $K$ @ CB   &(11.8,0.1,1.0)& (0.0,0.0, 0.0)& (0.8,0.6,0.7) & (0.0,180.0,0.0) & 3.9 & 0.0 & 0.0& 0.0 &H @ MX/XM\\\\

 & \multirow{3}{*}{ Se-Te-Te-Se $\bigg\vert$} & $\Gamma$ @ VB  &(8.9,1.1,0.2)& (0.0,180.0, 0.0)& (5.6,1.5,0.5) & (180.0,0.0,180.0) & -0.1 & 0.03 & 0.0& 0.0 &H @ MX/XM \\

 &  & $K$ @ CB    &(8.9,1.1,0.2)& (0.0,180.0, 0.0)& (1.75,0.6,0.8) & (0.0,0.0,0.0) & 4.6 & 0.02 & 180.0& 0.0 &T @AA\\\\

\multirow{9}{*}{\begin{sideways}\colorbox{orange!10}{\hspace*{5mm} 2H MoSeTe \hspace*{5mm}}\end{sideways}} & \multirow{3}{*}{ Se-Te-Se-Te $\bigg\vert$} & $\Gamma$ @ VB  &(11.5,0.7,0.6) & (146.2, -167.0, -61.9)& (32.5,4.0,3.9) & (-45.5,-178.0,105.4) & -0.02 & 0.09 & 143.4.0& 104.2 &T @ AB\\

  &  &  $K$ @ CB   &(11.5,0.7,0.6) & (146.2, -167.0, -61.9)& (18.1,9.2,4.4)& (130.1,1.7,-101.4)  & 4.5 & 0.03 & 145.7& 0.0&T @ XX\\\\

  & \multirow{3}{*}{ Te-Se-Se-Te $\bigg\vert$} & $\Gamma$ @ VB  &(11.9,0.0,0.9)& (0.0, 0.0, 0.0)& (6.5,9.5,14.5) & (180.0,180.0,180.0) & -0.8 & 0.06 & 0.0 & 0.0 &T @ MM\\

  &  &  $K$ @ CB  &(11.9,0.0,0.9)& (0.0, 0.0, 0.0)& (0.6, 0.4,0.1)& (-161.9,2.0,-138.5)  & 4.9 & 0.01 & -107.7& 0.0 &T @ MM\\\\

  & \multirow{3}{*}{ Se-Te-Te-Se $\bigg\vert$} & $\Gamma$ @ VB  &(10.6,0.9,0.2)& (157.7, 180.0, -53.0)& (7.9,1.6,0.5) & (13.0,2.0,-11.0) & -0.02 & 0.0 & 0.0& 0.0 &T @ AB\\

  &   &  K @ CB  &(10.6,0.9,0.2)& (157.7, 180.0, -53.0) &  (6.6,0.5,0.8)& (75.5,51.9,-41.5)  & 4.7 & 0.07 & 20.9& 0.0 & T @ AB\\\\

\end{longtable*}

}

\clearpage%
{\makeatletter
	\renewcommand\table@hook{\footnotesize} %
	\makeatother
\begin{longtable*}{p{0.5cm}p{1.75cm}Hccccccccccc}
\newpage
\caption{ Heterobilayer continuum model parameters.
All symbols have the same meaning as in Table \ref{tab:homo}.
For each material there are now three lines, corresponding to the values for the respective band edges: $\Gamma$ at the VBM, $K$ at the VBM, and $K$ at the CBM.
}
\label{tab:hetero} \\

\rotatebox{60}{Material} & Stack ordering & Band Edge & $W_{1,2,3}$& $\phi_W^{1,2,3}$ &$V_{1,2,3}$ & $\phi_V^{1,2,3}$ & $m_0^*$ &$U_1$ & $\phi_U^1$ & $\alpha_R$ & MSL \\ \hline

\multirow{10}{*}{\begin{sideways}\colorbox{blue!5}{\hspace*{8mm} 3R MoSSe/WSSe \hspace*{8mm}}\end{sideways}} & \multirow{3}{*}{ S-Se-S-Se $\bigg\vert$} & $\Gamma$ @ VB  &(7.4,1.5,0.6)& (-3.7,177.9, 54.6)& (5.3,6.6,2.8) & (121.3,3.8,-125.6) & -0.6 & 0.02 & -20.5& 25.4 & T @ XM \\

 &  & $K$ @ VB  &(7.4,1.5,0.6)& (-3.7,177.9, 54.6)& (3.7,0.6,0.4) & (104.6,166.0,-99.5) & -2.3 & 0.01 & -2.3& 0.0 &  T @ XM\\
 
 &  & $K$ @ CB  &(7.4,1.5,0.6)& (-3.7,177.9, 54.6)& (3.6,0.3,0.8) & (-106.2, -157.9, 61.5) & 2.1 & 0.02 & -173.0 & 0.0 &  T @ XM\\

\\

  & \multirow{3}{*}{ S-Se-Se-S $\bigg\vert$} &
 $\Gamma$ @ VB &(11.5,0.7,0.6) & (146.2, -167.0, -61.9)& ( 9.3 , 5.2 , 1.7 ) & ( -165.8 , 1.8 , -2.4 )  & -0.75 & 0.0 & 0.0& 60.8 &T @ MX\\

  & &  $K$ @ VB   &(11.5,0.7,0.6) & (146.2, -167.0, -61.9)& (4.0,0.5,1.0)& (-61.7,13.9,24.2) & -2.7 & 0.0 & 0.0 & 0.0&  T @ MX\\

  & &  $K$ @ VB   &(11.5,0.7,0.6) & (146.2, -167.0, -61.9)& (2.5,0.3,0.8)& (108.1,0.3,-23.0) & 1.6 & 0.03 & -12.9& 0.0& T @ MX\\

\\

  & \multirow{3}{*}{ Se-S-S-Se $\bigg\vert$}  &  $\Gamma$ @ VB  & (10.0,2.1,0.9) & (-0.7,180.0,179.2) 
& (38.3,14.8,11.1) & (-178.9,0.7,0.8) & -1.0 & 0.1 & 0.0& 44.5 & T @ MX\\

  &  &  $K$ @ VB   & (10.0,2.1,0.9) & (-0.7,180.0,179.2) 
  & (8.1,0.8,1.6)& (-59.9,180.0,140.4) & -2.68 & 0.0 & 0.0 & 0.0 & T @ MX\\

  &  &  $K$ @ CB   & (10.0,2.1,0.9) & (-0.7,180.0,179.2) 
  & (6.2,0.0,1.1)& (81.8,-164.0,-129.4) & 1.8 & 0.01 & 0.0 & 0.0 & T @ MX\\

\\

  & \multirow{3}{*}{ Se-S-Se-S $\bigg\vert$} &  $\Gamma$ @ VB  & ( 9.7 , 2.1 , 0.7 ) & ( -5.3 , -179.9 , 162.1 )
& ( 9.3 , 5.2 , 1.7 ) & ( -165.8 , 1.8 , -2.4 )  & -0.75 & 0.0 & 0.0& 60.8 & T @ MX\\

  &  &  $K$ @ VB   & ( 9.7 , 2.1 , 0.7 ) & ( -5.3 , -179.9 , 162.1 )
& (8.0,2.4,1.9)& (-54.5,180.0,141.4) & -2.7 & 0.0 & 0.0& 0.0 &  T @ MX\\

  &  &  $K$ @ CB   & ( 9.7 , 2.1 , 0.7 ) & ( -5.3 , -179.9 , 162.1 )
& (6.0,1.1,1.0)& (113.4,-3.9,-90.4) & 1.8 & 0.01 & -4.4& 0.0 &  T @ MX\\

\\

\multirow{12}{*}{\begin{sideways}\colorbox{orange!10}{\hspace*{8mm} 2H MoSSe/WSSe \hspace*{8mm}}\end{sideways}} & \multirow{3}{*}{ S-Se-S-Se $\bigg\vert$} &
$\Gamma$ @ VB  &( 4.8 , 0.1 , 0.1 ) & ( 132.1 , 0.5 , -78.1 )
& ( 2.8 , 0.1 , 0.2 ) & ( 116.9 , 22.2 , -102.9 ) & -0.55 & 0.0 & 0.0& 29.8 & T @ XX \\

 &  & $K$ @ VB   &( 4.8 , 0.1 , 0.1 ) & ( 132.1 , 0.5 , -78.1 )
 & ( 0.8 , 0.1 , 0.2 ) & ( -26.4 , 97.8 , -70.6 )& -2.28 & 0.0 & 0.0& 0.0 & T @ AB\\
 
&  & $K$ @ CB   &( 4.8 , 0.1 , 0.1 ) & ( 132.1 , 0.5 , -78.1 )
 & ( 1.2 , 0.0 , 0.1 ) & ( -70.9 , -28.5 , 109.7 )& -2.06 & 0.0 & 0.0& 0.0 & T @ MM\\
\\

& \multirow{3}{*}{ S-Se-Se-S $\bigg\vert$} &
$\Gamma$ @ VB  &( 7.9 , 1.3 , 0.6 ) & ( 143.7 , 179.7 , 123.6 )
& ( 8.9 , 1.8 , 2.9 ) & ( -26.5 , -0.6 , -3.9 )& -1.04 & 0.0 & 0.0& 64.5  &T @ AB \\

 &  & $K$ @ VB  &( 7.9 , 1.3 , 0.6 ) & ( 143.7 , 179.7 , 123.6 )
 & ( 0.8 , 0.1 , 0.2 ) & ( -26.4 , 97.8 , -70.6 )& -2.28 & 0.0 & 0.0& 0.0 & T @ AB\\
 
&  & $K$ @ CB   &( 7.9 , 1.3 , 0.6 ) & ( 143.7 , 179.7 , 123.6 )
 & ( 1.2 , 0.0 , 0.1 ) & ( -70.9 , -28.5 , 109.7 )& -2.06 & 0.0 & 0.0& 0.0 & T @ MM\\
\\

& \multirow{3}{*}{ Se-S-S-Se $\bigg\vert$} &
$\Gamma$ @ VB  &( 3.3 , 0.1 , 0.3 ) & ( 129.1 , 27.8 , -10.7 )
& ( 4.9 , 0.3 , 0.2 ) & ( 123.2 , -76.4 , -122.0 )& -0.73 & 0.0 & 0.0& 67.1  &T @ MM \\

 &  & $K$ @ VB  &( 3.3 , 0.1 , 0.3 ) & ( 129.1 , 27.8 , -10.7 )
 & ( 0.8 , 0.1 , 0.2 ) & ( -26.4 , 97.8 , -70.6 )& -2.28 & 0.0 & 0.0& 0.0 & T @ AB\\
 
&  & $K$ @ CB   &( 3.3 , 0.1 , 0.3 ) & ( 129.1 , 27.8 , -10.7 )
 & ( 1.2 , 0.0 , 0.1 ) & ( -70.9 , -28.5 , 109.7 )& -2.06 & 0.0 & 0.0& 0.0 & T @ MM\\
\\

& \multirow{3}{*}{ Se-S-Se-S $\bigg\vert$} &
$\Gamma$ @ VB  &( 8.3 , 1.8 , 0.7 ) & ( 137.1 , 180.0 , 34.5 )
& ( 6.8 , 5.0 , 2.4 ) & ( -32.8 , -0.9 , -145.8 )& -0.69 & 0.0 & 0.0& 70.1 &T @ XX \\

 &  & $K$ @ VB  &( 8.3 , 1.8 , 0.7 ) & ( 137.1 , 180.0 , 34.5 )
 & ( 0.8 , 0.1 , 0.2 ) & ( -26.4 , 97.8 , -70.6 )& -2.28 & 0.0 & 0.0& 0.0 & T @ AB\\
 
&  & $K$ @ CB   &( 8.3 , 1.8 , 0.7 ) & ( 137.1 , 180.0 , 34.5 )
 & ( 1.2 , 0.0 , 0.1 ) & ( -70.9 , -28.5 , 109.7 )& -2.06 & 0.0 & 0.0& 0.0 & T @ XX\\
\\

\multirow{12}{*}{\begin{sideways}\colorbox{blue!10}{\hspace*{8mm} 3R MoSTe/WSTe \hspace*{8mm}}\end{sideways}} & \multirow{3}{*}{ S-Te-S-Te $\bigg\vert$} &
$\Gamma$ @ VB  &( 10.6 , 2.0 , 0.7 ) & ( 8.3 , 180.0 , -150.6 )
& ( 5.1, 0.9 , 2.0 ) & ( -164.6 , 172.5 , 175.9 ) & -0.65 & 0.02& -4.5 & 68.1 & T @ MX \\

 &  & $K$ @ VB    &( 10.6 , 2.0 , 0.7 ) & ( 8.3 , 180.0 , -150.6 )
 & ( 4.9 , 0.3 , 1.0 ) & ( 42.4 , -24.0 , -52.8 )& -1.83 & 0.01 & -70.0& 0.0 & T @ AA\\
 
&  & $K$ @ CB   &( 10.6 , 2.0 , 0.7 ) & ( 8.3 , 180.0 , -150.6 )
 & ( 3.5 , 0.4 , 0.8 ) & ( -6.6 , 27.5 , -0.5 )& 1.64& 0.02 & 25.4 & 0.0 & T @ XM\\
\\

& \multirow{3}{*}{ S-Te-Te-S $\bigg\vert$} &
$\Gamma$ @ VB  &( 9.6 , 1.8 , 0.5 ) & ( 0.0 , 180.0 , 176.6 )
& ( 2.4 , 0.5 , 0.1 ) & ( -12.7 , 1.9 , 20.9 )& -0.55 & 0.01 & 180.0& 186.4 & T @ MX \\

 &  & $K$ @ VB  &( 9.6 , 1.8 , 0.5 ) & ( 0.0 , 180.0 , 176.6 )
 & ( 7.2 , 0.5 , 0.7 ) & ( -37.4 , 161.2 , 66.6 )& -2.1 & 0.01 &166.0& 0.0 & T @ MX\\
 
&  & $K$ @ CB   &( 9.6 , 1.8 , 0.5 ) & ( 0.0 , 180.0 , 176.6 )
 & ( 2.3 , 0.3 , 0.15 ) & ( 11.9 , -180.0 , -12.6 )& 0.6 & 0.07 & -3.0& 0.0 & T @ MX\\
\\

& \multirow{3}{*}{ Te-S-S-Te $\bigg\vert$} &
$\Gamma$ @ VB  &( 12.5 , 2.5 , 1.0 ) & ( -0.3 , 180.0 , 180.0 )
& ( 29.0  , 4.3 , 2.6 ) & ( -177.4 , 2.1 , -3.5 )& -1.6 & 0.15 & 0.0& 32.6 &T @ MX \\

 &  & $K$ @ VB  &( 12.5 , 2.5 , 1.0 ) & ( 0.4 , 180.0 , 180.0 )
 & ( 13.6 , 0.5 , 1.0 ) & ( -31.4 , -166.3 , 123.6 )& -2.13 & 0.01 & 130.0& 0.0& T @ MX\\
 
&  & $K$ @ CB   &( 12.5 , 2.5 , 1.0 ) & ( 0.4 , 180.0 , 180.0 )
 & ( 10.4, 1.8 , 1.1 ) & ( 44.7 , 2.5 , -74.7 )& 1.15 & 0.01 & 28.5& 0.0 & T @ MX\\
\\

& \multirow{3}{*}{ Te-S-Te-S $\bigg\vert$} &
$\Gamma$ @ VB  &( 10.6 , 2.0 , 0.7 ) & ( 8.1 , 180.0 , -151.7 )
& ( 5.1 , 0.9 , 2.0 ) & ( -166.8 , 175.0 , 175.8 )& -0.65 & 0.02 & -4.9& 69.1 &T @ MX \\

 &  & $K$ @ VB  &( 10.6 , 2.0 , 0.7 ) & ( 8.1 , 180.0 , -151.7 )
 & ( 3.2 , 0.9 , 1.3 ) & ( -66.4 , 0.0 , 31.33 )& -2.1 & 0.01 & 55.2& 0.0 & T @ MX\\
 
&  & $K$ @ CB   &( 10.6 , 2.0 , 0.7 ) & ( 8.1 , 180.0 , -151.7 )
 & ( 3.9 , 0.3 , 0.95 ) & ( 2.7 , -22.5 , -13.7 )& 1.1& 0.02 & 55.3& 0.0 &  T @ MX\\
\\

\multirow{12}{*}{\begin{sideways}\colorbox{orange!10}{\hspace*{8mm} 2H MoSTe/WSTe \hspace*{8mm}}\end{sideways}} & \multirow{3}{*}{ S-Te-S-Te $\bigg\vert$} &
$\Gamma$ @ VB  &( 10.1 , 1.1 , 0.9 ) & ( 132.3 , 180.0 , 61.6 )
& ( 1.8 , 0.9 , 0.9 ) & ( 55.0 , -4.1 , -102.1 ) & -0.4 & 0.0 & 0.0& 98.9  &T @ MM \\

 &  & $K$ @ VB    &( 10.1 , 1.1 , 0.9 ) & ( 132.3 , 180.0 , 61.6 )
 & ( 0.5 ,1.5 , 1.0 ) & ( -73.3 , 1.4 , -103.6 )& -1.83 & 0.01 &43.3& 0.0 & T @ XX\\
 
&  & $K$ @ CB    &( 10.1 , 1.1 , 0.9 ) & ( 132.3 , 180.0 , 61.6 )
 & ( 1.1 , 0.7 , 0.5 ) & ( -118.6 , 31.5 , -142.9 )& 1.66 & 0.03 & 24.0& 0.0 &  T @ MM\\
\\

& \multirow{3}{*}{ S-Te-Te-S $\bigg\vert$} &

$\Gamma$ @ VB  &( 9.7 , 1.6 , 0.8 ) & ( 153.3 , 180.0 , 130.6 )
& ( 3.0 , 0.5 , 0.2 ) & ( 134.6 , 1.2 , -19.5 )& -0.55 & 0.01 & 160.0& 227.4 &T @ XX \\

 &  & $K$ @ VB  &( 9.7 , 1.6 , 0.8 ) & ( 153.3 , 180.0 , 130.6 )
 & ( 7.1 , 0.7 , 0.9 ) & ( -36.4 , 171.2 , 81.6 )& -2.1 & 0.01 &166.0& 0.0 & T @ MM\\
 
&  & $K$ @ CB   &( 9.7 , 1.6 , 0.8 ) & ( 153.3 , 180.0 , 130.6 )
 & ( 2.3 , 0.3 , 0.15 ) & ( 11.9 , -180.0 , -12.6 )& 0.6 & 0.07 & -3.0& 0.0 & T @ MM\\
\\

& \multirow{3}{*}{ Te-S-S-Te $\bigg\vert$} &
$\Gamma$ @ VB  &( 10.9 , 1.7 , 0.9 ) & ( 117.3 , 178.0 , 46.6 )
& ( 29.0  , 5.7 , 3.6 ) & ( -177.4 , 1.0 , -2.3 )& -1.6 & 0.15 & 0.0& 32.6 &T @ MM \\
\color{red}
 &  & $K$ @ VB &( 10.9 , 1.7 , 0.9 ) & ( 117.3 , 178.0 , 46.6 )
 & ( 9.6, 0.3 , 1.5 ) & ( 98.0 , -113.5 , -47.1 )& -2.14 & 0.0 & 0.0& 0.0 & T @ XX\\
 
&  & $K$ @ CB   &( 10.9 , 1.7 , 0.9 ) & ( 117.3 , 178.0 , 46.6 )
 & ( 4.1, 1.4 , 0.2 ) & ( 115.0 , -9.4 , 60.2 )&  1.14 & 0.01 & 71.6& 0.0 & T @ MM\\
\\

& \multirow{3}{*}{ Te-S-Te-S $\bigg\vert$} &
$\Gamma$ @ VB  &( 9.9 , 1.8 , 0.7 ) & ( 137.3 , 180.0 , 100.6 )
& ( 27.8 , 20.5 , 13.7 ) & ( -1.8 , -33.0 , 8.7 )& -0.65 & 0.02 & -6.1& 69.8  & T @ AB \\

 &  & $K$ @ VB   &( 9.9 , 1.8 , 0.7 ) & ( 137.3 , 180.0 , 100.6 )
 & (5.0 , 0.2 , 1.3 ) & ( 167.0 , -171.0 , -157.8 )& -2.1 & 0.01 & 56.2& 0.0 & T @ XX\\
 
&  & $K$ @ CB    &( 9.9 , 1.8 , 0.7 ) & ( 137.3 , 180.0 , 100.6 )
 & ( 2.5 , 0.4 , 1.1 ) & ( 140.9 , 1.5 , -143.7 )& 1.1& 0.02 & -33.8& 0.0 & T @ AB\\
\\

\multirow{12}{*}{\begin{sideways}\colorbox{blue!10}{\hspace*{8mm} 3R MoSeTe/WSeTe \hspace*{8mm}}\end{sideways}} & \multirow{3}{*}{ Se-Te-Se-Te $\bigg\vert$} &
$\Gamma$ @ VB  &( 13.3 , 2.4 , 0.9 ) & ( 2.3 , 180.0 , -170.0 )
& ( 16.9 , 2.2 , 1.6 )  & ( 3.4 , 142.5 , 176.7 ) & -0.7 & 0.04 & 11.1& 109.0 &T @ XM \\

 &  & $K$ @ VB   &( 13.3 , 2.4 , 0.9 ) & ( 2.3 , 180.0 , -170.0 )
 & ( 12.6 , 2.2 , 0.7 ) & ( 82.3 , -5.6 , -39.6 )& -2.0 & 0.0 &0.0& 0.0 & T @ XM\\
 
&  & $K$ @ CB   &( 13.3 , 2.4 , 0.9 ) & ( 2.3 , 180.0 , -170.0 )
 & ( 2.6 , 1.8 , 0.6 ) & ( -17.8 , 2.8 , 5.0 )& 1.9& 0.03 & 4.9& 0.0 & T @ XM\\
\\

& \multirow{3}{*}{ Se-Te-Te-Se $\bigg\vert$} &
$\Gamma$ @ VB &( 12.0 , 2.2 , 0.7 ) & ( -0.5 , 180.0 , -178.6 )
& ( 22.2 , 16.2 , 9.9 ) & ( -3.7 , -26.9 , -4.6 )& -0.7 & 0.01 & -7.4 & 70.9 & T @ AA \\

 &  & $K$ @ VB  &( 12.0 , 2.2 , 0.7 ) & ( -0.5 , 180.0 , -178.6 )
 & ( 10.8 , 0.6 , 0.9 ) & ( -34.4 , 164.2 , 99.6 )& -2.35 & 0.01 &151.0& 0.0 & T @ AA\\
 
&  & $K$ @ CB  &( 12.0 , 2.2 , 0.7 ) & ( -0.5 , 180.0 , -178.6 )
 & ( 6.9 , 0.6 , 0.4 ) & ( 21.1 , -177.0 , -162.6 )& 1.0 & 0.06 & -11.0& 0.0 & T @ MX\\
\\

& \multirow{3}{*}{ Te-Se-Se-Te $\bigg\vert$} &
$\Gamma$ @ VB  &( 15.3 , 2.4 , 1.0 ) & ( 0.0 , 180.0 , 180.0 )
& ( 14.9 , 5.2 , 5.6 )  & ( 170.0 , 142.5 , 176.7 ) & -0.7 & 0.03 & 3.1& 73.0 & T @ AA \\

 &  & $K$ @ VB   &( 15.3 , 2.4 , 1.0 ) & ( 0.0 , 180.0 , 180.0 )
 & ( 10.4 , 0.9 , 0.2 ) & ( -53.3 , 148.6 , 16.6 )& -2.4 & 0.01 &141.7& 0.0 & T @ AA\\
 
&  & $K$ @ CB   &( 15.3 , 2.4 , 1.0 ) & ( 0.0 , 180.0 , 180.0 )
 & ( 9.9 , 1.5 , 1.1 ) & ( 60.6 , 180.0 , -70.0 )& 1.57& 0.01 & 20.0& 0.0 & T @ XM\\
\\

& \multirow{3}{*}{ Te-Se-Te-Se $\bigg\vert$} &
$\Gamma$ @ VB  &( 15.3 , 3.5 , 0.9 ) & ( -2.1 , 180.0 , 168.7 )
& ( 17.1 , 7.9 , 2.0 ) & ( -3.0 , 5.0 , 5.8 )& -0.65 & 0.01 & -8.9& 72.1 &T @AA \\

 &  & $K$ @ VB  &( 15.3 , 3.5 , 0.9 ) & ( -2.1 , 180.0 , 168.7 )
 & ( 7.5 , 1.1, 0.3 ) & ( 135.4 , 0.5 , 20.33 )& -2.4 & 0.0 & 0.0 & 0.0 &  T @ XM\\
 
&  & $K$ @ CB   &( 15.3 , 3.5 , 0.9 ) & ( -2.1 , 180.0 , 168.7 )
 & ( 8.0 , 1.3 , 0.3 ) & (134.7 , 0.5 , -4.7 )& 1.5& 0.02 & -32.3& 0.0 & T @ AA\\
\\

\multirow{12}{*}{\begin{sideways}\colorbox{orange!10}{\hspace*{8mm} 2H MoSeTe/WSeTe \hspace*{8mm}}\end{sideways}} & \multirow{3}{*}{ Se-Te-Se-Te $\bigg\vert$} &
$\Gamma$ @ VB  &( 13.4 , 2.3 , 0.9 ) & ( 2.3 , 180.0 , -169.0 )
& ( 16.9 , 2.2 , 1.6 )  & ( 3.4 , 142.5 , 176.7 ) & -0.7 & 0.04 & 11.1& 109.0 &T @ XX \\

 &  & $K$ @ VB  &( 13.4 , 2.3 , 0.9 ) & ( 2.3 , 180.0 , -169.0 )
 & ( 12.6 , 2.2 , 0.7 ) & ( 82.3 , -5.6 , -39.6 )& -2.0 & 0.0 &0.0& 0.0 & T @ XX\\
 
&  & $K$ @ CB  &( 13.4 , 2.3 , 0.9 ) & ( 2.3 , 180.0 , -169.0 )
 & ( 2.6 , 1.8 , 0.6 ) & ( -17.8 , 2.8 , 5.0 )& 1.9& 0.03 & 4.9& 0.0 &  T @ XX\\
\\

& \multirow{3}{*}{ Se-Te-Te-Se $\bigg\vert$} &
$\Gamma$ @ VB &( 12.0 , 2.0 , 0.9 ) & ( 148.6 , 180.0 , 122.5 )
& ( 3.3 , 2.4 , 1.4 ) & ( 25.6 , 2.0 , -40.6 )& -1.24 & 0.04 & 155.4 & 29.0 & T @ AB \\

 &  & $K$ @ VB  &( 12.0 , 2.0 , 0.9 ) & ( 148.6 , 180.0 , 122.5 )
 & ( 8.0 , 0.1 , 0.3 ) & ( 148.7 , 164.9 , -55.6 )& -2.36 & 0.01 &23.0& 0.0 & T @ XX\\
 
&  & $K$ @ CB  &( 12.0 , 2.0 , 0.9 ) & ( 148.6 , 180.0 , 122.5 )
 & ( 5.3 , 0.1 , 0.4 ) & ( 132.9 , 30.9 , 4.2 )& 1.0 & 0.07 & 164.0& 0.0 &  T @ AB\\
\\

& \multirow{3}{*}{ Te-Se-Se-Te $\bigg\vert$} &
$\Gamma$ @ VB  &( 9.9 , 2.0 , 0.8 ) & ( 126.0 , 180.0 , 51.2 )
& ( 25.7 , 6.6 , 5.1 )  & ( -62.8 , -4.8 , 164.7 ) & -0.9 & 0.1 & 127.1& 65.5 & T @ MM \\

 &  & $K$ @ VB   &( 9.9 , 2.0 , 0.8 ) & ( 126.0 , 180.0 , 51.2 )
 & ( 4.5 , 0.4 , 0.4 ) & ( 130.3 , 145.6 , -16.2 )& -2.4 & 0.01 &107.7& 0.0 & T @ XX\\
 
&  & $K$ @ CB   &( 9.9 , 2.0 , 0.8 ) & ( 126.0 , 180.0 , 51.2 )
 & ( 2.6 , 0.2 , 0.4 ) & ( 127.6 , -2.7 , -58.0 )& 1.44& 0.01 & 67.0& 0.0 &  T @ MM\\
\\

& \multirow{3}{*}{ Te-Se-Te-Se $\bigg\vert$} &
$\Gamma$ @ VB  &( 12.0 , 3.0 , 1.8 ) & ( 135.8 , 180.0 , 59.8 )
& ( 2.1 , 1.3 , 0.5 ) & ( 21.7 , 172.0 , 55.8 )& -0.61 & 0.03 & 120.3& 153.0 & T @ XX \\

 &  & $K$ @ VB &( 12.0 , 3.0 , 1.8 ) & ( 135.8 , 180.0 , 59.8 )
 & ( 7.4 , 4.2, 2.3 ) & ( 112.4 , 3.2 , -166.3 )& -2.4 & 0.01 & 46.9 & 0.0 & T @ XX\\
 
&  & $K$ @ CB   &( 12.0 , 3.0 , 1.8 ) & ( 135.8 , 180.0 , 59.8 )
 & ( 4.5 , 1.4 , 1.2 ) & (160.7 , -3.3 , -82.7 )& 1.45& 0.02 & 94.0& 0.0 &  T @ AB\\
\\

\end{longtable*}
}

\section{Acknowledgements}
We are grateful to Daniel T. Larson, Stephen Carr, Hao Tang and Ziyan Zhu for helpful discussions and comments.\\
This research used resources of the National Energy Research Scientific Computing Center (NERSC), a U.S. Department of Energy Office of Science User Facility located at Lawrence Berkeley National Laboratory, operated under Contract No. DE-AC02-05CH11231 using NERSC award BES-ERCAP0020773.
This work used the Extreme Science and Engineering Discovery Environment (XSEDE) Stampede2 at the Texas Advanced Computing Center through allocation TG-DMR120073, which is supported by National Science Foundation grant number ACI-1548562.
MA and GRS acknowledge funding from the National Science Foundation under Award No. DMR-1922172 and the Army Research Office under Cooperative Agreement Number W911NF-21-2-0147.
EK acknowledges funding from the STC Center for Integrated Quantum Materials, NSF Grant No. DMR-1231319; NSF DMREF Award No. 1922172; and the Army Research Office under Cooperative Agreement Number W911NF-21-2-0147.

\vspace{-0.5cm}


\begin{thebibliography}{98}%
\makeatletter
\providecommand \@ifxundefined [1]{%
 \@ifx{#1\undefined}
}%
\providecommand \@ifnum [1]{%
 \ifnum #1\expandafter \@firstoftwo
 \else \expandafter \@secondoftwo
 \fi
}%
\providecommand \@ifx [1]{%
 \ifx #1\expandafter \@firstoftwo
 \else \expandafter \@secondoftwo
 \fi
}%
\providecommand \natexlab [1]{#1}%
\providecommand \enquote  [1]{``#1''}%
\providecommand \bibnamefont  [1]{#1}%
\providecommand \bibfnamefont [1]{#1}%
\providecommand \citenamefont [1]{#1}%
\providecommand \href@noop [0]{\@secondoftwo}%
\providecommand \href [0]{\begingroup \@sanitize@url \@href}%
\providecommand \@href[1]{\@@startlink{#1}\@@href}%
\providecommand \@@href[1]{\endgroup#1\@@endlink}%
\providecommand \@sanitize@url [0]{\catcode `\\12\catcode `\$12\catcode
  `\&12\catcode `\#12\catcode `\^12\catcode `\_12\catcode `\%12\relax}%
\providecommand \@@startlink[1]{}%
\providecommand \@@endlink[0]{}%
\providecommand \url  [0]{\begingroup\@sanitize@url \@url }%
\providecommand \@url [1]{\endgroup\@href {#1}{\urlprefix }}%
\providecommand \urlprefix  [0]{URL }%
\providecommand \Eprint [0]{\href }%
\providecommand \doibase [0]{https://doi.org/}%
\providecommand \selectlanguage [0]{\@gobble}%
\providecommand \bibinfo  [0]{\@secondoftwo}%
\providecommand \bibfield  [0]{\@secondoftwo}%
\providecommand \translation [1]{[#1]}%
\providecommand \BibitemOpen [0]{}%
\providecommand \bibitemStop [0]{}%
\providecommand \bibitemNoStop [0]{.\EOS\space}%
\providecommand \EOS [0]{\spacefactor3000\relax}%
\providecommand \BibitemShut  [1]{\csname bibitem#1\endcsname}%
\let\auto@bib@innerbib\@empty
\bibitem [{\citenamefont {Ginzburg}(2004)}]{Ginzburg2004}%
  \BibitemOpen
  \bibfield  {author} {\bibinfo {author} {\bibfnamefont {V.~L.}\ \bibnamefont
  {Ginzburg}},\ }\bibfield  {title} {\bibinfo {title} {Nobel lecture: On
  superconductivity and superfluidity (what i have and have not managed to do)
  as well as on the {\textquotedblleft}physical minimum{\textquotedblright} at
  the beginning of the {XXI} century},\ }\href
  {https://doi.org/10.1103/revmodphys.76.981} {\bibfield  {journal} {\bibinfo
  {journal} {Reviews of Modern Physics}\ }\textbf {\bibinfo {volume} {76}},\
  \bibinfo {pages} {981} (\bibinfo {year} {2004})}\BibitemShut {NoStop}%
\bibitem [{\citenamefont {Stormer}(1999)}]{Stormer1999}%
  \BibitemOpen
  \bibfield  {author} {\bibinfo {author} {\bibfnamefont {H.~L.}\ \bibnamefont
  {Stormer}},\ }\bibfield  {title} {\bibinfo {title} {Nobel lecture: The
  fractional quantum hall effect},\ }\href
  {https://doi.org/10.1103/revmodphys.71.875} {\bibfield  {journal} {\bibinfo
  {journal} {Reviews of Modern Physics}\ }\textbf {\bibinfo {volume} {71}},\
  \bibinfo {pages} {875} (\bibinfo {year} {1999})}\BibitemShut {NoStop}%
\bibitem [{\citenamefont {Keimer}\ and\ \citenamefont
  {Moore}(2017)}]{Keimer2017}%
  \BibitemOpen
  \bibfield  {author} {\bibinfo {author} {\bibfnamefont {B.}~\bibnamefont
  {Keimer}}\ and\ \bibinfo {author} {\bibfnamefont {J.~E.}\ \bibnamefont
  {Moore}},\ }\bibfield  {title} {\bibinfo {title} {The physics of quantum
  materials},\ }\href {https://doi.org/10.1038/nphys4302} {\bibfield  {journal}
  {\bibinfo  {journal} {Nature Physics}\ }\textbf {\bibinfo {volume} {13}},\
  \bibinfo {pages} {1045} (\bibinfo {year} {2017})}\BibitemShut {NoStop}%
\bibitem [{\citenamefont {Basov}\ \emph {et~al.}(2017)\citenamefont {Basov},
  \citenamefont {Averitt},\ and\ \citenamefont {Hsieh}}]{Basov2017}%
  \BibitemOpen
  \bibfield  {author} {\bibinfo {author} {\bibfnamefont {D.~N.}\ \bibnamefont
  {Basov}}, \bibinfo {author} {\bibfnamefont {R.~D.}\ \bibnamefont {Averitt}},\
  and\ \bibinfo {author} {\bibfnamefont {D.}~\bibnamefont {Hsieh}},\ }\bibfield
   {title} {\bibinfo {title} {Towards properties on demand in quantum
  materials},\ }\href {https://doi.org/10.1038/nmat5017} {\bibfield  {journal}
  {\bibinfo  {journal} {Nature Materials}\ }\textbf {\bibinfo {volume} {16}},\
  \bibinfo {pages} {1077} (\bibinfo {year} {2017})}\BibitemShut {NoStop}%
\bibitem [{\citenamefont {Giustino}\ \emph {et~al.}(2020)\citenamefont
  {Giustino}, \citenamefont {Lee}, \citenamefont {Trier}, \citenamefont
  {Bibes}, \citenamefont {Winter}, \citenamefont {Valent{\'{\i}}},
  \citenamefont {Son}, \citenamefont {Taillefer}, \citenamefont {Heil},
  \citenamefont {Figueroa}, \citenamefont {Pla{\c{c}}ais}, \citenamefont {Wu},
  \citenamefont {Yazyev}, \citenamefont {Bakkers}, \citenamefont {Nyg{\aa}rd},
  \citenamefont {Forn-D{\'{\i}}az}, \citenamefont {Franceschi}, \citenamefont
  {McIver}, \citenamefont {Torres}, \citenamefont {Low}, \citenamefont {Kumar},
  \citenamefont {Galceran}, \citenamefont {Valenzuela}, \citenamefont
  {Costache}, \citenamefont {Manchon}, \citenamefont {Kim}, \citenamefont
  {Schleder}, \citenamefont {Fazzio},\ and\ \citenamefont
  {Roche}}]{Giustino2020}%
  \BibitemOpen
  \bibfield  {author} {\bibinfo {author} {\bibfnamefont {F.}~\bibnamefont
  {Giustino}}, \bibinfo {author} {\bibfnamefont {J.~H.}\ \bibnamefont {Lee}},
  \bibinfo {author} {\bibfnamefont {F.}~\bibnamefont {Trier}}, \bibinfo
  {author} {\bibfnamefont {M.}~\bibnamefont {Bibes}}, \bibinfo {author}
  {\bibfnamefont {S.~M.}\ \bibnamefont {Winter}}, \bibinfo {author}
  {\bibfnamefont {R.}~\bibnamefont {Valent{\'{\i}}}}, \bibinfo {author}
  {\bibfnamefont {Y.-W.}\ \bibnamefont {Son}}, \bibinfo {author} {\bibfnamefont
  {L.}~\bibnamefont {Taillefer}}, \bibinfo {author} {\bibfnamefont
  {C.}~\bibnamefont {Heil}}, \bibinfo {author} {\bibfnamefont {A.~I.}\
  \bibnamefont {Figueroa}}, \bibinfo {author} {\bibfnamefont {B.}~\bibnamefont
  {Pla{\c{c}}ais}}, \bibinfo {author} {\bibfnamefont {Q.}~\bibnamefont {Wu}},
  \bibinfo {author} {\bibfnamefont {O.~V.}\ \bibnamefont {Yazyev}}, \bibinfo
  {author} {\bibfnamefont {E.~P. A.~M.}\ \bibnamefont {Bakkers}}, \bibinfo
  {author} {\bibfnamefont {J.}~\bibnamefont {Nyg{\aa}rd}}, \bibinfo {author}
  {\bibfnamefont {P.}~\bibnamefont {Forn-D{\'{\i}}az}}, \bibinfo {author}
  {\bibfnamefont {S.~D.}\ \bibnamefont {Franceschi}}, \bibinfo {author}
  {\bibfnamefont {J.~W.}\ \bibnamefont {McIver}}, \bibinfo {author}
  {\bibfnamefont {L.~E. F.~F.}\ \bibnamefont {Torres}}, \bibinfo {author}
  {\bibfnamefont {T.}~\bibnamefont {Low}}, \bibinfo {author} {\bibfnamefont
  {A.}~\bibnamefont {Kumar}}, \bibinfo {author} {\bibfnamefont
  {R.}~\bibnamefont {Galceran}}, \bibinfo {author} {\bibfnamefont {S.~O.}\
  \bibnamefont {Valenzuela}}, \bibinfo {author} {\bibfnamefont {M.~V.}\
  \bibnamefont {Costache}}, \bibinfo {author} {\bibfnamefont {A.}~\bibnamefont
  {Manchon}}, \bibinfo {author} {\bibfnamefont {E.-A.}\ \bibnamefont {Kim}},
  \bibinfo {author} {\bibfnamefont {G.~R.}\ \bibnamefont {Schleder}}, \bibinfo
  {author} {\bibfnamefont {A.}~\bibnamefont {Fazzio}},\ and\ \bibinfo {author}
  {\bibfnamefont {S.}~\bibnamefont {Roche}},\ }\bibfield  {title} {\bibinfo
  {title} {The 2021 quantum materials roadmap},\ }\href
  {https://doi.org/10.1088/2515-7639/abb74e} {\bibfield  {journal} {\bibinfo
  {journal} {Journal of Physics: Materials}\ }\textbf {\bibinfo {volume} {3}},\
  \bibinfo {pages} {042006} (\bibinfo {year} {2020})}\BibitemShut {NoStop}%
\bibitem [{\citenamefont {Cao}\ \emph {et~al.}(2018{\natexlab{a}})\citenamefont
  {Cao}, \citenamefont {Fatemi}, \citenamefont {Demir}, \citenamefont {Fang},
  \citenamefont {Tomarken}, \citenamefont {Luo}, \citenamefont
  {Sanchez-Yamagishi}, \citenamefont {Watanabe}, \citenamefont {Taniguchi},
  \citenamefont {Kaxiras}, \citenamefont {Ashoori},\ and\ \citenamefont
  {Jarillo-Herrero}}]{Herrero-1}%
  \BibitemOpen
  \bibfield  {author} {\bibinfo {author} {\bibfnamefont {Y.}~\bibnamefont
  {Cao}}, \bibinfo {author} {\bibfnamefont {V.}~\bibnamefont {Fatemi}},
  \bibinfo {author} {\bibfnamefont {A.}~\bibnamefont {Demir}}, \bibinfo
  {author} {\bibfnamefont {S.}~\bibnamefont {Fang}}, \bibinfo {author}
  {\bibfnamefont {S.~L.}\ \bibnamefont {Tomarken}}, \bibinfo {author}
  {\bibfnamefont {J.~Y.}\ \bibnamefont {Luo}}, \bibinfo {author} {\bibfnamefont
  {J.~D.}\ \bibnamefont {Sanchez-Yamagishi}}, \bibinfo {author} {\bibfnamefont
  {K.}~\bibnamefont {Watanabe}}, \bibinfo {author} {\bibfnamefont
  {T.}~\bibnamefont {Taniguchi}}, \bibinfo {author} {\bibfnamefont
  {E.}~\bibnamefont {Kaxiras}}, \bibinfo {author} {\bibfnamefont {R.~C.}\
  \bibnamefont {Ashoori}},\ and\ \bibinfo {author} {\bibfnamefont
  {P.}~\bibnamefont {Jarillo-Herrero}},\ }\bibfield  {title} {\bibinfo {title}
  {Correlated insulator behaviour at half-filling in magic-angle graphene
  superlattices},\ }\href {https://doi.org/10.1038/nature26154} {\bibfield
  {journal} {\bibinfo  {journal} {Nature}\ }\textbf {\bibinfo {volume} {556}},\
  \bibinfo {pages} {80} (\bibinfo {year} {2018}{\natexlab{a}})}\BibitemShut
  {NoStop}%
\bibitem [{\citenamefont {Cao}\ \emph {et~al.}(2018{\natexlab{b}})\citenamefont
  {Cao}, \citenamefont {Fatemi}, \citenamefont {Fang}, \citenamefont
  {Watanabe}, \citenamefont {Taniguchi}, \citenamefont {Kaxiras},\ and\
  \citenamefont {Jarillo-Herrero}}]{Herrero-2}%
  \BibitemOpen
  \bibfield  {author} {\bibinfo {author} {\bibfnamefont {Y.}~\bibnamefont
  {Cao}}, \bibinfo {author} {\bibfnamefont {V.}~\bibnamefont {Fatemi}},
  \bibinfo {author} {\bibfnamefont {S.}~\bibnamefont {Fang}}, \bibinfo {author}
  {\bibfnamefont {K.}~\bibnamefont {Watanabe}}, \bibinfo {author}
  {\bibfnamefont {T.}~\bibnamefont {Taniguchi}}, \bibinfo {author}
  {\bibfnamefont {E.}~\bibnamefont {Kaxiras}},\ and\ \bibinfo {author}
  {\bibfnamefont {P.}~\bibnamefont {Jarillo-Herrero}},\ }\bibfield  {title}
  {\bibinfo {title} {Unconventional superconductivity in magic-angle graphene
  superlattices},\ }\href {https://doi.org/10.1038/nature26160} {\bibfield
  {journal} {\bibinfo  {journal} {Nature}\ }\textbf {\bibinfo {volume} {556}},\
  \bibinfo {pages} {43} (\bibinfo {year} {2018}{\natexlab{b}})}\BibitemShut
  {NoStop}%
\bibitem [{\citenamefont {Lu}\ \emph {et~al.}(2019)\citenamefont {Lu},
  \citenamefont {Stepanov}, \citenamefont {Yang}, \citenamefont {Xie},
  \citenamefont {Aamir}, \citenamefont {Das}, \citenamefont {Urgell},
  \citenamefont {Watanabe}, \citenamefont {Taniguchi}, \citenamefont {Zhang},
  \citenamefont {Bachtold}, \citenamefont {MacDonald},\ and\ \citenamefont
  {Efetov}}]{Lu2019}%
  \BibitemOpen
  \bibfield  {author} {\bibinfo {author} {\bibfnamefont {X.}~\bibnamefont
  {Lu}}, \bibinfo {author} {\bibfnamefont {P.}~\bibnamefont {Stepanov}},
  \bibinfo {author} {\bibfnamefont {W.}~\bibnamefont {Yang}}, \bibinfo {author}
  {\bibfnamefont {M.}~\bibnamefont {Xie}}, \bibinfo {author} {\bibfnamefont
  {M.~A.}\ \bibnamefont {Aamir}}, \bibinfo {author} {\bibfnamefont
  {I.}~\bibnamefont {Das}}, \bibinfo {author} {\bibfnamefont {C.}~\bibnamefont
  {Urgell}}, \bibinfo {author} {\bibfnamefont {K.}~\bibnamefont {Watanabe}},
  \bibinfo {author} {\bibfnamefont {T.}~\bibnamefont {Taniguchi}}, \bibinfo
  {author} {\bibfnamefont {G.}~\bibnamefont {Zhang}}, \bibinfo {author}
  {\bibfnamefont {A.}~\bibnamefont {Bachtold}}, \bibinfo {author}
  {\bibfnamefont {A.~H.}\ \bibnamefont {MacDonald}},\ and\ \bibinfo {author}
  {\bibfnamefont {D.~K.}\ \bibnamefont {Efetov}},\ }\bibfield  {title}
  {\bibinfo {title} {Superconductors, orbital magnets and correlated states in
  magic-angle bilayer graphene},\ }\href
  {https://doi.org/10.1038/s41586-019-1695-0} {\bibfield  {journal} {\bibinfo
  {journal} {Nature}\ }\textbf {\bibinfo {volume} {574}},\ \bibinfo {pages}
  {653} (\bibinfo {year} {2019})}\BibitemShut {NoStop}%
\bibitem [{\citenamefont {He}\ \emph {et~al.}(2021)\citenamefont {He},
  \citenamefont {Zhang}, \citenamefont {Li}, \citenamefont {Fei}, \citenamefont
  {Watanabe}, \citenamefont {Taniguchi}, \citenamefont {Xu},\ and\
  \citenamefont {Yankowitz}}]{He2021}%
  \BibitemOpen
  \bibfield  {author} {\bibinfo {author} {\bibfnamefont {M.}~\bibnamefont
  {He}}, \bibinfo {author} {\bibfnamefont {Y.-H.}\ \bibnamefont {Zhang}},
  \bibinfo {author} {\bibfnamefont {Y.}~\bibnamefont {Li}}, \bibinfo {author}
  {\bibfnamefont {Z.}~\bibnamefont {Fei}}, \bibinfo {author} {\bibfnamefont
  {K.}~\bibnamefont {Watanabe}}, \bibinfo {author} {\bibfnamefont
  {T.}~\bibnamefont {Taniguchi}}, \bibinfo {author} {\bibfnamefont
  {X.}~\bibnamefont {Xu}},\ and\ \bibinfo {author} {\bibfnamefont
  {M.}~\bibnamefont {Yankowitz}},\ }\bibfield  {title} {\bibinfo {title}
  {Competing correlated states and abundant orbital magnetism in twisted
  monolayer-bilayer graphene},\ }\href
  {https://doi.org/10.1038/s41467-021-25044-1} {\bibfield  {journal} {\bibinfo
  {journal} {Nature Communications}\ }\textbf {\bibinfo {volume} {12}},\
  \bibinfo {pages} {4727} (\bibinfo {year} {2021})}\BibitemShut {NoStop}%
\bibitem [{\citenamefont {Yankowitz}\ \emph {et~al.}(2019)\citenamefont
  {Yankowitz}, \citenamefont {Chen}, \citenamefont {Polshyn}, \citenamefont
  {Zhang}, \citenamefont {Watanabe}, \citenamefont {Taniguchi}, \citenamefont
  {Graf}, \citenamefont {Young},\ and\ \citenamefont {Dean}}]{Yanko_science}%
  \BibitemOpen
  \bibfield  {author} {\bibinfo {author} {\bibfnamefont {M.}~\bibnamefont
  {Yankowitz}}, \bibinfo {author} {\bibfnamefont {S.}~\bibnamefont {Chen}},
  \bibinfo {author} {\bibfnamefont {H.}~\bibnamefont {Polshyn}}, \bibinfo
  {author} {\bibfnamefont {Y.}~\bibnamefont {Zhang}}, \bibinfo {author}
  {\bibfnamefont {K.}~\bibnamefont {Watanabe}}, \bibinfo {author}
  {\bibfnamefont {T.}~\bibnamefont {Taniguchi}}, \bibinfo {author}
  {\bibfnamefont {D.}~\bibnamefont {Graf}}, \bibinfo {author} {\bibfnamefont
  {A.~F.}\ \bibnamefont {Young}},\ and\ \bibinfo {author} {\bibfnamefont
  {C.~R.}\ \bibnamefont {Dean}},\ }\bibfield  {title} {\bibinfo {title} {Tuning
  superconductivity in twisted bilayer graphene},\ }\href
  {https://doi.org/10.1126/science.aav1910} {\bibfield  {journal} {\bibinfo
  {journal} {Science}\ }\textbf {\bibinfo {volume} {363}},\ \bibinfo {pages}
  {1059} (\bibinfo {year} {2019})}\BibitemShut {NoStop}%
\bibitem [{\citenamefont {Zhou}\ \emph {et~al.}(2021)\citenamefont {Zhou},
  \citenamefont {Xie}, \citenamefont {Taniguchi}, \citenamefont {Watanabe},\
  and\ \citenamefont {Young}}]{Zhou2021}%
  \BibitemOpen
  \bibfield  {author} {\bibinfo {author} {\bibfnamefont {H.}~\bibnamefont
  {Zhou}}, \bibinfo {author} {\bibfnamefont {T.}~\bibnamefont {Xie}}, \bibinfo
  {author} {\bibfnamefont {T.}~\bibnamefont {Taniguchi}}, \bibinfo {author}
  {\bibfnamefont {K.}~\bibnamefont {Watanabe}},\ and\ \bibinfo {author}
  {\bibfnamefont {A.~F.}\ \bibnamefont {Young}},\ }\bibfield  {title} {\bibinfo
  {title} {Superconductivity in rhombohedral trilayer graphene},\ }\href
  {https://doi.org/10.1038/s41586-021-03926-0} {\bibfield  {journal} {\bibinfo
  {journal} {Nature}\ }\textbf {\bibinfo {volume} {598}},\ \bibinfo {pages}
  {434} (\bibinfo {year} {2021})}\BibitemShut {NoStop}%
\bibitem [{\citenamefont {Park}\ \emph {et~al.}(2021)\citenamefont {Park},
  \citenamefont {Cao}, \citenamefont {Watanabe}, \citenamefont {Taniguchi},\
  and\ \citenamefont {Jarillo-Herrero}}]{Park2021}%
  \BibitemOpen
  \bibfield  {author} {\bibinfo {author} {\bibfnamefont {J.~M.}\ \bibnamefont
  {Park}}, \bibinfo {author} {\bibfnamefont {Y.}~\bibnamefont {Cao}}, \bibinfo
  {author} {\bibfnamefont {K.}~\bibnamefont {Watanabe}}, \bibinfo {author}
  {\bibfnamefont {T.}~\bibnamefont {Taniguchi}},\ and\ \bibinfo {author}
  {\bibfnamefont {P.}~\bibnamefont {Jarillo-Herrero}},\ }\bibfield  {title}
  {\bibinfo {title} {Tunable strongly coupled superconductivity in magic-angle
  twisted trilayer graphene},\ }\href
  {https://doi.org/10.1038/s41586-021-03192-0} {\bibfield  {journal} {\bibinfo
  {journal} {Nature}\ }\textbf {\bibinfo {volume} {590}},\ \bibinfo {pages}
  {249} (\bibinfo {year} {2021})}\BibitemShut {NoStop}%
\bibitem [{\citenamefont {Li}\ \emph {et~al.}(2018)\citenamefont {Li},
  \citenamefont {Cheng},\ and\ \citenamefont {Huang}}]{Li_2018}%
  \BibitemOpen
  \bibfield  {author} {\bibinfo {author} {\bibfnamefont {R.}~\bibnamefont
  {Li}}, \bibinfo {author} {\bibfnamefont {Y.}~\bibnamefont {Cheng}},\ and\
  \bibinfo {author} {\bibfnamefont {W.}~\bibnamefont {Huang}},\ }\bibfield
  {title} {\bibinfo {title} {Recent progress of janus 2d transition metal
  chalcogenides: From theory to experiments},\ }\href
  {https://doi.org/https://doi.org/10.1002/smll.201802091} {\bibfield
  {journal} {\bibinfo  {journal} {Small}\ }\textbf {\bibinfo {volume} {14}},\
  \bibinfo {pages} {1802091} (\bibinfo {year} {2018})}\BibitemShut {NoStop}%
\bibitem [{\citenamefont {Gadelha}\ \emph {et~al.}(2021)\citenamefont
  {Gadelha}, \citenamefont {Ohlberg}, \citenamefont {Rabelo}, \citenamefont
  {Neto}, \citenamefont {Vasconcelos}, \citenamefont {Campos}, \citenamefont
  {Lemos}, \citenamefont {Ornelas}, \citenamefont {Miranda}, \citenamefont
  {Nadas}, \citenamefont {Santana}, \citenamefont {Watanabe}, \citenamefont
  {Taniguchi}, \citenamefont {van Troeye}, \citenamefont {Lamparski},
  \citenamefont {Meunier}, \citenamefont {Nguyen}, \citenamefont {Paszko},
  \citenamefont {Charlier}, \citenamefont {Campos}, \citenamefont
  {Can{\c{c}}ado}, \citenamefont {Medeiros-Ribeiro},\ and\ \citenamefont
  {Jorio}}]{Gadelha2021}%
  \BibitemOpen
  \bibfield  {author} {\bibinfo {author} {\bibfnamefont {A.~C.}\ \bibnamefont
  {Gadelha}}, \bibinfo {author} {\bibfnamefont {D.~A.~A.}\ \bibnamefont
  {Ohlberg}}, \bibinfo {author} {\bibfnamefont {C.}~\bibnamefont {Rabelo}},
  \bibinfo {author} {\bibfnamefont {E.~G.~S.}\ \bibnamefont {Neto}}, \bibinfo
  {author} {\bibfnamefont {T.~L.}\ \bibnamefont {Vasconcelos}}, \bibinfo
  {author} {\bibfnamefont {J.~L.}\ \bibnamefont {Campos}}, \bibinfo {author}
  {\bibfnamefont {J.~S.}\ \bibnamefont {Lemos}}, \bibinfo {author}
  {\bibfnamefont {V.}~\bibnamefont {Ornelas}}, \bibinfo {author} {\bibfnamefont
  {D.}~\bibnamefont {Miranda}}, \bibinfo {author} {\bibfnamefont
  {R.}~\bibnamefont {Nadas}}, \bibinfo {author} {\bibfnamefont {F.~C.}\
  \bibnamefont {Santana}}, \bibinfo {author} {\bibfnamefont {K.}~\bibnamefont
  {Watanabe}}, \bibinfo {author} {\bibfnamefont {T.}~\bibnamefont {Taniguchi}},
  \bibinfo {author} {\bibfnamefont {B.}~\bibnamefont {van Troeye}}, \bibinfo
  {author} {\bibfnamefont {M.}~\bibnamefont {Lamparski}}, \bibinfo {author}
  {\bibfnamefont {V.}~\bibnamefont {Meunier}}, \bibinfo {author} {\bibfnamefont
  {V.-H.}\ \bibnamefont {Nguyen}}, \bibinfo {author} {\bibfnamefont
  {D.}~\bibnamefont {Paszko}}, \bibinfo {author} {\bibfnamefont {J.-C.}\
  \bibnamefont {Charlier}}, \bibinfo {author} {\bibfnamefont {L.~C.}\
  \bibnamefont {Campos}}, \bibinfo {author} {\bibfnamefont {L.~G.}\
  \bibnamefont {Can{\c{c}}ado}}, \bibinfo {author} {\bibfnamefont
  {G.}~\bibnamefont {Medeiros-Ribeiro}},\ and\ \bibinfo {author} {\bibfnamefont
  {A.}~\bibnamefont {Jorio}},\ }\bibfield  {title} {\bibinfo {title}
  {Localization of lattice dynamics in low-angle twisted bilayer graphene},\
  }\href {https://doi.org/10.1038/s41586-021-03252-5} {\bibfield  {journal}
  {\bibinfo  {journal} {Nature}\ }\textbf {\bibinfo {volume} {590}},\ \bibinfo
  {pages} {405} (\bibinfo {year} {2021})}\BibitemShut {NoStop}%
\bibitem [{\citenamefont {Saito}\ \emph {et~al.}(2020)\citenamefont {Saito},
  \citenamefont {Ge}, \citenamefont {Watanabe}, \citenamefont {Taniguchi},\
  and\ \citenamefont {Young}}]{Saito2020}%
  \BibitemOpen
  \bibfield  {author} {\bibinfo {author} {\bibfnamefont {Y.}~\bibnamefont
  {Saito}}, \bibinfo {author} {\bibfnamefont {J.}~\bibnamefont {Ge}}, \bibinfo
  {author} {\bibfnamefont {K.}~\bibnamefont {Watanabe}}, \bibinfo {author}
  {\bibfnamefont {T.}~\bibnamefont {Taniguchi}},\ and\ \bibinfo {author}
  {\bibfnamefont {A.~F.}\ \bibnamefont {Young}},\ }\bibfield  {title} {\bibinfo
  {title} {Independent superconductors and correlated insulators in twisted
  bilayer graphene},\ }\href {https://doi.org/10.1038/s41567-020-0928-3}
  {\bibfield  {journal} {\bibinfo  {journal} {Nature Physics}\ }\textbf
  {\bibinfo {volume} {16}},\ \bibinfo {pages} {926} (\bibinfo {year}
  {2020})}\BibitemShut {NoStop}%
\bibitem [{\citenamefont {Wong}\ \emph {et~al.}(2020)\citenamefont {Wong},
  \citenamefont {Nuckolls}, \citenamefont {Oh}, \citenamefont {Lian},
  \citenamefont {Xie}, \citenamefont {Jeon}, \citenamefont {Watanabe},
  \citenamefont {Taniguchi}, \citenamefont {Bernevig},\ and\ \citenamefont
  {Yazdani}}]{Wong2020}%
  \BibitemOpen
  \bibfield  {author} {\bibinfo {author} {\bibfnamefont {D.}~\bibnamefont
  {Wong}}, \bibinfo {author} {\bibfnamefont {K.~P.}\ \bibnamefont {Nuckolls}},
  \bibinfo {author} {\bibfnamefont {M.}~\bibnamefont {Oh}}, \bibinfo {author}
  {\bibfnamefont {B.}~\bibnamefont {Lian}}, \bibinfo {author} {\bibfnamefont
  {Y.}~\bibnamefont {Xie}}, \bibinfo {author} {\bibfnamefont {S.}~\bibnamefont
  {Jeon}}, \bibinfo {author} {\bibfnamefont {K.}~\bibnamefont {Watanabe}},
  \bibinfo {author} {\bibfnamefont {T.}~\bibnamefont {Taniguchi}}, \bibinfo
  {author} {\bibfnamefont {B.~A.}\ \bibnamefont {Bernevig}},\ and\ \bibinfo
  {author} {\bibfnamefont {A.}~\bibnamefont {Yazdani}},\ }\bibfield  {title}
  {\bibinfo {title} {Cascade of electronic transitions in magic-angle twisted
  bilayer graphene},\ }\href {https://doi.org/10.1038/s41586-020-2339-0}
  {\bibfield  {journal} {\bibinfo  {journal} {Nature}\ }\textbf {\bibinfo
  {volume} {582}},\ \bibinfo {pages} {198} (\bibinfo {year}
  {2020})}\BibitemShut {NoStop}%
\bibitem [{\citenamefont {Andrei}\ \emph {et~al.}(2021)\citenamefont {Andrei},
  \citenamefont {Efetov}, \citenamefont {Jarillo-Herrero}, \citenamefont
  {MacDonald}, \citenamefont {Mak}, \citenamefont {Senthil}, \citenamefont
  {Tutuc}, \citenamefont {Yazdani},\ and\ \citenamefont {Young}}]{Andrei2021}%
  \BibitemOpen
  \bibfield  {author} {\bibinfo {author} {\bibfnamefont {E.~Y.}\ \bibnamefont
  {Andrei}}, \bibinfo {author} {\bibfnamefont {D.~K.}\ \bibnamefont {Efetov}},
  \bibinfo {author} {\bibfnamefont {P.}~\bibnamefont {Jarillo-Herrero}},
  \bibinfo {author} {\bibfnamefont {A.~H.}\ \bibnamefont {MacDonald}}, \bibinfo
  {author} {\bibfnamefont {K.~F.}\ \bibnamefont {Mak}}, \bibinfo {author}
  {\bibfnamefont {T.}~\bibnamefont {Senthil}}, \bibinfo {author} {\bibfnamefont
  {E.}~\bibnamefont {Tutuc}}, \bibinfo {author} {\bibfnamefont
  {A.}~\bibnamefont {Yazdani}},\ and\ \bibinfo {author} {\bibfnamefont {A.~F.}\
  \bibnamefont {Young}},\ }\bibfield  {title} {\bibinfo {title} {The marvels of
  moir{\'e} materials},\ }\href {https://doi.org/10.1038/s41578-021-00284-1}
  {\bibfield  {journal} {\bibinfo  {journal} {Nature Reviews Materials}\
  }\textbf {\bibinfo {volume} {6}},\ \bibinfo {pages} {201} (\bibinfo {year}
  {2021})}\BibitemShut {NoStop}%
\bibitem [{\citenamefont {Carr}\ \emph {et~al.}(2020)\citenamefont {Carr},
  \citenamefont {Fang},\ and\ \citenamefont {Kaxiras}}]{Carr2020_review}%
  \BibitemOpen
  \bibfield  {author} {\bibinfo {author} {\bibfnamefont {S.}~\bibnamefont
  {Carr}}, \bibinfo {author} {\bibfnamefont {S.}~\bibnamefont {Fang}},\ and\
  \bibinfo {author} {\bibfnamefont {E.}~\bibnamefont {Kaxiras}},\ }\bibfield
  {title} {\bibinfo {title} {Electronic-structure methods for twisted
  moir{\'{e}} layers},\ }\href {https://doi.org/10.1038/s41578-020-0214-0}
  {\bibfield  {journal} {\bibinfo  {journal} {Nature Reviews Materials}\
  }\textbf {\bibinfo {volume} {5}},\ \bibinfo {pages} {748} (\bibinfo {year}
  {2020})}\BibitemShut {NoStop}%
\bibitem [{\citenamefont {Lau}\ \emph {et~al.}(2022)\citenamefont {Lau},
  \citenamefont {Bockrath}, \citenamefont {Mak},\ and\ \citenamefont
  {Zhang}}]{Lau2022}%
  \BibitemOpen
  \bibfield  {author} {\bibinfo {author} {\bibfnamefont {C.~N.}\ \bibnamefont
  {Lau}}, \bibinfo {author} {\bibfnamefont {M.~W.}\ \bibnamefont {Bockrath}},
  \bibinfo {author} {\bibfnamefont {K.~F.}\ \bibnamefont {Mak}},\ and\ \bibinfo
  {author} {\bibfnamefont {F.}~\bibnamefont {Zhang}},\ }\bibfield  {title}
  {\bibinfo {title} {Reproducibility in the fabrication and physics of
  moir{\'{e}} materials},\ }\href {https://doi.org/10.1038/s41586-021-04173-z}
  {\bibfield  {journal} {\bibinfo  {journal} {Nature}\ }\textbf {\bibinfo
  {volume} {602}},\ \bibinfo {pages} {41} (\bibinfo {year} {2022})}\BibitemShut
  {NoStop}%
\bibitem [{\citenamefont {Bistritzer}\ and\ \citenamefont
  {MacDonald}(2011)}]{Bistritzer_PNAS}%
  \BibitemOpen
  \bibfield  {author} {\bibinfo {author} {\bibfnamefont {R.}~\bibnamefont
  {Bistritzer}}\ and\ \bibinfo {author} {\bibfnamefont {A.~H.}\ \bibnamefont
  {MacDonald}},\ }\bibfield  {title} {\bibinfo {title} {Moiré bands in twisted
  double-layer graphene},\ }\href {https://doi.org/10.1073/pnas.1108174108}
  {\bibfield  {journal} {\bibinfo  {journal} {Proceedings of the National
  Academy of Sciences}\ }\textbf {\bibinfo {volume} {108}},\ \bibinfo {pages}
  {12233} (\bibinfo {year} {2011})}\BibitemShut {NoStop}%
\bibitem [{\citenamefont {Carr}\ \emph {et~al.}(2017)\citenamefont {Carr},
  \citenamefont {Massatt}, \citenamefont {Fang}, \citenamefont {Cazeaux},
  \citenamefont {Luskin},\ and\ \citenamefont
  {Kaxiras}}]{Carr2017_twistronics}%
  \BibitemOpen
  \bibfield  {author} {\bibinfo {author} {\bibfnamefont {S.}~\bibnamefont
  {Carr}}, \bibinfo {author} {\bibfnamefont {D.}~\bibnamefont {Massatt}},
  \bibinfo {author} {\bibfnamefont {S.}~\bibnamefont {Fang}}, \bibinfo {author}
  {\bibfnamefont {P.}~\bibnamefont {Cazeaux}}, \bibinfo {author} {\bibfnamefont
  {M.}~\bibnamefont {Luskin}},\ and\ \bibinfo {author} {\bibfnamefont
  {E.}~\bibnamefont {Kaxiras}},\ }\bibfield  {title} {\bibinfo {title}
  {Twistronics: Manipulating the electronic properties of two-dimensional
  layered structures through their twist angle},\ }\href
  {https://doi.org/10.1103/PhysRevB.95.075420} {\bibfield  {journal} {\bibinfo
  {journal} {Physical Review B}\ }\textbf {\bibinfo {volume} {95}},\ \bibinfo
  {pages} {075420} (\bibinfo {year} {2017})}\BibitemShut {NoStop}%
\bibitem [{\citenamefont {Song}\ \emph {et~al.}(2019)\citenamefont {Song},
  \citenamefont {Wang}, \citenamefont {Shi}, \citenamefont {Li}, \citenamefont
  {Fang},\ and\ \citenamefont {Bernevig}}]{Song_PRL}%
  \BibitemOpen
  \bibfield  {author} {\bibinfo {author} {\bibfnamefont {Z.}~\bibnamefont
  {Song}}, \bibinfo {author} {\bibfnamefont {Z.}~\bibnamefont {Wang}}, \bibinfo
  {author} {\bibfnamefont {W.}~\bibnamefont {Shi}}, \bibinfo {author}
  {\bibfnamefont {G.}~\bibnamefont {Li}}, \bibinfo {author} {\bibfnamefont
  {C.}~\bibnamefont {Fang}},\ and\ \bibinfo {author} {\bibfnamefont {B.~A.}\
  \bibnamefont {Bernevig}},\ }\bibfield  {title} {\bibinfo {title} {All magic
  angles in twisted bilayer graphene are topological},\ }\href
  {https://doi.org/10.1103/PhysRevLett.123.036401} {\bibfield  {journal}
  {\bibinfo  {journal} {Phys. Rev. Lett.}\ }\textbf {\bibinfo {volume} {123}},\
  \bibinfo {pages} {036401} (\bibinfo {year} {2019})}\BibitemShut {NoStop}%
\bibitem [{\citenamefont {Angeli}\ \emph {et~al.}(2018)\citenamefont {Angeli},
  \citenamefont {Mandelli}, \citenamefont {Valli}, \citenamefont {Amaricci},
  \citenamefont {Capone}, \citenamefont {Tosatti},\ and\ \citenamefont
  {Fabrizio}}]{Fabrizio_PRB}%
  \BibitemOpen
  \bibfield  {author} {\bibinfo {author} {\bibfnamefont {M.}~\bibnamefont
  {Angeli}}, \bibinfo {author} {\bibfnamefont {D.}~\bibnamefont {Mandelli}},
  \bibinfo {author} {\bibfnamefont {A.}~\bibnamefont {Valli}}, \bibinfo
  {author} {\bibfnamefont {A.}~\bibnamefont {Amaricci}}, \bibinfo {author}
  {\bibfnamefont {M.}~\bibnamefont {Capone}}, \bibinfo {author} {\bibfnamefont
  {E.}~\bibnamefont {Tosatti}},\ and\ \bibinfo {author} {\bibfnamefont
  {M.}~\bibnamefont {Fabrizio}},\ }\bibfield  {title} {\bibinfo {title}
  {Emergent ${D}_{6}$ symmetry in fully relaxed magic-angle twisted bilayer
  graphene},\ }\href {https://doi.org/10.1103/PhysRevB.98.235137} {\bibfield
  {journal} {\bibinfo  {journal} {Phys. Rev. B}\ }\textbf {\bibinfo {volume}
  {98}},\ \bibinfo {pages} {235137} (\bibinfo {year} {2018})}\BibitemShut
  {NoStop}%
\bibitem [{\citenamefont {Po}\ \emph {et~al.}(2018)\citenamefont {Po},
  \citenamefont {Zou}, \citenamefont {Vishwanath},\ and\ \citenamefont
  {Senthil}}]{Po_PRX}%
  \BibitemOpen
  \bibfield  {author} {\bibinfo {author} {\bibfnamefont {H.~C.}\ \bibnamefont
  {Po}}, \bibinfo {author} {\bibfnamefont {L.}~\bibnamefont {Zou}}, \bibinfo
  {author} {\bibfnamefont {A.}~\bibnamefont {Vishwanath}},\ and\ \bibinfo
  {author} {\bibfnamefont {T.}~\bibnamefont {Senthil}},\ }\bibfield  {title}
  {\bibinfo {title} {Origin of mott insulating behavior and superconductivity
  in twisted bilayer graphene},\ }\href
  {https://doi.org/10.1103/PhysRevX.8.031089} {\bibfield  {journal} {\bibinfo
  {journal} {Phys. Rev. X}\ }\textbf {\bibinfo {volume} {8}},\ \bibinfo {pages}
  {031089} (\bibinfo {year} {2018})}\BibitemShut {NoStop}%
\bibitem [{\citenamefont {Angeli}\ \emph {et~al.}(2019)\citenamefont {Angeli},
  \citenamefont {Tosatti},\ and\ \citenamefont {Fabrizio}}]{Fabrizio_PRX}%
  \BibitemOpen
  \bibfield  {author} {\bibinfo {author} {\bibfnamefont {M.}~\bibnamefont
  {Angeli}}, \bibinfo {author} {\bibfnamefont {E.}~\bibnamefont {Tosatti}},\
  and\ \bibinfo {author} {\bibfnamefont {M.}~\bibnamefont {Fabrizio}},\
  }\bibfield  {title} {\bibinfo {title} {Valley jahn-teller effect in twisted
  bilayer graphene},\ }\href {https://doi.org/10.1103/PhysRevX.9.041010}
  {\bibfield  {journal} {\bibinfo  {journal} {Phys. Rev. X}\ }\textbf {\bibinfo
  {volume} {9}},\ \bibinfo {pages} {041010} (\bibinfo {year}
  {2019})}\BibitemShut {NoStop}%
\bibitem [{\citenamefont {Yoo}\ \emph {et~al.}(2019)\citenamefont {Yoo},
  \citenamefont {Engelke}, \citenamefont {Carr}, \citenamefont {Fang},
  \citenamefont {Zhang}, \citenamefont {Cazeaux}, \citenamefont {Sung},
  \citenamefont {Hovden}, \citenamefont {Tsen}, \citenamefont {Taniguchi},
  \citenamefont {Watanabe}, \citenamefont {Yi}, \citenamefont {Kim},
  \citenamefont {Luskin}, \citenamefont {Tadmor}, \citenamefont {Kaxiras},\
  and\ \citenamefont {Kim}}]{Yoo2019}%
  \BibitemOpen
  \bibfield  {author} {\bibinfo {author} {\bibfnamefont {H.}~\bibnamefont
  {Yoo}}, \bibinfo {author} {\bibfnamefont {R.}~\bibnamefont {Engelke}},
  \bibinfo {author} {\bibfnamefont {S.}~\bibnamefont {Carr}}, \bibinfo {author}
  {\bibfnamefont {S.}~\bibnamefont {Fang}}, \bibinfo {author} {\bibfnamefont
  {K.}~\bibnamefont {Zhang}}, \bibinfo {author} {\bibfnamefont
  {P.}~\bibnamefont {Cazeaux}}, \bibinfo {author} {\bibfnamefont {S.~H.}\
  \bibnamefont {Sung}}, \bibinfo {author} {\bibfnamefont {R.}~\bibnamefont
  {Hovden}}, \bibinfo {author} {\bibfnamefont {A.~W.}\ \bibnamefont {Tsen}},
  \bibinfo {author} {\bibfnamefont {T.}~\bibnamefont {Taniguchi}}, \bibinfo
  {author} {\bibfnamefont {K.}~\bibnamefont {Watanabe}}, \bibinfo {author}
  {\bibfnamefont {G.-C.}\ \bibnamefont {Yi}}, \bibinfo {author} {\bibfnamefont
  {M.}~\bibnamefont {Kim}}, \bibinfo {author} {\bibfnamefont {M.}~\bibnamefont
  {Luskin}}, \bibinfo {author} {\bibfnamefont {E.~B.}\ \bibnamefont {Tadmor}},
  \bibinfo {author} {\bibfnamefont {E.}~\bibnamefont {Kaxiras}},\ and\ \bibinfo
  {author} {\bibfnamefont {P.}~\bibnamefont {Kim}},\ }\bibfield  {title}
  {\bibinfo {title} {Atomic and electronic reconstruction at the van der waals
  interface in twisted bilayer graphene},\ }\href
  {https://doi.org/10.1038/s41563-019-0346-z} {\bibfield  {journal} {\bibinfo
  {journal} {Nature Materials}\ }\textbf {\bibinfo {volume} {18}},\ \bibinfo
  {pages} {448} (\bibinfo {year} {2019})}\BibitemShut {NoStop}%
\bibitem [{\citenamefont {Lian}\ \emph {et~al.}(2019)\citenamefont {Lian},
  \citenamefont {Wang},\ and\ \citenamefont {Bernevig}}]{Bernevig_PRL}%
  \BibitemOpen
  \bibfield  {author} {\bibinfo {author} {\bibfnamefont {B.}~\bibnamefont
  {Lian}}, \bibinfo {author} {\bibfnamefont {Z.}~\bibnamefont {Wang}},\ and\
  \bibinfo {author} {\bibfnamefont {B.~A.}\ \bibnamefont {Bernevig}},\
  }\bibfield  {title} {\bibinfo {title} {Twisted bilayer graphene: A
  phonon-driven superconductor},\ }\href
  {https://doi.org/10.1103/PhysRevLett.122.257002} {\bibfield  {journal}
  {\bibinfo  {journal} {Phys. Rev. Lett.}\ }\textbf {\bibinfo {volume} {122}},\
  \bibinfo {pages} {257002} (\bibinfo {year} {2019})}\BibitemShut {NoStop}%
\bibitem [{\citenamefont {Blason}\ and\ \citenamefont
  {Fabrizio}(2022)}]{Blason2021}%
  \BibitemOpen
  \bibfield  {author} {\bibinfo {author} {\bibfnamefont {A.}~\bibnamefont
  {Blason}}\ and\ \bibinfo {author} {\bibfnamefont {M.}~\bibnamefont
  {Fabrizio}},\ }\bibfield  {title} {\bibinfo {title} {Topological jahn-teller
  mott insulators in twisted bilayer graphene},\ }\Eprint
  {https://arxiv.org/abs/2204.05190} {arXiv:2204.05190}  (\bibinfo {year}
  {2022})\BibitemShut {NoStop}%
\bibitem [{\citenamefont {Ghiotto}\ \emph {et~al.}(2021)\citenamefont
  {Ghiotto}, \citenamefont {Shih}, \citenamefont {Pereira}, \citenamefont
  {Rhodes}, \citenamefont {Kim}, \citenamefont {Zang}, \citenamefont {Millis},
  \citenamefont {Watanabe}, \citenamefont {Taniguchi}, \citenamefont {Hone},
  \citenamefont {Wang}, \citenamefont {Dean},\ and\ \citenamefont
  {Pasupathy}}]{Ghiotto_2021}%
  \BibitemOpen
  \bibfield  {author} {\bibinfo {author} {\bibfnamefont {A.}~\bibnamefont
  {Ghiotto}}, \bibinfo {author} {\bibfnamefont {E.-M.}\ \bibnamefont {Shih}},
  \bibinfo {author} {\bibfnamefont {G.~S. S.~G.}\ \bibnamefont {Pereira}},
  \bibinfo {author} {\bibfnamefont {D.~A.}\ \bibnamefont {Rhodes}}, \bibinfo
  {author} {\bibfnamefont {B.}~\bibnamefont {Kim}}, \bibinfo {author}
  {\bibfnamefont {J.}~\bibnamefont {Zang}}, \bibinfo {author} {\bibfnamefont
  {A.~J.}\ \bibnamefont {Millis}}, \bibinfo {author} {\bibfnamefont
  {K.}~\bibnamefont {Watanabe}}, \bibinfo {author} {\bibfnamefont
  {T.}~\bibnamefont {Taniguchi}}, \bibinfo {author} {\bibfnamefont {J.~C.}\
  \bibnamefont {Hone}}, \bibinfo {author} {\bibfnamefont {L.}~\bibnamefont
  {Wang}}, \bibinfo {author} {\bibfnamefont {C.~R.}\ \bibnamefont {Dean}},\
  and\ \bibinfo {author} {\bibfnamefont {A.~N.}\ \bibnamefont {Pasupathy}},\
  }\bibfield  {title} {\bibinfo {title} {Quantum criticality in twisted
  transition metal dichalcogenides},\ }\href
  {https://doi.org/10.1038/s41586-021-03815-6} {\bibfield  {journal} {\bibinfo
  {journal} {Nature}\ }\textbf {\bibinfo {volume} {597}},\ \bibinfo {pages}
  {345} (\bibinfo {year} {2021})}\BibitemShut {NoStop}%
\bibitem [{\citenamefont {Grzeszczyk}\ \emph {et~al.}(2021)\citenamefont
  {Grzeszczyk}, \citenamefont {Szpakowski}, \citenamefont {Slobodeniuk},
  \citenamefont {Kazimierczuk}, \citenamefont {Bhatnagar}, \citenamefont
  {Taniguchi}, \citenamefont {Watanabe}, \citenamefont {Kossacki},
  \citenamefont {Potemski}, \citenamefont {Babi{\'{n}}ski},\ and\ \citenamefont
  {Molas}}]{Grzeszczyk_2021}%
  \BibitemOpen
  \bibfield  {author} {\bibinfo {author} {\bibfnamefont {M.}~\bibnamefont
  {Grzeszczyk}}, \bibinfo {author} {\bibfnamefont {J.}~\bibnamefont
  {Szpakowski}}, \bibinfo {author} {\bibfnamefont {A.~O.}\ \bibnamefont
  {Slobodeniuk}}, \bibinfo {author} {\bibfnamefont {T.}~\bibnamefont
  {Kazimierczuk}}, \bibinfo {author} {\bibfnamefont {M.}~\bibnamefont
  {Bhatnagar}}, \bibinfo {author} {\bibfnamefont {T.}~\bibnamefont
  {Taniguchi}}, \bibinfo {author} {\bibfnamefont {K.}~\bibnamefont {Watanabe}},
  \bibinfo {author} {\bibfnamefont {P.}~\bibnamefont {Kossacki}}, \bibinfo
  {author} {\bibfnamefont {M.}~\bibnamefont {Potemski}}, \bibinfo {author}
  {\bibfnamefont {A.}~\bibnamefont {Babi{\'{n}}ski}},\ and\ \bibinfo {author}
  {\bibfnamefont {M.~R.}\ \bibnamefont {Molas}},\ }\bibfield  {title} {\bibinfo
  {title} {The optical response of artificially twisted mos$_2$ bilayers},\
  }\href {https://doi.org/10.1038/s41598-021-95700-5} {\bibfield  {journal}
  {\bibinfo  {journal} {Scientific Reports}\ }\textbf {\bibinfo {volume}
  {11}},\ \bibinfo {pages} {17037} (\bibinfo {year} {2021})}\BibitemShut
  {NoStop}%
\bibitem [{\citenamefont {Xu}\ \emph {et~al.}(2022)\citenamefont {Xu},
  \citenamefont {Kang}, \citenamefont {Watanabe}, \citenamefont {Taniguchi},
  \citenamefont {Mak},\ and\ \citenamefont {Shan}}]{Jie_2022}%
  \BibitemOpen
  \bibfield  {author} {\bibinfo {author} {\bibfnamefont {Y.}~\bibnamefont
  {Xu}}, \bibinfo {author} {\bibfnamefont {K.}~\bibnamefont {Kang}}, \bibinfo
  {author} {\bibfnamefont {K.}~\bibnamefont {Watanabe}}, \bibinfo {author}
  {\bibfnamefont {T.}~\bibnamefont {Taniguchi}}, \bibinfo {author}
  {\bibfnamefont {K.~F.}\ \bibnamefont {Mak}},\ and\ \bibinfo {author}
  {\bibfnamefont {J.}~\bibnamefont {Shan}},\ }\bibfield  {title} {\bibinfo
  {title} {Tunable bilayer hubbard model physics in twisted wse2},\ }\Eprint
  {https://arxiv.org/abs/2202.02055} {arXiv:2202.02055}  (\bibinfo {year}
  {2022})\BibitemShut {NoStop}%
\bibitem [{\citenamefont {Tang}\ \emph {et~al.}(2020)\citenamefont {Tang},
  \citenamefont {Li}, \citenamefont {Li}, \citenamefont {Xu}, \citenamefont
  {Liu}, \citenamefont {Barmak}, \citenamefont {Watanabe}, \citenamefont
  {Taniguchi}, \citenamefont {MacDonald}, \citenamefont {Shan},\ and\
  \citenamefont {Mak}}]{Tang_2020}%
  \BibitemOpen
  \bibfield  {author} {\bibinfo {author} {\bibfnamefont {Y.}~\bibnamefont
  {Tang}}, \bibinfo {author} {\bibfnamefont {L.}~\bibnamefont {Li}}, \bibinfo
  {author} {\bibfnamefont {T.}~\bibnamefont {Li}}, \bibinfo {author}
  {\bibfnamefont {Y.}~\bibnamefont {Xu}}, \bibinfo {author} {\bibfnamefont
  {S.}~\bibnamefont {Liu}}, \bibinfo {author} {\bibfnamefont {K.}~\bibnamefont
  {Barmak}}, \bibinfo {author} {\bibfnamefont {K.}~\bibnamefont {Watanabe}},
  \bibinfo {author} {\bibfnamefont {T.}~\bibnamefont {Taniguchi}}, \bibinfo
  {author} {\bibfnamefont {A.~H.}\ \bibnamefont {MacDonald}}, \bibinfo {author}
  {\bibfnamefont {J.}~\bibnamefont {Shan}},\ and\ \bibinfo {author}
  {\bibfnamefont {K.~F.}\ \bibnamefont {Mak}},\ }\bibfield  {title} {\bibinfo
  {title} {Simulation of hubbard model physics in wse2/ws2 moir{\'e}
  superlattices},\ }\href {https://doi.org/10.1038/s41586-020-2085-3}
  {\bibfield  {journal} {\bibinfo  {journal} {Nature}\ }\textbf {\bibinfo
  {volume} {579}},\ \bibinfo {pages} {353} (\bibinfo {year}
  {2020})}\BibitemShut {NoStop}%
\bibitem [{\citenamefont {Wang}\ \emph {et~al.}(2020)\citenamefont {Wang},
  \citenamefont {Shih}, \citenamefont {Ghiotto}, \citenamefont {Xian},
  \citenamefont {Rhodes}, \citenamefont {Tan}, \citenamefont {Claassen},
  \citenamefont {Kennes}, \citenamefont {Bai}, \citenamefont {Kim},
  \citenamefont {Watanabe}, \citenamefont {Taniguchi}, \citenamefont {Zhu},
  \citenamefont {Hone}, \citenamefont {Rubio}, \citenamefont {Pasupathy},\ and\
  \citenamefont {Dean}}]{Wang_2020}%
  \BibitemOpen
  \bibfield  {author} {\bibinfo {author} {\bibfnamefont {L.}~\bibnamefont
  {Wang}}, \bibinfo {author} {\bibfnamefont {E.-M.}\ \bibnamefont {Shih}},
  \bibinfo {author} {\bibfnamefont {A.}~\bibnamefont {Ghiotto}}, \bibinfo
  {author} {\bibfnamefont {L.}~\bibnamefont {Xian}}, \bibinfo {author}
  {\bibfnamefont {D.~A.}\ \bibnamefont {Rhodes}}, \bibinfo {author}
  {\bibfnamefont {C.}~\bibnamefont {Tan}}, \bibinfo {author} {\bibfnamefont
  {M.}~\bibnamefont {Claassen}}, \bibinfo {author} {\bibfnamefont {D.~M.}\
  \bibnamefont {Kennes}}, \bibinfo {author} {\bibfnamefont {Y.}~\bibnamefont
  {Bai}}, \bibinfo {author} {\bibfnamefont {B.}~\bibnamefont {Kim}}, \bibinfo
  {author} {\bibfnamefont {K.}~\bibnamefont {Watanabe}}, \bibinfo {author}
  {\bibfnamefont {T.}~\bibnamefont {Taniguchi}}, \bibinfo {author}
  {\bibfnamefont {X.}~\bibnamefont {Zhu}}, \bibinfo {author} {\bibfnamefont
  {J.}~\bibnamefont {Hone}}, \bibinfo {author} {\bibfnamefont {A.}~\bibnamefont
  {Rubio}}, \bibinfo {author} {\bibfnamefont {A.~N.}\ \bibnamefont
  {Pasupathy}},\ and\ \bibinfo {author} {\bibfnamefont {C.~R.}\ \bibnamefont
  {Dean}},\ }\bibfield  {title} {\bibinfo {title} {Correlated electronic phases
  in twisted bilayer transition metal dichalcogenides},\ }\href
  {https://doi.org/10.1038/s41563-020-0708-6} {\bibfield  {journal} {\bibinfo
  {journal} {Nature Materials}\ }\textbf {\bibinfo {volume} {19}},\ \bibinfo
  {pages} {861} (\bibinfo {year} {2020})}\BibitemShut {NoStop}%
\bibitem [{\citenamefont {Tran}\ \emph {et~al.}(2019)\citenamefont {Tran},
  \citenamefont {Moody}, \citenamefont {Wu}, \citenamefont {Lu}, \citenamefont
  {Choi}, \citenamefont {Kim}, \citenamefont {Rai}, \citenamefont {Sanchez},
  \citenamefont {Quan}, \citenamefont {Singh}, \citenamefont {Embley},
  \citenamefont {Zepeda}, \citenamefont {Campbell}, \citenamefont {Autry},
  \citenamefont {Taniguchi}, \citenamefont {Watanabe}, \citenamefont {Lu},
  \citenamefont {Banerjee}, \citenamefont {Silverman}, \citenamefont {Kim},
  \citenamefont {Tutuc}, \citenamefont {Yang}, \citenamefont {MacDonald},\ and\
  \citenamefont {Li}}]{Tran_2019}%
  \BibitemOpen
  \bibfield  {author} {\bibinfo {author} {\bibfnamefont {K.}~\bibnamefont
  {Tran}}, \bibinfo {author} {\bibfnamefont {G.}~\bibnamefont {Moody}},
  \bibinfo {author} {\bibfnamefont {F.}~\bibnamefont {Wu}}, \bibinfo {author}
  {\bibfnamefont {X.}~\bibnamefont {Lu}}, \bibinfo {author} {\bibfnamefont
  {J.}~\bibnamefont {Choi}}, \bibinfo {author} {\bibfnamefont {K.}~\bibnamefont
  {Kim}}, \bibinfo {author} {\bibfnamefont {A.}~\bibnamefont {Rai}}, \bibinfo
  {author} {\bibfnamefont {D.~A.}\ \bibnamefont {Sanchez}}, \bibinfo {author}
  {\bibfnamefont {J.}~\bibnamefont {Quan}}, \bibinfo {author} {\bibfnamefont
  {A.}~\bibnamefont {Singh}}, \bibinfo {author} {\bibfnamefont
  {J.}~\bibnamefont {Embley}}, \bibinfo {author} {\bibfnamefont
  {A.}~\bibnamefont {Zepeda}}, \bibinfo {author} {\bibfnamefont
  {M.}~\bibnamefont {Campbell}}, \bibinfo {author} {\bibfnamefont
  {T.}~\bibnamefont {Autry}}, \bibinfo {author} {\bibfnamefont
  {T.}~\bibnamefont {Taniguchi}}, \bibinfo {author} {\bibfnamefont
  {K.}~\bibnamefont {Watanabe}}, \bibinfo {author} {\bibfnamefont
  {N.}~\bibnamefont {Lu}}, \bibinfo {author} {\bibfnamefont {S.~K.}\
  \bibnamefont {Banerjee}}, \bibinfo {author} {\bibfnamefont {K.~L.}\
  \bibnamefont {Silverman}}, \bibinfo {author} {\bibfnamefont {S.}~\bibnamefont
  {Kim}}, \bibinfo {author} {\bibfnamefont {E.}~\bibnamefont {Tutuc}}, \bibinfo
  {author} {\bibfnamefont {L.}~\bibnamefont {Yang}}, \bibinfo {author}
  {\bibfnamefont {A.~H.}\ \bibnamefont {MacDonald}},\ and\ \bibinfo {author}
  {\bibfnamefont {X.}~\bibnamefont {Li}},\ }\bibfield  {title} {\bibinfo
  {title} {Evidence for moir{\'e} excitons in van der waals heterostructures},\
  }\href {https://doi.org/10.1038/s41586-019-0975-z} {\bibfield  {journal}
  {\bibinfo  {journal} {Nature}\ }\textbf {\bibinfo {volume} {567}},\ \bibinfo
  {pages} {71} (\bibinfo {year} {2019})}\BibitemShut {NoStop}%
\bibitem [{\citenamefont {Kennes}\ \emph {et~al.}(2021)\citenamefont {Kennes},
  \citenamefont {Claassen}, \citenamefont {Xian}, \citenamefont {Georges},
  \citenamefont {Millis}, \citenamefont {Hone}, \citenamefont {Dean},
  \citenamefont {Basov}, \citenamefont {Pasupathy},\ and\ \citenamefont
  {Rubio}}]{Kennes2021}%
  \BibitemOpen
  \bibfield  {author} {\bibinfo {author} {\bibfnamefont {D.~M.}\ \bibnamefont
  {Kennes}}, \bibinfo {author} {\bibfnamefont {M.}~\bibnamefont {Claassen}},
  \bibinfo {author} {\bibfnamefont {L.}~\bibnamefont {Xian}}, \bibinfo {author}
  {\bibfnamefont {A.}~\bibnamefont {Georges}}, \bibinfo {author} {\bibfnamefont
  {A.~J.}\ \bibnamefont {Millis}}, \bibinfo {author} {\bibfnamefont
  {J.}~\bibnamefont {Hone}}, \bibinfo {author} {\bibfnamefont {C.~R.}\
  \bibnamefont {Dean}}, \bibinfo {author} {\bibfnamefont {D.~N.}\ \bibnamefont
  {Basov}}, \bibinfo {author} {\bibfnamefont {A.~N.}\ \bibnamefont
  {Pasupathy}},\ and\ \bibinfo {author} {\bibfnamefont {A.}~\bibnamefont
  {Rubio}},\ }\bibfield  {title} {\bibinfo {title} {Moir{\'{e}}
  heterostructures as a condensed-matter quantum simulator},\ }\href
  {https://doi.org/10.1038/s41567-020-01154-3} {\bibfield  {journal} {\bibinfo
  {journal} {Nature Physics}\ }\textbf {\bibinfo {volume} {17}},\ \bibinfo
  {pages} {155} (\bibinfo {year} {2021})}\BibitemShut {NoStop}%
\bibitem [{\citenamefont {Tritsaris}\ \emph {et~al.}(2021)\citenamefont
  {Tritsaris}, \citenamefont {Carr},\ and\ \citenamefont
  {Schleder}}]{Tritsaris2021}%
  \BibitemOpen
  \bibfield  {author} {\bibinfo {author} {\bibfnamefont {G.~A.}\ \bibnamefont
  {Tritsaris}}, \bibinfo {author} {\bibfnamefont {S.}~\bibnamefont {Carr}},\
  and\ \bibinfo {author} {\bibfnamefont {G.~R.}\ \bibnamefont {Schleder}},\
  }\bibfield  {title} {\bibinfo {title} {Computational design of moir{\'{e}}
  assemblies aided by artificial intelligence},\ }\href
  {https://doi.org/10.1063/5.0044511} {\bibfield  {journal} {\bibinfo
  {journal} {Applied Physics Reviews}\ }\textbf {\bibinfo {volume} {8}},\
  \bibinfo {pages} {031401} (\bibinfo {year} {2021})}\BibitemShut {NoStop}%
\bibitem [{\citenamefont {Defo}\ \emph {et~al.}(2016)\citenamefont {Defo},
  \citenamefont {Fang}, \citenamefont {Shirodkar}, \citenamefont {Tritsaris},
  \citenamefont {Dimoulas},\ and\ \citenamefont {Kaxiras}}]{Rodrick_PRB}%
  \BibitemOpen
  \bibfield  {author} {\bibinfo {author} {\bibfnamefont {R.~K.}\ \bibnamefont
  {Defo}}, \bibinfo {author} {\bibfnamefont {S.}~\bibnamefont {Fang}}, \bibinfo
  {author} {\bibfnamefont {S.~N.}\ \bibnamefont {Shirodkar}}, \bibinfo {author}
  {\bibfnamefont {G.~A.}\ \bibnamefont {Tritsaris}}, \bibinfo {author}
  {\bibfnamefont {A.}~\bibnamefont {Dimoulas}},\ and\ \bibinfo {author}
  {\bibfnamefont {E.}~\bibnamefont {Kaxiras}},\ }\bibfield  {title} {\bibinfo
  {title} {Strain dependence of band gaps and exciton energies in pure and
  mixed transition-metal dichalcogenides},\ }\href
  {https://doi.org/10.1103/PhysRevB.94.155310} {\bibfield  {journal} {\bibinfo
  {journal} {Phys. Rev. B}\ }\textbf {\bibinfo {volume} {94}},\ \bibinfo
  {pages} {155310} (\bibinfo {year} {2016})}\BibitemShut {NoStop}%
\bibitem [{\citenamefont {Zhang}\ \emph {et~al.}(2017)\citenamefont {Zhang},
  \citenamefont {Jia}, \citenamefont {Kholmanov}, \citenamefont {Dong},
  \citenamefont {Er}, \citenamefont {Chen}, \citenamefont {Guo}, \citenamefont
  {Jin}, \citenamefont {Shenoy}, \citenamefont {Shi},\ and\ \citenamefont
  {Lou}}]{Zhang_2017}%
  \BibitemOpen
  \bibfield  {author} {\bibinfo {author} {\bibfnamefont {J.}~\bibnamefont
  {Zhang}}, \bibinfo {author} {\bibfnamefont {S.}~\bibnamefont {Jia}}, \bibinfo
  {author} {\bibfnamefont {I.}~\bibnamefont {Kholmanov}}, \bibinfo {author}
  {\bibfnamefont {L.}~\bibnamefont {Dong}}, \bibinfo {author} {\bibfnamefont
  {D.}~\bibnamefont {Er}}, \bibinfo {author} {\bibfnamefont {W.}~\bibnamefont
  {Chen}}, \bibinfo {author} {\bibfnamefont {H.}~\bibnamefont {Guo}}, \bibinfo
  {author} {\bibfnamefont {Z.}~\bibnamefont {Jin}}, \bibinfo {author}
  {\bibfnamefont {V.~B.}\ \bibnamefont {Shenoy}}, \bibinfo {author}
  {\bibfnamefont {L.}~\bibnamefont {Shi}},\ and\ \bibinfo {author}
  {\bibfnamefont {J.}~\bibnamefont {Lou}},\ }\bibfield  {title} {\bibinfo
  {title} {Janus monolayer transition-metal dichalcogenides},\ }\href
  {https://doi.org/10.1021/acsnano.7b03186} {\bibfield  {journal} {\bibinfo
  {journal} {ACS Nano}\ }\textbf {\bibinfo {volume} {11}},\ \bibinfo {pages}
  {8192} (\bibinfo {year} {2017})}\BibitemShut {NoStop}%
\bibitem [{\citenamefont {Wu}\ \emph {et~al.}(2018)\citenamefont {Wu},
  \citenamefont {Lovorn}, \citenamefont {Tutuc},\ and\ \citenamefont
  {MacDonald}}]{Wu_PRL}%
  \BibitemOpen
  \bibfield  {author} {\bibinfo {author} {\bibfnamefont {F.}~\bibnamefont
  {Wu}}, \bibinfo {author} {\bibfnamefont {T.}~\bibnamefont {Lovorn}}, \bibinfo
  {author} {\bibfnamefont {E.}~\bibnamefont {Tutuc}},\ and\ \bibinfo {author}
  {\bibfnamefont {A.~H.}\ \bibnamefont {MacDonald}},\ }\bibfield  {title}
  {\bibinfo {title} {Hubbard model physics in transition metal dichalcogenide
  moir\'e bands},\ }\href {https://doi.org/10.1103/PhysRevLett.121.026402}
  {\bibfield  {journal} {\bibinfo  {journal} {Phys. Rev. Lett.}\ }\textbf
  {\bibinfo {volume} {121}},\ \bibinfo {pages} {026402} (\bibinfo {year}
  {2018})}\BibitemShut {NoStop}%
\bibitem [{\citenamefont {Angeli}\ and\ \citenamefont
  {MacDonald}(2021)}]{Angeli_PNAS}%
  \BibitemOpen
  \bibfield  {author} {\bibinfo {author} {\bibfnamefont {M.}~\bibnamefont
  {Angeli}}\ and\ \bibinfo {author} {\bibfnamefont {A.~H.}\ \bibnamefont
  {MacDonald}},\ }\bibfield  {title} {\bibinfo {title} {$\gamma$-valley
  transition metal dichalcogenide moiré bands},\ }\href
  {https://doi.org/10.1073/pnas.2021826118} {\bibfield  {journal} {\bibinfo
  {journal} {Proceedings of the National Academy of Sciences}\ }\textbf
  {\bibinfo {volume} {118}},\ \bibinfo {pages} {e2021826118} (\bibinfo {year}
  {2021})}\BibitemShut {NoStop}%
\bibitem [{\citenamefont {Hu}\ and\ \citenamefont {MacDonald}(2021)}]{HF1}%
  \BibitemOpen
  \bibfield  {author} {\bibinfo {author} {\bibfnamefont {N.~C.}\ \bibnamefont
  {Hu}}\ and\ \bibinfo {author} {\bibfnamefont {A.~H.}\ \bibnamefont
  {MacDonald}},\ }\bibfield  {title} {\bibinfo {title} {Competing magnetic
  states in transition metal dichalcogenide moir\'e materials},\ }\href
  {https://doi.org/10.1103/PhysRevB.104.214403} {\bibfield  {journal} {\bibinfo
   {journal} {Phys. Rev. B}\ }\textbf {\bibinfo {volume} {104}},\ \bibinfo
  {pages} {214403} (\bibinfo {year} {2021})}\BibitemShut {NoStop}%
\bibitem [{\citenamefont {Zang}\ \emph {et~al.}(2021)\citenamefont {Zang},
  \citenamefont {Wang}, \citenamefont {Cano},\ and\ \citenamefont
  {Millis}}]{HF2}%
  \BibitemOpen
  \bibfield  {author} {\bibinfo {author} {\bibfnamefont {J.}~\bibnamefont
  {Zang}}, \bibinfo {author} {\bibfnamefont {J.}~\bibnamefont {Wang}}, \bibinfo
  {author} {\bibfnamefont {J.}~\bibnamefont {Cano}},\ and\ \bibinfo {author}
  {\bibfnamefont {A.~J.}\ \bibnamefont {Millis}},\ }\bibfield  {title}
  {\bibinfo {title} {Hartree-fock study of the moir\'e hubbard model for
  twisted bilayer transition metal dichalcogenides},\ }\href
  {https://doi.org/10.1103/PhysRevB.104.075150} {\bibfield  {journal} {\bibinfo
   {journal} {Phys. Rev. B}\ }\textbf {\bibinfo {volume} {104}},\ \bibinfo
  {pages} {075150} (\bibinfo {year} {2021})}\BibitemShut {NoStop}%
\bibitem [{\citenamefont {Pan}\ \emph {et~al.}(2020{\natexlab{a}})\citenamefont
  {Pan}, \citenamefont {Wu},\ and\ \citenamefont {Das~Sarma}}]{HF3}%
  \BibitemOpen
  \bibfield  {author} {\bibinfo {author} {\bibfnamefont {H.}~\bibnamefont
  {Pan}}, \bibinfo {author} {\bibfnamefont {F.}~\bibnamefont {Wu}},\ and\
  \bibinfo {author} {\bibfnamefont {S.}~\bibnamefont {Das~Sarma}},\ }\bibfield
  {title} {\bibinfo {title} {Band topology, hubbard model, heisenberg model,
  and dzyaloshinskii-moriya interaction in twisted bilayer
  ${\mathrm{wse}}_{2}$},\ }\href
  {https://doi.org/10.1103/PhysRevResearch.2.033087} {\bibfield  {journal}
  {\bibinfo  {journal} {Phys. Rev. Research}\ }\textbf {\bibinfo {volume}
  {2}},\ \bibinfo {pages} {033087} (\bibinfo {year}
  {2020}{\natexlab{a}})}\BibitemShut {NoStop}%
\bibitem [{\citenamefont {Pan}\ \emph {et~al.}(2020{\natexlab{b}})\citenamefont
  {Pan}, \citenamefont {Wu},\ and\ \citenamefont {Das~Sarma}}]{HF4}%
  \BibitemOpen
  \bibfield  {author} {\bibinfo {author} {\bibfnamefont {H.}~\bibnamefont
  {Pan}}, \bibinfo {author} {\bibfnamefont {F.}~\bibnamefont {Wu}},\ and\
  \bibinfo {author} {\bibfnamefont {S.}~\bibnamefont {Das~Sarma}},\ }\bibfield
  {title} {\bibinfo {title} {Band topology, hubbard model, heisenberg model,
  and dzyaloshinskii-moriya interaction in twisted bilayer
  ${\mathrm{wse}}_{2}$},\ }\href
  {https://doi.org/10.1103/PhysRevResearch.2.033087} {\bibfield  {journal}
  {\bibinfo  {journal} {Phys. Rev. Research}\ }\textbf {\bibinfo {volume}
  {2}},\ \bibinfo {pages} {033087} (\bibinfo {year}
  {2020}{\natexlab{b}})}\BibitemShut {NoStop}%
\bibitem [{\citenamefont {Liu}\ \emph {et~al.}(2021)\citenamefont {Liu},
  \citenamefont {Khalaf}, \citenamefont {Lee},\ and\ \citenamefont
  {Vishwanath}}]{Liu_PRR}%
  \BibitemOpen
  \bibfield  {author} {\bibinfo {author} {\bibfnamefont {S.}~\bibnamefont
  {Liu}}, \bibinfo {author} {\bibfnamefont {E.}~\bibnamefont {Khalaf}},
  \bibinfo {author} {\bibfnamefont {J.~Y.}\ \bibnamefont {Lee}},\ and\ \bibinfo
  {author} {\bibfnamefont {A.}~\bibnamefont {Vishwanath}},\ }\bibfield  {title}
  {\bibinfo {title} {Nematic topological semimetal and insulator in magic-angle
  bilayer graphene at charge neutrality},\ }\href
  {https://doi.org/10.1103/PhysRevResearch.3.013033} {\bibfield  {journal}
  {\bibinfo  {journal} {Phys. Rev. Research}\ }\textbf {\bibinfo {volume}
  {3}},\ \bibinfo {pages} {013033} (\bibinfo {year} {2021})}\BibitemShut
  {NoStop}%
\bibitem [{\citenamefont {Bultinck}\ \emph {et~al.}(2020)\citenamefont
  {Bultinck}, \citenamefont {Khalaf}, \citenamefont {Liu}, \citenamefont
  {Chatterjee}, \citenamefont {Vishwanath},\ and\ \citenamefont
  {Zaletel}}]{HF5}%
  \BibitemOpen
  \bibfield  {author} {\bibinfo {author} {\bibfnamefont {N.}~\bibnamefont
  {Bultinck}}, \bibinfo {author} {\bibfnamefont {E.}~\bibnamefont {Khalaf}},
  \bibinfo {author} {\bibfnamefont {S.}~\bibnamefont {Liu}}, \bibinfo {author}
  {\bibfnamefont {S.}~\bibnamefont {Chatterjee}}, \bibinfo {author}
  {\bibfnamefont {A.}~\bibnamefont {Vishwanath}},\ and\ \bibinfo {author}
  {\bibfnamefont {M.~P.}\ \bibnamefont {Zaletel}},\ }\bibfield  {title}
  {\bibinfo {title} {Ground state and hidden symmetry of magic-angle graphene
  at even integer filling},\ }\href
  {https://doi.org/10.1103/PhysRevX.10.031034} {\bibfield  {journal} {\bibinfo
  {journal} {Phys. Rev. X}\ }\textbf {\bibinfo {volume} {10}},\ \bibinfo
  {pages} {031034} (\bibinfo {year} {2020})}\BibitemShut {NoStop}%
\bibitem [{\citenamefont {Xie}\ \emph {et~al.}(2021)\citenamefont {Xie},
  \citenamefont {Regnault}, \citenamefont {C\ifmmode \u{a}\else
  \u{a}\fi{}lug\ifmmode~\u{a}\else \u{a}\fi{}ru}, \citenamefont {Bernevig},\
  and\ \citenamefont {Lian}}]{HF6}%
  \BibitemOpen
  \bibfield  {author} {\bibinfo {author} {\bibfnamefont {F.}~\bibnamefont
  {Xie}}, \bibinfo {author} {\bibfnamefont {N.}~\bibnamefont {Regnault}},
  \bibinfo {author} {\bibfnamefont {D.}~\bibnamefont {C\ifmmode \u{a}\else
  \u{a}\fi{}lug\ifmmode~\u{a}\else \u{a}\fi{}ru}}, \bibinfo {author}
  {\bibfnamefont {B.~A.}\ \bibnamefont {Bernevig}},\ and\ \bibinfo {author}
  {\bibfnamefont {B.}~\bibnamefont {Lian}},\ }\bibfield  {title} {\bibinfo
  {title} {Twisted symmetric trilayer graphene. ii. projected hartree-fock
  study},\ }\href {https://doi.org/10.1103/PhysRevB.104.115167} {\bibfield
  {journal} {\bibinfo  {journal} {Phys. Rev. B}\ }\textbf {\bibinfo {volume}
  {104}},\ \bibinfo {pages} {115167} (\bibinfo {year} {2021})}\BibitemShut
  {NoStop}%
\bibitem [{\citenamefont {Zhang}\ \emph {et~al.}(2020)\citenamefont {Zhang},
  \citenamefont {Jiang}, \citenamefont {Wang},\ and\ \citenamefont
  {Zhang}}]{HF7}%
  \BibitemOpen
  \bibfield  {author} {\bibinfo {author} {\bibfnamefont {Y.}~\bibnamefont
  {Zhang}}, \bibinfo {author} {\bibfnamefont {K.}~\bibnamefont {Jiang}},
  \bibinfo {author} {\bibfnamefont {Z.}~\bibnamefont {Wang}},\ and\ \bibinfo
  {author} {\bibfnamefont {F.}~\bibnamefont {Zhang}},\ }\bibfield  {title}
  {\bibinfo {title} {Correlated insulating phases of twisted bilayer graphene
  at commensurate filling fractions: A hartree-fock study},\ }\href
  {https://doi.org/10.1103/PhysRevB.102.035136} {\bibfield  {journal} {\bibinfo
   {journal} {Phys. Rev. B}\ }\textbf {\bibinfo {volume} {102}},\ \bibinfo
  {pages} {035136} (\bibinfo {year} {2020})}\BibitemShut {NoStop}%
\bibitem [{\citenamefont {Morales-Dur\'an}\ \emph {et~al.}(2021)\citenamefont
  {Morales-Dur\'an}, \citenamefont {MacDonald},\ and\ \citenamefont
  {Potasz}}]{Duran_PRB}%
  \BibitemOpen
  \bibfield  {author} {\bibinfo {author} {\bibfnamefont {N.}~\bibnamefont
  {Morales-Dur\'an}}, \bibinfo {author} {\bibfnamefont {A.~H.}\ \bibnamefont
  {MacDonald}},\ and\ \bibinfo {author} {\bibfnamefont {P.}~\bibnamefont
  {Potasz}},\ }\bibfield  {title} {\bibinfo {title} {Metal-insulator transition
  in transition metal dichalcogenide heterobilayer moir\'e superlattices},\
  }\href {https://doi.org/10.1103/PhysRevB.103.L241110} {\bibfield  {journal}
  {\bibinfo  {journal} {Phys. Rev. B}\ }\textbf {\bibinfo {volume} {103}},\
  \bibinfo {pages} {L241110} (\bibinfo {year} {2021})}\BibitemShut {NoStop}%
\bibitem [{\citenamefont {Morales-Dur\'an}\ \emph {et~al.}(2022)\citenamefont
  {Morales-Dur\'an}, \citenamefont {Hu}, \citenamefont {Potasz},\ and\
  \citenamefont {MacDonald}}]{Duran_PRL}%
  \BibitemOpen
  \bibfield  {author} {\bibinfo {author} {\bibfnamefont {N.}~\bibnamefont
  {Morales-Dur\'an}}, \bibinfo {author} {\bibfnamefont {N.~C.}\ \bibnamefont
  {Hu}}, \bibinfo {author} {\bibfnamefont {P.}~\bibnamefont {Potasz}},\ and\
  \bibinfo {author} {\bibfnamefont {A.~H.}\ \bibnamefont {MacDonald}},\
  }\bibfield  {title} {\bibinfo {title} {Nonlocal interactions in moir\'e
  hubbard systems},\ }\href {https://doi.org/10.1103/PhysRevLett.128.217202}
  {\bibfield  {journal} {\bibinfo  {journal} {Phys. Rev. Lett.}\ }\textbf
  {\bibinfo {volume} {128}},\ \bibinfo {pages} {217202} (\bibinfo {year}
  {2022})}\BibitemShut {NoStop}%
\bibitem [{\citenamefont {Carr}\ \emph {et~al.}(2019)\citenamefont {Carr},
  \citenamefont {Fang}, \citenamefont {Po}, \citenamefont {Vishwanath},\ and\
  \citenamefont {Kaxiras}}]{Carr_wannier}%
  \BibitemOpen
  \bibfield  {author} {\bibinfo {author} {\bibfnamefont {S.}~\bibnamefont
  {Carr}}, \bibinfo {author} {\bibfnamefont {S.}~\bibnamefont {Fang}}, \bibinfo
  {author} {\bibfnamefont {H.~C.}\ \bibnamefont {Po}}, \bibinfo {author}
  {\bibfnamefont {A.}~\bibnamefont {Vishwanath}},\ and\ \bibinfo {author}
  {\bibfnamefont {E.}~\bibnamefont {Kaxiras}},\ }\bibfield  {title} {\bibinfo
  {title} {Derivation of wannier orbitals and minimal-basis tight-binding
  hamiltonians for twisted bilayer graphene: First-principles approach},\
  }\href {https://doi.org/10.1103/PhysRevResearch.1.033072} {\bibfield
  {journal} {\bibinfo  {journal} {Phys. Rev. Research}\ }\textbf {\bibinfo
  {volume} {1}},\ \bibinfo {pages} {033072} (\bibinfo {year}
  {2019})}\BibitemShut {NoStop}%
\bibitem [{\citenamefont {Kennes}\ \emph {et~al.}(2018)\citenamefont {Kennes},
  \citenamefont {Lischner},\ and\ \citenamefont {Karrasch}}]{Kennes_PRB}%
  \BibitemOpen
  \bibfield  {author} {\bibinfo {author} {\bibfnamefont {D.~M.}\ \bibnamefont
  {Kennes}}, \bibinfo {author} {\bibfnamefont {J.}~\bibnamefont {Lischner}},\
  and\ \bibinfo {author} {\bibfnamefont {C.}~\bibnamefont {Karrasch}},\
  }\bibfield  {title} {\bibinfo {title} {Strong correlations and
  $d+\mathit{id}$ superconductivity in twisted bilayer graphene},\ }\href
  {https://doi.org/10.1103/PhysRevB.98.241407} {\bibfield  {journal} {\bibinfo
  {journal} {Phys. Rev. B}\ }\textbf {\bibinfo {volume} {98}},\ \bibinfo
  {pages} {241407} (\bibinfo {year} {2018})}\BibitemShut {NoStop}%
\bibitem [{\citenamefont {Xian}\ \emph {et~al.}(2021)\citenamefont {Xian},
  \citenamefont {Claassen}, \citenamefont {Kiese}, \citenamefont {Scherer},
  \citenamefont {Trebst}, \citenamefont {Kennes},\ and\ \citenamefont
  {Rubio}}]{Xian2021}%
  \BibitemOpen
  \bibfield  {author} {\bibinfo {author} {\bibfnamefont {L.}~\bibnamefont
  {Xian}}, \bibinfo {author} {\bibfnamefont {M.}~\bibnamefont {Claassen}},
  \bibinfo {author} {\bibfnamefont {D.}~\bibnamefont {Kiese}}, \bibinfo
  {author} {\bibfnamefont {M.~M.}\ \bibnamefont {Scherer}}, \bibinfo {author}
  {\bibfnamefont {S.}~\bibnamefont {Trebst}}, \bibinfo {author} {\bibfnamefont
  {D.~M.}\ \bibnamefont {Kennes}},\ and\ \bibinfo {author} {\bibfnamefont
  {A.}~\bibnamefont {Rubio}},\ }\bibfield  {title} {\bibinfo {title}
  {Realization of nearly dispersionless bands with strong orbital anisotropy
  from destructive interference in twisted bilayer mos2},\ }\href
  {https://doi.org/10.1038/s41467-021-25922-8} {\bibfield  {journal} {\bibinfo
  {journal} {Nature Communications}\ }\textbf {\bibinfo {volume} {12}},\
  \bibinfo {pages} {5644} (\bibinfo {year} {2021})}\BibitemShut {NoStop}%
\bibitem [{\citenamefont {Kennes}\ \emph {et~al.}(2020)\citenamefont {Kennes},
  \citenamefont {Xian}, \citenamefont {Claassen},\ and\ \citenamefont
  {Rubio}}]{Kennes2020}%
  \BibitemOpen
  \bibfield  {author} {\bibinfo {author} {\bibfnamefont {D.~M.}\ \bibnamefont
  {Kennes}}, \bibinfo {author} {\bibfnamefont {L.}~\bibnamefont {Xian}},
  \bibinfo {author} {\bibfnamefont {M.}~\bibnamefont {Claassen}},\ and\
  \bibinfo {author} {\bibfnamefont {A.}~\bibnamefont {Rubio}},\ }\bibfield
  {title} {\bibinfo {title} {One-dimensional flat bands in twisted bilayer
  germanium selenide},\ }\href {https://doi.org/10.1038/s41467-020-14947-0}
  {\bibfield  {journal} {\bibinfo  {journal} {Nature Communications}\ }\textbf
  {\bibinfo {volume} {11}},\ \bibinfo {pages} {1124} (\bibinfo {year}
  {2020})}\BibitemShut {NoStop}%
\bibitem [{\citenamefont {Claassen}\ \emph {et~al.}(2022)\citenamefont
  {Claassen}, \citenamefont {Xian}, \citenamefont {Kennes},\ and\ \citenamefont
  {Rubio}}]{Claassen2022}%
  \BibitemOpen
  \bibfield  {author} {\bibinfo {author} {\bibfnamefont {M.}~\bibnamefont
  {Claassen}}, \bibinfo {author} {\bibfnamefont {L.}~\bibnamefont {Xian}},
  \bibinfo {author} {\bibfnamefont {D.~M.}\ \bibnamefont {Kennes}},\ and\
  \bibinfo {author} {\bibfnamefont {A.}~\bibnamefont {Rubio}},\ }\bibfield
  {title} {\bibinfo {title} {Ultra-strong spin--orbit coupling and topological
  moir{\'e} engineering in twisted zrs2 bilayers},\ }\href
  {https://doi.org/10.1038/s41467-022-31604-w} {\bibfield  {journal} {\bibinfo
  {journal} {Nature Communications}\ }\textbf {\bibinfo {volume} {13}},\
  \bibinfo {pages} {4915} (\bibinfo {year} {2022})}\BibitemShut {NoStop}%
\bibitem [{\citenamefont {Xian}\ \emph {et~al.}(2019)\citenamefont {Xian},
  \citenamefont {Kennes}, \citenamefont {Tancogne-Dejean}, \citenamefont
  {Altarelli},\ and\ \citenamefont {Rubio}}]{Xian2019}%
  \BibitemOpen
  \bibfield  {author} {\bibinfo {author} {\bibfnamefont {L.}~\bibnamefont
  {Xian}}, \bibinfo {author} {\bibfnamefont {D.~M.}\ \bibnamefont {Kennes}},
  \bibinfo {author} {\bibfnamefont {N.}~\bibnamefont {Tancogne-Dejean}},
  \bibinfo {author} {\bibfnamefont {M.}~\bibnamefont {Altarelli}},\ and\
  \bibinfo {author} {\bibfnamefont {A.}~\bibnamefont {Rubio}},\ }\bibfield
  {title} {\bibinfo {title} {Multiflat bands and strong correlations in twisted
  bilayer boron nitride: Doping-induced correlated insulator and
  superconductor},\ }\href {https://doi.org/10.1021/acs.nanolett.9b00986}
  {\bibfield  {journal} {\bibinfo  {journal} {Nano Letters}\ }\textbf {\bibinfo
  {volume} {19}},\ \bibinfo {pages} {4934} (\bibinfo {year}
  {2019})}\BibitemShut {NoStop}%
\bibitem [{\citenamefont {Soejima}\ \emph {et~al.}(2020)\citenamefont
  {Soejima}, \citenamefont {Parker}, \citenamefont {Bultinck}, \citenamefont
  {Hauschild},\ and\ \citenamefont {Zaletel}}]{Zaletel_PRB}%
  \BibitemOpen
  \bibfield  {author} {\bibinfo {author} {\bibfnamefont {T.}~\bibnamefont
  {Soejima}}, \bibinfo {author} {\bibfnamefont {D.~E.}\ \bibnamefont {Parker}},
  \bibinfo {author} {\bibfnamefont {N.}~\bibnamefont {Bultinck}}, \bibinfo
  {author} {\bibfnamefont {J.}~\bibnamefont {Hauschild}},\ and\ \bibinfo
  {author} {\bibfnamefont {M.~P.}\ \bibnamefont {Zaletel}},\ }\bibfield
  {title} {\bibinfo {title} {Efficient simulation of moir\'e materials using
  the density matrix renormalization group},\ }\href
  {https://doi.org/10.1103/PhysRevB.102.205111} {\bibfield  {journal} {\bibinfo
   {journal} {Phys. Rev. B}\ }\textbf {\bibinfo {volume} {102}},\ \bibinfo
  {pages} {205111} (\bibinfo {year} {2020})}\BibitemShut {NoStop}%
\bibitem [{\citenamefont {Weston}\ \emph {et~al.}(2020)\citenamefont {Weston},
  \citenamefont {Zou}, \citenamefont {Enaldiev}, \citenamefont {Summerfield},
  \citenamefont {Clark}, \citenamefont {Z{\'o}lyomi}, \citenamefont {Graham},
  \citenamefont {Yelgel}, \citenamefont {Magorrian}, \citenamefont {Zhou},
  \citenamefont {Zultak}, \citenamefont {Hopkinson}, \citenamefont {Barinov},
  \citenamefont {Bointon}, \citenamefont {Kretinin}, \citenamefont {Wilson},
  \citenamefont {Beton}, \citenamefont {Fal'ko}, \citenamefont {Haigh},\ and\
  \citenamefont {Gorbachev}}]{Weston_Nature}%
  \BibitemOpen
  \bibfield  {author} {\bibinfo {author} {\bibfnamefont {A.}~\bibnamefont
  {Weston}}, \bibinfo {author} {\bibfnamefont {Y.}~\bibnamefont {Zou}},
  \bibinfo {author} {\bibfnamefont {V.}~\bibnamefont {Enaldiev}}, \bibinfo
  {author} {\bibfnamefont {A.}~\bibnamefont {Summerfield}}, \bibinfo {author}
  {\bibfnamefont {N.}~\bibnamefont {Clark}}, \bibinfo {author} {\bibfnamefont
  {V.}~\bibnamefont {Z{\'o}lyomi}}, \bibinfo {author} {\bibfnamefont
  {A.}~\bibnamefont {Graham}}, \bibinfo {author} {\bibfnamefont
  {C.}~\bibnamefont {Yelgel}}, \bibinfo {author} {\bibfnamefont
  {S.}~\bibnamefont {Magorrian}}, \bibinfo {author} {\bibfnamefont
  {M.}~\bibnamefont {Zhou}}, \bibinfo {author} {\bibfnamefont {J.}~\bibnamefont
  {Zultak}}, \bibinfo {author} {\bibfnamefont {D.}~\bibnamefont {Hopkinson}},
  \bibinfo {author} {\bibfnamefont {A.}~\bibnamefont {Barinov}}, \bibinfo
  {author} {\bibfnamefont {T.~H.}\ \bibnamefont {Bointon}}, \bibinfo {author}
  {\bibfnamefont {A.}~\bibnamefont {Kretinin}}, \bibinfo {author}
  {\bibfnamefont {N.~R.}\ \bibnamefont {Wilson}}, \bibinfo {author}
  {\bibfnamefont {P.~H.}\ \bibnamefont {Beton}}, \bibinfo {author}
  {\bibfnamefont {V.~I.}\ \bibnamefont {Fal'ko}}, \bibinfo {author}
  {\bibfnamefont {S.~J.}\ \bibnamefont {Haigh}},\ and\ \bibinfo {author}
  {\bibfnamefont {R.}~\bibnamefont {Gorbachev}},\ }\bibfield  {title} {\bibinfo
  {title} {Atomic reconstruction in twisted bilayers of transition metal
  dichalcogenides},\ }\href {https://doi.org/10.1038/s41565-020-0682-9}
  {\bibfield  {journal} {\bibinfo  {journal} {Nature Nanotechnology}\ }\textbf
  {\bibinfo {volume} {15}},\ \bibinfo {pages} {592} (\bibinfo {year}
  {2020})}\BibitemShut {NoStop}%
\bibitem [{\citenamefont {Nowick}(1995)}]{nowick_1995}%
  \BibitemOpen
  \bibfield  {author} {\bibinfo {author} {\bibfnamefont {A.~S.}\ \bibnamefont
  {Nowick}},\ }\href {https://doi.org/10.1017/CBO9780511524318} {\emph
  {\bibinfo {title} {Crystal Properties via Group Theory}}}\ (\bibinfo
  {publisher} {Cambridge University Press},\ \bibinfo {year}
  {1995})\BibitemShut {NoStop}%
\bibitem [{\citenamefont {Zhou}\ \emph {et~al.}(2019)\citenamefont {Zhou},
  \citenamefont {Chen}, \citenamefont {Yang}, \citenamefont {Liu},\ and\
  \citenamefont {Ouyang}}]{Zhou_PRB}%
  \BibitemOpen
  \bibfield  {author} {\bibinfo {author} {\bibfnamefont {W.}~\bibnamefont
  {Zhou}}, \bibinfo {author} {\bibfnamefont {J.}~\bibnamefont {Chen}}, \bibinfo
  {author} {\bibfnamefont {Z.}~\bibnamefont {Yang}}, \bibinfo {author}
  {\bibfnamefont {J.}~\bibnamefont {Liu}},\ and\ \bibinfo {author}
  {\bibfnamefont {F.}~\bibnamefont {Ouyang}},\ }\bibfield  {title} {\bibinfo
  {title} {Geometry and electronic structure of monolayer, bilayer, and
  multilayer janus wsse},\ }\href {https://doi.org/10.1103/PhysRevB.99.075160}
  {\bibfield  {journal} {\bibinfo  {journal} {Phys. Rev. B}\ }\textbf {\bibinfo
  {volume} {99}},\ \bibinfo {pages} {075160} (\bibinfo {year}
  {2019})}\BibitemShut {NoStop}%
\bibitem [{\citenamefont {Jung}\ \emph {et~al.}(2014)\citenamefont {Jung},
  \citenamefont {Raoux}, \citenamefont {Qiao},\ and\ \citenamefont
  {MacDonald}}]{Jung_PRB}%
  \BibitemOpen
  \bibfield  {author} {\bibinfo {author} {\bibfnamefont {J.}~\bibnamefont
  {Jung}}, \bibinfo {author} {\bibfnamefont {A.}~\bibnamefont {Raoux}},
  \bibinfo {author} {\bibfnamefont {Z.}~\bibnamefont {Qiao}},\ and\ \bibinfo
  {author} {\bibfnamefont {A.~H.}\ \bibnamefont {MacDonald}},\ }\bibfield
  {title} {\bibinfo {title} {Ab initio theory of moir\'e superlattice bands in
  layered two-dimensional materials},\ }\href
  {https://doi.org/10.1103/PhysRevB.89.205414} {\bibfield  {journal} {\bibinfo
  {journal} {Phys. Rev. B}\ }\textbf {\bibinfo {volume} {89}},\ \bibinfo
  {pages} {205414} (\bibinfo {year} {2014})}\BibitemShut {NoStop}%
\bibitem [{\citenamefont {Carr}\ \emph {et~al.}(2018)\citenamefont {Carr},
  \citenamefont {Massatt}, \citenamefont {Torrisi}, \citenamefont {Cazeaux},
  \citenamefont {Luskin},\ and\ \citenamefont {Kaxiras}}]{Carr_PRB}%
  \BibitemOpen
  \bibfield  {author} {\bibinfo {author} {\bibfnamefont {S.}~\bibnamefont
  {Carr}}, \bibinfo {author} {\bibfnamefont {D.}~\bibnamefont {Massatt}},
  \bibinfo {author} {\bibfnamefont {S.~B.}\ \bibnamefont {Torrisi}}, \bibinfo
  {author} {\bibfnamefont {P.}~\bibnamefont {Cazeaux}}, \bibinfo {author}
  {\bibfnamefont {M.}~\bibnamefont {Luskin}},\ and\ \bibinfo {author}
  {\bibfnamefont {E.}~\bibnamefont {Kaxiras}},\ }\bibfield  {title} {\bibinfo
  {title} {Relaxation and domain formation in incommensurate two-dimensional
  heterostructures},\ }\href {https://doi.org/10.1103/PhysRevB.98.224102}
  {\bibfield  {journal} {\bibinfo  {journal} {Phys. Rev. B}\ }\textbf {\bibinfo
  {volume} {98}},\ \bibinfo {pages} {224102} (\bibinfo {year}
  {2018})}\BibitemShut {NoStop}%
\bibitem [{\citenamefont {Korm{\'{a}}nyos}\ \emph {et~al.}(2015)\citenamefont
  {Korm{\'{a}}nyos}, \citenamefont {Burkard}, \citenamefont {Gmitra},
  \citenamefont {Fabian}, \citenamefont {Z{\'{o}}lyomi}, \citenamefont
  {Drummond},\ and\ \citenamefont {Fal'ko}}]{TMDs_falko}%
  \BibitemOpen
  \bibfield  {author} {\bibinfo {author} {\bibfnamefont {A.}~\bibnamefont
  {Korm{\'{a}}nyos}}, \bibinfo {author} {\bibfnamefont {G.}~\bibnamefont
  {Burkard}}, \bibinfo {author} {\bibfnamefont {M.}~\bibnamefont {Gmitra}},
  \bibinfo {author} {\bibfnamefont {J.}~\bibnamefont {Fabian}}, \bibinfo
  {author} {\bibfnamefont {V.}~\bibnamefont {Z{\'{o}}lyomi}}, \bibinfo {author}
  {\bibfnamefont {N.~D.}\ \bibnamefont {Drummond}},\ and\ \bibinfo {author}
  {\bibfnamefont {V.}~\bibnamefont {Fal'ko}},\ }\bibfield  {title} {\bibinfo
  {title} {k$\cdotp$p theory for two-dimensional transition metal
  dichalcogenide semiconductors},\ }\href
  {https://doi.org/10.1088/2053-1583/2/2/022001} {\bibfield  {journal}
  {\bibinfo  {journal} {2D Materials}\ }\textbf {\bibinfo {volume} {2}},\
  \bibinfo {pages} {022001} (\bibinfo {year} {2015})}\BibitemShut {NoStop}%
\bibitem [{\citenamefont {Giannozzi}\ \emph {et~al.}(2009)\citenamefont
  {Giannozzi}, \citenamefont {Baroni}, \citenamefont {Bonini}, \citenamefont
  {Calandra}, \citenamefont {Car}, \citenamefont {Cavazzoni}, \citenamefont
  {Ceresoli}, \citenamefont {Chiarotti}, \citenamefont {Cococcioni},
  \citenamefont {Dabo}, \citenamefont {Corso}, \citenamefont {de~Gironcoli},
  \citenamefont {Fabris}, \citenamefont {Fratesi}, \citenamefont {Gebauer},
  \citenamefont {Gerstmann}, \citenamefont {Gougoussis}, \citenamefont
  {Kokalj}, \citenamefont {Lazzeri}, \citenamefont {Martin-Samos},
  \citenamefont {Marzari}, \citenamefont {Mauri}, \citenamefont {Mazzarello},
  \citenamefont {Paolini}, \citenamefont {Pasquarello}, \citenamefont
  {Paulatto}, \citenamefont {Sbraccia}, \citenamefont {Scandolo}, \citenamefont
  {Sclauzero}, \citenamefont {Seitsonen}, \citenamefont {Smogunov},
  \citenamefont {Umari},\ and\ \citenamefont {Wentzcovitch}}]{QE1}%
  \BibitemOpen
  \bibfield  {author} {\bibinfo {author} {\bibfnamefont {P.}~\bibnamefont
  {Giannozzi}}, \bibinfo {author} {\bibfnamefont {S.}~\bibnamefont {Baroni}},
  \bibinfo {author} {\bibfnamefont {N.}~\bibnamefont {Bonini}}, \bibinfo
  {author} {\bibfnamefont {M.}~\bibnamefont {Calandra}}, \bibinfo {author}
  {\bibfnamefont {R.}~\bibnamefont {Car}}, \bibinfo {author} {\bibfnamefont
  {C.}~\bibnamefont {Cavazzoni}}, \bibinfo {author} {\bibfnamefont
  {D.}~\bibnamefont {Ceresoli}}, \bibinfo {author} {\bibfnamefont {G.~L.}\
  \bibnamefont {Chiarotti}}, \bibinfo {author} {\bibfnamefont {M.}~\bibnamefont
  {Cococcioni}}, \bibinfo {author} {\bibfnamefont {I.}~\bibnamefont {Dabo}},
  \bibinfo {author} {\bibfnamefont {A.~D.}\ \bibnamefont {Corso}}, \bibinfo
  {author} {\bibfnamefont {S.}~\bibnamefont {de~Gironcoli}}, \bibinfo {author}
  {\bibfnamefont {S.}~\bibnamefont {Fabris}}, \bibinfo {author} {\bibfnamefont
  {G.}~\bibnamefont {Fratesi}}, \bibinfo {author} {\bibfnamefont
  {R.}~\bibnamefont {Gebauer}}, \bibinfo {author} {\bibfnamefont
  {U.}~\bibnamefont {Gerstmann}}, \bibinfo {author} {\bibfnamefont
  {C.}~\bibnamefont {Gougoussis}}, \bibinfo {author} {\bibfnamefont
  {A.}~\bibnamefont {Kokalj}}, \bibinfo {author} {\bibfnamefont
  {M.}~\bibnamefont {Lazzeri}}, \bibinfo {author} {\bibfnamefont
  {L.}~\bibnamefont {Martin-Samos}}, \bibinfo {author} {\bibfnamefont
  {N.}~\bibnamefont {Marzari}}, \bibinfo {author} {\bibfnamefont
  {F.}~\bibnamefont {Mauri}}, \bibinfo {author} {\bibfnamefont
  {R.}~\bibnamefont {Mazzarello}}, \bibinfo {author} {\bibfnamefont
  {S.}~\bibnamefont {Paolini}}, \bibinfo {author} {\bibfnamefont
  {A.}~\bibnamefont {Pasquarello}}, \bibinfo {author} {\bibfnamefont
  {L.}~\bibnamefont {Paulatto}}, \bibinfo {author} {\bibfnamefont
  {C.}~\bibnamefont {Sbraccia}}, \bibinfo {author} {\bibfnamefont
  {S.}~\bibnamefont {Scandolo}}, \bibinfo {author} {\bibfnamefont
  {G.}~\bibnamefont {Sclauzero}}, \bibinfo {author} {\bibfnamefont {A.~P.}\
  \bibnamefont {Seitsonen}}, \bibinfo {author} {\bibfnamefont {A.}~\bibnamefont
  {Smogunov}}, \bibinfo {author} {\bibfnamefont {P.}~\bibnamefont {Umari}},\
  and\ \bibinfo {author} {\bibfnamefont {R.~M.}\ \bibnamefont {Wentzcovitch}},\
  }\bibfield  {title} {\bibinfo {title} {{QUANTUM} {ESPRESSO}: a modular and
  open-source software project for quantum simulations of materials},\ }\href
  {https://doi.org/10.1088/0953-8984/21/39/395502} {\bibfield  {journal}
  {\bibinfo  {journal} {Journal of Physics: Condensed Matter}\ }\textbf
  {\bibinfo {volume} {21}},\ \bibinfo {pages} {395502} (\bibinfo {year}
  {2009})}\BibitemShut {NoStop}%
\bibitem [{\citenamefont {Giannozzi}\ \emph {et~al.}(2017)\citenamefont
  {Giannozzi}, \citenamefont {Andreussi}, \citenamefont {Brumme}, \citenamefont
  {Bunau}, \citenamefont {Nardelli}, \citenamefont {Calandra}, \citenamefont
  {Car}, \citenamefont {Cavazzoni}, \citenamefont {Ceresoli}, \citenamefont
  {Cococcioni}, \citenamefont {Colonna}, \citenamefont {Carnimeo},
  \citenamefont {Corso}, \citenamefont {de~Gironcoli}, \citenamefont {Delugas},
  \citenamefont {DiStasio}, \citenamefont {Ferretti}, \citenamefont {Floris},
  \citenamefont {Fratesi}, \citenamefont {Fugallo}, \citenamefont {Gebauer},
  \citenamefont {Gerstmann}, \citenamefont {Giustino}, \citenamefont {Gorni},
  \citenamefont {Jia}, \citenamefont {Kawamura}, \citenamefont {Ko},
  \citenamefont {Kokalj}, \citenamefont {K\"{u}{\c{c}}\"{u}kbenli},
  \citenamefont {Lazzeri}, \citenamefont {Marsili}, \citenamefont {Marzari},
  \citenamefont {Mauri}, \citenamefont {Nguyen}, \citenamefont {Nguyen},
  \citenamefont {de-la Roza}, \citenamefont {Paulatto}, \citenamefont
  {Ponc{\'{e}}}, \citenamefont {Rocca}, \citenamefont {Sabatini}, \citenamefont
  {Santra}, \citenamefont {Schlipf}, \citenamefont {Seitsonen}, \citenamefont
  {Smogunov}, \citenamefont {Timrov}, \citenamefont {Thonhauser}, \citenamefont
  {Umari}, \citenamefont {Vast}, \citenamefont {Wu},\ and\ \citenamefont
  {Baroni}}]{QE2}%
  \BibitemOpen
  \bibfield  {author} {\bibinfo {author} {\bibfnamefont {P.}~\bibnamefont
  {Giannozzi}}, \bibinfo {author} {\bibfnamefont {O.}~\bibnamefont
  {Andreussi}}, \bibinfo {author} {\bibfnamefont {T.}~\bibnamefont {Brumme}},
  \bibinfo {author} {\bibfnamefont {O.}~\bibnamefont {Bunau}}, \bibinfo
  {author} {\bibfnamefont {M.~B.}\ \bibnamefont {Nardelli}}, \bibinfo {author}
  {\bibfnamefont {M.}~\bibnamefont {Calandra}}, \bibinfo {author}
  {\bibfnamefont {R.}~\bibnamefont {Car}}, \bibinfo {author} {\bibfnamefont
  {C.}~\bibnamefont {Cavazzoni}}, \bibinfo {author} {\bibfnamefont
  {D.}~\bibnamefont {Ceresoli}}, \bibinfo {author} {\bibfnamefont
  {M.}~\bibnamefont {Cococcioni}}, \bibinfo {author} {\bibfnamefont
  {N.}~\bibnamefont {Colonna}}, \bibinfo {author} {\bibfnamefont
  {I.}~\bibnamefont {Carnimeo}}, \bibinfo {author} {\bibfnamefont {A.~D.}\
  \bibnamefont {Corso}}, \bibinfo {author} {\bibfnamefont {S.}~\bibnamefont
  {de~Gironcoli}}, \bibinfo {author} {\bibfnamefont {P.}~\bibnamefont
  {Delugas}}, \bibinfo {author} {\bibfnamefont {R.~A.}\ \bibnamefont
  {DiStasio}}, \bibinfo {author} {\bibfnamefont {A.}~\bibnamefont {Ferretti}},
  \bibinfo {author} {\bibfnamefont {A.}~\bibnamefont {Floris}}, \bibinfo
  {author} {\bibfnamefont {G.}~\bibnamefont {Fratesi}}, \bibinfo {author}
  {\bibfnamefont {G.}~\bibnamefont {Fugallo}}, \bibinfo {author} {\bibfnamefont
  {R.}~\bibnamefont {Gebauer}}, \bibinfo {author} {\bibfnamefont
  {U.}~\bibnamefont {Gerstmann}}, \bibinfo {author} {\bibfnamefont
  {F.}~\bibnamefont {Giustino}}, \bibinfo {author} {\bibfnamefont
  {T.}~\bibnamefont {Gorni}}, \bibinfo {author} {\bibfnamefont
  {J.}~\bibnamefont {Jia}}, \bibinfo {author} {\bibfnamefont {M.}~\bibnamefont
  {Kawamura}}, \bibinfo {author} {\bibfnamefont {H.-Y.}\ \bibnamefont {Ko}},
  \bibinfo {author} {\bibfnamefont {A.}~\bibnamefont {Kokalj}}, \bibinfo
  {author} {\bibfnamefont {E.}~\bibnamefont {K\"{u}{\c{c}}\"{u}kbenli}},
  \bibinfo {author} {\bibfnamefont {M.}~\bibnamefont {Lazzeri}}, \bibinfo
  {author} {\bibfnamefont {M.}~\bibnamefont {Marsili}}, \bibinfo {author}
  {\bibfnamefont {N.}~\bibnamefont {Marzari}}, \bibinfo {author} {\bibfnamefont
  {F.}~\bibnamefont {Mauri}}, \bibinfo {author} {\bibfnamefont {N.~L.}\
  \bibnamefont {Nguyen}}, \bibinfo {author} {\bibfnamefont {H.-V.}\
  \bibnamefont {Nguyen}}, \bibinfo {author} {\bibfnamefont {A.~O.}\
  \bibnamefont {de-la Roza}}, \bibinfo {author} {\bibfnamefont
  {L.}~\bibnamefont {Paulatto}}, \bibinfo {author} {\bibfnamefont
  {S.}~\bibnamefont {Ponc{\'{e}}}}, \bibinfo {author} {\bibfnamefont
  {D.}~\bibnamefont {Rocca}}, \bibinfo {author} {\bibfnamefont
  {R.}~\bibnamefont {Sabatini}}, \bibinfo {author} {\bibfnamefont
  {B.}~\bibnamefont {Santra}}, \bibinfo {author} {\bibfnamefont
  {M.}~\bibnamefont {Schlipf}}, \bibinfo {author} {\bibfnamefont {A.~P.}\
  \bibnamefont {Seitsonen}}, \bibinfo {author} {\bibfnamefont {A.}~\bibnamefont
  {Smogunov}}, \bibinfo {author} {\bibfnamefont {I.}~\bibnamefont {Timrov}},
  \bibinfo {author} {\bibfnamefont {T.}~\bibnamefont {Thonhauser}}, \bibinfo
  {author} {\bibfnamefont {P.}~\bibnamefont {Umari}}, \bibinfo {author}
  {\bibfnamefont {N.}~\bibnamefont {Vast}}, \bibinfo {author} {\bibfnamefont
  {X.}~\bibnamefont {Wu}},\ and\ \bibinfo {author} {\bibfnamefont
  {S.}~\bibnamefont {Baroni}},\ }\bibfield  {title} {\bibinfo {title} {Advanced
  capabilities for materials modelling with quantum {ESPRESSO}},\ }\href
  {https://doi.org/10.1088/1361-648x/aa8f79} {\bibfield  {journal} {\bibinfo
  {journal} {Journal of Physics: Condensed Matter}\ }\textbf {\bibinfo {volume}
  {29}},\ \bibinfo {pages} {465901} (\bibinfo {year} {2017})}\BibitemShut
  {NoStop}%
\bibitem [{\citenamefont {Hamada}\ and\ \citenamefont
  {Otani}(2010)}]{Hamada_PRB}%
  \BibitemOpen
  \bibfield  {author} {\bibinfo {author} {\bibfnamefont {I.}~\bibnamefont
  {Hamada}}\ and\ \bibinfo {author} {\bibfnamefont {M.}~\bibnamefont {Otani}},\
  }\bibfield  {title} {\bibinfo {title} {Comparative van der waals
  density-functional study of graphene on metal surfaces},\ }\href
  {https://doi.org/10.1103/PhysRevB.82.153412} {\bibfield  {journal} {\bibinfo
  {journal} {Phys. Rev. B}\ }\textbf {\bibinfo {volume} {82}},\ \bibinfo
  {pages} {153412} (\bibinfo {year} {2010})}\BibitemShut {NoStop}%
\bibitem [{\citenamefont {Berland}\ \emph {et~al.}(2015)\citenamefont
  {Berland}, \citenamefont {Cooper}, \citenamefont {Lee}, \citenamefont
  {Schröder}, \citenamefont {Thonhauser}, \citenamefont {Hyldgaard},\ and\
  \citenamefont {Lundqvist}}]{Berland_2015}%
  \BibitemOpen
  \bibfield  {author} {\bibinfo {author} {\bibfnamefont {K.}~\bibnamefont
  {Berland}}, \bibinfo {author} {\bibfnamefont {V.~R.}\ \bibnamefont {Cooper}},
  \bibinfo {author} {\bibfnamefont {K.}~\bibnamefont {Lee}}, \bibinfo {author}
  {\bibfnamefont {E.}~\bibnamefont {Schröder}}, \bibinfo {author}
  {\bibfnamefont {T.}~\bibnamefont {Thonhauser}}, \bibinfo {author}
  {\bibfnamefont {P.}~\bibnamefont {Hyldgaard}},\ and\ \bibinfo {author}
  {\bibfnamefont {B.~I.}\ \bibnamefont {Lundqvist}},\ }\bibfield  {title}
  {\bibinfo {title} {van der waals forces in density functional theory: a
  review of the {vdW}-{DF} method},\ }\href
  {https://doi.org/10.1088/0034-4885/78/6/066501} {\bibfield  {journal}
  {\bibinfo  {journal} {Reports on Progress in Physics}\ }\textbf {\bibinfo
  {volume} {78}},\ \bibinfo {pages} {066501} (\bibinfo {year}
  {2015})}\BibitemShut {NoStop}%
\bibitem [{SM()}]{SM}%
  \BibitemOpen
  \href@noop {} {}\bibinfo {note} {See Supplemental Material at
  http://link.aps.org/supplemental/XXXXXXX for comparison between continuum
  model and DFT band structures, for details on the effect of relaxation on the
  Generalized stacking fault energy (GSFE), for band structures of selected
  twisted 2H homobilayers with symmetric ordering, and for band structures of
  selected asymmetrically ordered twisted bilayers showing their spin
  polarization.}\BibitemShut {Stop}%
\bibitem [{\citenamefont {Zhou}\ \emph {et~al.}(2015)\citenamefont {Zhou},
  \citenamefont {Han}, \citenamefont {Dai}, \citenamefont {Sun},\ and\
  \citenamefont {Srolovitz}}]{GSFE_1}%
  \BibitemOpen
  \bibfield  {author} {\bibinfo {author} {\bibfnamefont {S.}~\bibnamefont
  {Zhou}}, \bibinfo {author} {\bibfnamefont {J.}~\bibnamefont {Han}}, \bibinfo
  {author} {\bibfnamefont {S.}~\bibnamefont {Dai}}, \bibinfo {author}
  {\bibfnamefont {J.}~\bibnamefont {Sun}},\ and\ \bibinfo {author}
  {\bibfnamefont {D.~J.}\ \bibnamefont {Srolovitz}},\ }\bibfield  {title}
  {\bibinfo {title} {van der waals bilayer energetics: Generalized
  stacking-fault energy of graphene, boron nitride, and graphene/boron nitride
  bilayers},\ }\href {https://doi.org/10.1103/PhysRevB.92.155438} {\bibfield
  {journal} {\bibinfo  {journal} {Phys. Rev. B}\ }\textbf {\bibinfo {volume}
  {92}},\ \bibinfo {pages} {155438} (\bibinfo {year} {2015})}\BibitemShut
  {NoStop}%
\bibitem [{\citenamefont {Kaxiras}\ and\ \citenamefont
  {Duesbery}(1993)}]{GSFE_2}%
  \BibitemOpen
  \bibfield  {author} {\bibinfo {author} {\bibfnamefont {E.}~\bibnamefont
  {Kaxiras}}\ and\ \bibinfo {author} {\bibfnamefont {M.~S.}\ \bibnamefont
  {Duesbery}},\ }\bibfield  {title} {\bibinfo {title} {Free energies of
  generalized stacking faults in si and implications for the brittle-ductile
  transition},\ }\href {https://doi.org/10.1103/PhysRevLett.70.3752} {\bibfield
   {journal} {\bibinfo  {journal} {Phys. Rev. Lett.}\ }\textbf {\bibinfo
  {volume} {70}},\ \bibinfo {pages} {3752} (\bibinfo {year}
  {1993})}\BibitemShut {NoStop}%
\bibitem [{\citenamefont {Tang}\ \emph {et~al.}(2021)\citenamefont {Tang},
  \citenamefont {Carr},\ and\ \citenamefont {Kaxiras}}]{Tang_PRB}%
  \BibitemOpen
  \bibfield  {author} {\bibinfo {author} {\bibfnamefont {H.}~\bibnamefont
  {Tang}}, \bibinfo {author} {\bibfnamefont {S.}~\bibnamefont {Carr}},\ and\
  \bibinfo {author} {\bibfnamefont {E.}~\bibnamefont {Kaxiras}},\ }\bibfield
  {title} {\bibinfo {title} {Geometric origins of topological insulation in
  twisted layered semiconductors},\ }\href
  {https://doi.org/10.1103/PhysRevB.104.155415} {\bibfield  {journal} {\bibinfo
   {journal} {Phys. Rev. B}\ }\textbf {\bibinfo {volume} {104}},\ \bibinfo
  {pages} {155415} (\bibinfo {year} {2021})}\BibitemShut {NoStop}%
\bibitem [{\citenamefont {Zou}\ \emph {et~al.}(2018)\citenamefont {Zou},
  \citenamefont {Po}, \citenamefont {Vishwanath},\ and\ \citenamefont
  {Senthil}}]{Senthil_PRB}%
  \BibitemOpen
  \bibfield  {author} {\bibinfo {author} {\bibfnamefont {L.}~\bibnamefont
  {Zou}}, \bibinfo {author} {\bibfnamefont {H.~C.}\ \bibnamefont {Po}},
  \bibinfo {author} {\bibfnamefont {A.}~\bibnamefont {Vishwanath}},\ and\
  \bibinfo {author} {\bibfnamefont {T.}~\bibnamefont {Senthil}},\ }\bibfield
  {title} {\bibinfo {title} {Band structure of twisted bilayer graphene:
  Emergent symmetries, commensurate approximants, and wannier obstructions},\
  }\href {https://doi.org/10.1103/PhysRevB.98.085435} {\bibfield  {journal}
  {\bibinfo  {journal} {Phys. Rev. B}\ }\textbf {\bibinfo {volume} {98}},\
  \bibinfo {pages} {085435} (\bibinfo {year} {2018})}\BibitemShut {NoStop}%
\bibitem [{\citenamefont {{Angeli, Mattia}}\ and\ \citenamefont {{Fabrizio,
  Michele}}(2020)}]{Fabrizio_EPJ}%
  \BibitemOpen
  \bibfield  {author} {\bibinfo {author} {\bibnamefont {{Angeli, Mattia}}}\
  and\ \bibinfo {author} {\bibnamefont {{Fabrizio, Michele}}},\ }\bibfield
  {title} {\bibinfo {title} {Jahn-teller coupling to moir\'e phonons in the
  continuum model formalism for small-angle twisted bilayer graphene},\ }\href
  {https://doi.org/10.1140/epjp/s13360-020-00647-7} {\bibfield  {journal}
  {\bibinfo  {journal} {Eur. Phys. J. Plus}\ }\textbf {\bibinfo {volume}
  {135}},\ \bibinfo {pages} {630} (\bibinfo {year} {2020})}\BibitemShut
  {NoStop}%
\bibitem [{\citenamefont {Bradlyn}\ \emph {et~al.}(2017)\citenamefont
  {Bradlyn}, \citenamefont {Elcoro}, \citenamefont {Cano}, \citenamefont
  {Vergniory}, \citenamefont {Wang}, \citenamefont {Felser}, \citenamefont
  {Aroyo},\ and\ \citenamefont {Bernevig}}]{Bernevig_Nature}%
  \BibitemOpen
  \bibfield  {author} {\bibinfo {author} {\bibfnamefont {B.}~\bibnamefont
  {Bradlyn}}, \bibinfo {author} {\bibfnamefont {L.}~\bibnamefont {Elcoro}},
  \bibinfo {author} {\bibfnamefont {J.}~\bibnamefont {Cano}}, \bibinfo {author}
  {\bibfnamefont {M.~G.}\ \bibnamefont {Vergniory}}, \bibinfo {author}
  {\bibfnamefont {Z.}~\bibnamefont {Wang}}, \bibinfo {author} {\bibfnamefont
  {C.}~\bibnamefont {Felser}}, \bibinfo {author} {\bibfnamefont {M.~I.}\
  \bibnamefont {Aroyo}},\ and\ \bibinfo {author} {\bibfnamefont {B.~A.}\
  \bibnamefont {Bernevig}},\ }\bibfield  {title} {\bibinfo {title} {Topological
  quantum chemistry},\ }\href {https://doi.org/10.1038/nature23268} {\bibfield
  {journal} {\bibinfo  {journal} {Nature}\ }\textbf {\bibinfo {volume} {547}},\
  \bibinfo {pages} {298} (\bibinfo {year} {2017})}\BibitemShut {NoStop}%
\bibitem [{\citenamefont {Elcoro}\ \emph {et~al.}(2017)\citenamefont {Elcoro},
  \citenamefont {Bradlyn}, \citenamefont {Wang}, \citenamefont {Vergniory},
  \citenamefont {Cano}, \citenamefont {Felser}, \citenamefont {Bernevig},
  \citenamefont {Orobengoa}, \citenamefont {de~la Flor},\ and\ \citenamefont
  {Aroyo}}]{Bilbao_server}%
  \BibitemOpen
  \bibfield  {author} {\bibinfo {author} {\bibfnamefont {L.}~\bibnamefont
  {Elcoro}}, \bibinfo {author} {\bibfnamefont {B.}~\bibnamefont {Bradlyn}},
  \bibinfo {author} {\bibfnamefont {Z.}~\bibnamefont {Wang}}, \bibinfo {author}
  {\bibfnamefont {M.~G.}\ \bibnamefont {Vergniory}}, \bibinfo {author}
  {\bibfnamefont {J.}~\bibnamefont {Cano}}, \bibinfo {author} {\bibfnamefont
  {C.}~\bibnamefont {Felser}}, \bibinfo {author} {\bibfnamefont {B.~A.}\
  \bibnamefont {Bernevig}}, \bibinfo {author} {\bibfnamefont {D.}~\bibnamefont
  {Orobengoa}}, \bibinfo {author} {\bibfnamefont {G.}~\bibnamefont {de~la
  Flor}},\ and\ \bibinfo {author} {\bibfnamefont {M.~I.}\ \bibnamefont
  {Aroyo}},\ }\bibfield  {title} {\bibinfo {title} {Double crystallographic
  groups and their representations on the bilbao crystallographic server},\
  }\href {https://doi.org/10.1107/s1600576717011712} {\bibfield  {journal}
  {\bibinfo  {journal} {Journal of Applied Crystallography}\ }\textbf {\bibinfo
  {volume} {50}},\ \bibinfo {pages} {1457} (\bibinfo {year}
  {2017})}\BibitemShut {NoStop}%
\bibitem [{\citenamefont {Han}\ \emph {et~al.}(2012)\citenamefont {Han},
  \citenamefont {Helton}, \citenamefont {Chu}, \citenamefont {Nocera},
  \citenamefont {Rodriguez-Rivera}, \citenamefont {Broholm},\ and\
  \citenamefont {Lee}}]{SpinLiquid}%
  \BibitemOpen
  \bibfield  {author} {\bibinfo {author} {\bibfnamefont {T.-H.}\ \bibnamefont
  {Han}}, \bibinfo {author} {\bibfnamefont {J.~S.}\ \bibnamefont {Helton}},
  \bibinfo {author} {\bibfnamefont {S.}~\bibnamefont {Chu}}, \bibinfo {author}
  {\bibfnamefont {D.~G.}\ \bibnamefont {Nocera}}, \bibinfo {author}
  {\bibfnamefont {J.~A.}\ \bibnamefont {Rodriguez-Rivera}}, \bibinfo {author}
  {\bibfnamefont {C.}~\bibnamefont {Broholm}},\ and\ \bibinfo {author}
  {\bibfnamefont {Y.~S.}\ \bibnamefont {Lee}},\ }\bibfield  {title} {\bibinfo
  {title} {Fractionalized excitations in the spin-liquid state of a
  kagome-lattice antiferromagnet},\ }\href
  {https://doi.org/10.1038/nature11659} {\bibfield  {journal} {\bibinfo
  {journal} {Nature}\ }\textbf {\bibinfo {volume} {492}},\ \bibinfo {pages}
  {406} (\bibinfo {year} {2012})}\BibitemShut {NoStop}%
\bibitem [{\citenamefont {Broholm}\ \emph {et~al.}(2020)\citenamefont
  {Broholm}, \citenamefont {Cava}, \citenamefont {Kivelson}, \citenamefont
  {Nocera}, \citenamefont {Norman},\ and\ \citenamefont
  {Senthil}}]{Quantum_spin_liquids}%
  \BibitemOpen
  \bibfield  {author} {\bibinfo {author} {\bibfnamefont {C.}~\bibnamefont
  {Broholm}}, \bibinfo {author} {\bibfnamefont {R.~J.}\ \bibnamefont {Cava}},
  \bibinfo {author} {\bibfnamefont {S.~A.}\ \bibnamefont {Kivelson}}, \bibinfo
  {author} {\bibfnamefont {D.~G.}\ \bibnamefont {Nocera}}, \bibinfo {author}
  {\bibfnamefont {M.~R.}\ \bibnamefont {Norman}},\ and\ \bibinfo {author}
  {\bibfnamefont {T.}~\bibnamefont {Senthil}},\ }\bibfield  {title} {\bibinfo
  {title} {Quantum spin liquids},\ }\href
  {https://doi.org/10.1126/science.aay0668} {\bibfield  {journal} {\bibinfo
  {journal} {Science}\ }\textbf {\bibinfo {volume} {367}},\ \bibinfo {pages}
  {eaay0668} (\bibinfo {year} {2020})}\BibitemShut {NoStop}%
\bibitem [{\citenamefont {Castro~Neto}\ \emph {et~al.}(2009)\citenamefont
  {Castro~Neto}, \citenamefont {Guinea}, \citenamefont {Peres}, \citenamefont
  {Novoselov},\ and\ \citenamefont {Geim}}]{graphene_review}%
  \BibitemOpen
  \bibfield  {author} {\bibinfo {author} {\bibfnamefont {A.~H.}\ \bibnamefont
  {Castro~Neto}}, \bibinfo {author} {\bibfnamefont {F.}~\bibnamefont {Guinea}},
  \bibinfo {author} {\bibfnamefont {N.~M.~R.}\ \bibnamefont {Peres}}, \bibinfo
  {author} {\bibfnamefont {K.~S.}\ \bibnamefont {Novoselov}},\ and\ \bibinfo
  {author} {\bibfnamefont {A.~K.}\ \bibnamefont {Geim}},\ }\bibfield  {title}
  {\bibinfo {title} {The electronic properties of graphene},\ }\href
  {https://doi.org/10.1103/RevModPhys.81.109} {\bibfield  {journal} {\bibinfo
  {journal} {Rev. Mod. Phys.}\ }\textbf {\bibinfo {volume} {81}},\ \bibinfo
  {pages} {109} (\bibinfo {year} {2009})}\BibitemShut {NoStop}%
\bibitem [{\citenamefont {Crasto~de Lima}\ \emph {et~al.}(2019)\citenamefont
  {Crasto~de Lima}, \citenamefont {Miwa},\ and\ \citenamefont
  {Su\'arez~Morell}}]{Morell2019}%
  \BibitemOpen
  \bibfield  {author} {\bibinfo {author} {\bibfnamefont {F.}~\bibnamefont
  {Crasto~de Lima}}, \bibinfo {author} {\bibfnamefont {R.~H.}\ \bibnamefont
  {Miwa}},\ and\ \bibinfo {author} {\bibfnamefont {E.}~\bibnamefont
  {Su\'arez~Morell}},\ }\bibfield  {title} {\bibinfo {title} {Double flat bands
  in kagome twisted bilayers},\ }\href
  {https://doi.org/10.1103/PhysRevB.100.155421} {\bibfield  {journal} {\bibinfo
   {journal} {Phys. Rev. B}\ }\textbf {\bibinfo {volume} {100}},\ \bibinfo
  {pages} {155421} (\bibinfo {year} {2019})}\BibitemShut {NoStop}%
\bibitem [{\citenamefont {Semenoff}(2012)}]{Chiral}%
  \BibitemOpen
  \bibfield  {author} {\bibinfo {author} {\bibfnamefont {G.~W.}\ \bibnamefont
  {Semenoff}},\ }\bibfield  {title} {\bibinfo {title} {Chiral symmetry breaking
  in graphene},\ }\href {https://doi.org/10.1088/0031-8949/2012/t146/014016}
  {\bibfield  {journal} {\bibinfo  {journal} {Physica Scripta}\ }\textbf
  {\bibinfo {volume} {T146}},\ \bibinfo {pages} {014016} (\bibinfo {year}
  {2012})}\BibitemShut {NoStop}%
\bibitem [{\citenamefont {Tang}\ \emph {et~al.}(2011)\citenamefont {Tang},
  \citenamefont {Mei},\ and\ \citenamefont {Wen}}]{Tang}%
  \BibitemOpen
  \bibfield  {author} {\bibinfo {author} {\bibfnamefont {E.}~\bibnamefont
  {Tang}}, \bibinfo {author} {\bibfnamefont {J.-W.}\ \bibnamefont {Mei}},\ and\
  \bibinfo {author} {\bibfnamefont {X.-G.}\ \bibnamefont {Wen}},\ }\bibfield
  {title} {\bibinfo {title} {High-temperature fractional quantum hall states},\
  }\href {https://doi.org/10.1103/PhysRevLett.106.236802} {\bibfield  {journal}
  {\bibinfo  {journal} {Phys. Rev. Lett.}\ }\textbf {\bibinfo {volume} {106}},\
  \bibinfo {pages} {236802} (\bibinfo {year} {2011})}\BibitemShut {NoStop}%
\bibitem [{\citenamefont {Iimura}\ and\ \citenamefont {Imai}(2018)}]{Iimura}%
  \BibitemOpen
  \bibfield  {author} {\bibinfo {author} {\bibfnamefont {S.}~\bibnamefont
  {Iimura}}\ and\ \bibinfo {author} {\bibfnamefont {Y.}~\bibnamefont {Imai}},\
  }\bibfield  {title} {\bibinfo {title} {Thermal hall conductivity in
  superconducting phase on kagome lattice},\ }\href
  {https://doi.org/10.7566/JPSJ.87.094715} {\bibfield  {journal} {\bibinfo
  {journal} {Journal of the Physical Society of Japan}\ }\textbf {\bibinfo
  {volume} {87}},\ \bibinfo {pages} {094715} (\bibinfo {year}
  {2018})}\BibitemShut {NoStop}%
\bibitem [{\citenamefont {Kane}\ and\ \citenamefont {Mele}(2005)}]{Kane}%
  \BibitemOpen
  \bibfield  {author} {\bibinfo {author} {\bibfnamefont {C.~L.}\ \bibnamefont
  {Kane}}\ and\ \bibinfo {author} {\bibfnamefont {E.~J.}\ \bibnamefont
  {Mele}},\ }\bibfield  {title} {\bibinfo {title} {Quantum spin hall effect in
  graphene},\ }\href {https://doi.org/10.1103/PhysRevLett.95.226801} {\bibfield
   {journal} {\bibinfo  {journal} {Phys. Rev. Lett.}\ }\textbf {\bibinfo
  {volume} {95}},\ \bibinfo {pages} {226801} (\bibinfo {year}
  {2005})}\BibitemShut {NoStop}%
\bibitem [{\citenamefont {Wang}\ \emph {et~al.}(2013)\citenamefont {Wang},
  \citenamefont {Liu},\ and\ \citenamefont {Liu}}]{Wang}%
  \BibitemOpen
  \bibfield  {author} {\bibinfo {author} {\bibfnamefont {Z.~F.}\ \bibnamefont
  {Wang}}, \bibinfo {author} {\bibfnamefont {Z.}~\bibnamefont {Liu}},\ and\
  \bibinfo {author} {\bibfnamefont {F.}~\bibnamefont {Liu}},\ }\bibfield
  {title} {\bibinfo {title} {Quantum anomalous hall effect in 2d organic
  topological insulators},\ }\href
  {https://doi.org/10.1103/PhysRevLett.110.196801} {\bibfield  {journal}
  {\bibinfo  {journal} {Phys. Rev. Lett.}\ }\textbf {\bibinfo {volume} {110}},\
  \bibinfo {pages} {196801} (\bibinfo {year} {2013})}\BibitemShut {NoStop}%
\bibitem [{\citenamefont {Wu}(2008)}]{WU2}%
  \BibitemOpen
  \bibfield  {author} {\bibinfo {author} {\bibfnamefont {C.}~\bibnamefont
  {Wu}},\ }\bibfield  {title} {\bibinfo {title} {Orbital ordering and
  frustration of $p$-band mott insulators},\ }\href
  {https://doi.org/10.1103/PhysRevLett.100.200406} {\bibfield  {journal}
  {\bibinfo  {journal} {Phys. Rev. Lett.}\ }\textbf {\bibinfo {volume} {100}},\
  \bibinfo {pages} {200406} (\bibinfo {year} {2008})}\BibitemShut {NoStop}%
\bibitem [{\citenamefont {Zhang}\ \emph {et~al.}(2014)\citenamefont {Zhang},
  \citenamefont {Li},\ and\ \citenamefont {Wu}}]{WU3}%
  \BibitemOpen
  \bibfield  {author} {\bibinfo {author} {\bibfnamefont {G.-F.}\ \bibnamefont
  {Zhang}}, \bibinfo {author} {\bibfnamefont {Y.}~\bibnamefont {Li}},\ and\
  \bibinfo {author} {\bibfnamefont {C.}~\bibnamefont {Wu}},\ }\bibfield
  {title} {\bibinfo {title} {Honeycomb lattice with multiorbital structure:
  Topological and quantum anomalous hall insulators with large gaps},\ }\href
  {https://doi.org/10.1103/PhysRevB.90.075114} {\bibfield  {journal} {\bibinfo
  {journal} {Phys. Rev. B}\ }\textbf {\bibinfo {volume} {90}},\ \bibinfo
  {pages} {075114} (\bibinfo {year} {2014})}\BibitemShut {NoStop}%
\bibitem [{\citenamefont {Wen}\ \emph {et~al.}(2010)\citenamefont {Wen},
  \citenamefont {R\"uegg}, \citenamefont {Wang},\ and\ \citenamefont
  {Fiete}}]{Gregory}%
  \BibitemOpen
  \bibfield  {author} {\bibinfo {author} {\bibfnamefont {J.}~\bibnamefont
  {Wen}}, \bibinfo {author} {\bibfnamefont {A.}~\bibnamefont {R\"uegg}},
  \bibinfo {author} {\bibfnamefont {C.-C.~J.}\ \bibnamefont {Wang}},\ and\
  \bibinfo {author} {\bibfnamefont {G.~A.}\ \bibnamefont {Fiete}},\ }\bibfield
  {title} {\bibinfo {title} {Interaction-driven topological insulators on the
  kagome and the decorated honeycomb lattices},\ }\href
  {https://doi.org/10.1103/PhysRevB.82.075125} {\bibfield  {journal} {\bibinfo
  {journal} {Phys. Rev. B}\ }\textbf {\bibinfo {volume} {82}},\ \bibinfo
  {pages} {075125} (\bibinfo {year} {2010})}\BibitemShut {NoStop}%
\bibitem [{\citenamefont {Natori}\ \emph {et~al.}(2019)\citenamefont {Natori},
  \citenamefont {Nutakki}, \citenamefont {Pereira},\ and\ \citenamefont
  {Andrade}}]{Natori}%
  \BibitemOpen
  \bibfield  {author} {\bibinfo {author} {\bibfnamefont {W.~M.~H.}\
  \bibnamefont {Natori}}, \bibinfo {author} {\bibfnamefont {R.}~\bibnamefont
  {Nutakki}}, \bibinfo {author} {\bibfnamefont {R.~G.}\ \bibnamefont
  {Pereira}},\ and\ \bibinfo {author} {\bibfnamefont {E.~C.}\ \bibnamefont
  {Andrade}},\ }\bibfield  {title} {\bibinfo {title} {Su(4) heisenberg model on
  the honeycomb lattice with exchange-frustrated perturbations: Implications
  for twistronics and mott insulators},\ }\href
  {https://doi.org/10.1103/PhysRevB.100.205131} {\bibfield  {journal} {\bibinfo
   {journal} {Phys. Rev. B}\ }\textbf {\bibinfo {volume} {100}},\ \bibinfo
  {pages} {205131} (\bibinfo {year} {2019})}\BibitemShut {NoStop}%
\bibitem [{\citenamefont {Kiesel}\ \emph {et~al.}(2013)\citenamefont {Kiesel},
  \citenamefont {Platt},\ and\ \citenamefont {Thomale}}]{Kiesel}%
  \BibitemOpen
  \bibfield  {author} {\bibinfo {author} {\bibfnamefont {M.~L.}\ \bibnamefont
  {Kiesel}}, \bibinfo {author} {\bibfnamefont {C.}~\bibnamefont {Platt}},\ and\
  \bibinfo {author} {\bibfnamefont {R.}~\bibnamefont {Thomale}},\ }\bibfield
  {title} {\bibinfo {title} {Unconventional fermi surface instabilities in the
  kagome hubbard model},\ }\href
  {https://doi.org/10.1103/PhysRevLett.110.126405} {\bibfield  {journal}
  {\bibinfo  {journal} {Phys. Rev. Lett.}\ }\textbf {\bibinfo {volume} {110}},\
  \bibinfo {pages} {126405} (\bibinfo {year} {2013})}\BibitemShut {NoStop}%
\bibitem [{\citenamefont {Sinova}\ \emph {et~al.}(2004)\citenamefont {Sinova},
  \citenamefont {Culcer}, \citenamefont {Niu}, \citenamefont {Sinitsyn},
  \citenamefont {Jungwirth},\ and\ \citenamefont {MacDonald}}]{Sinova_PRL}%
  \BibitemOpen
  \bibfield  {author} {\bibinfo {author} {\bibfnamefont {J.}~\bibnamefont
  {Sinova}}, \bibinfo {author} {\bibfnamefont {D.}~\bibnamefont {Culcer}},
  \bibinfo {author} {\bibfnamefont {Q.}~\bibnamefont {Niu}}, \bibinfo {author}
  {\bibfnamefont {N.~A.}\ \bibnamefont {Sinitsyn}}, \bibinfo {author}
  {\bibfnamefont {T.}~\bibnamefont {Jungwirth}},\ and\ \bibinfo {author}
  {\bibfnamefont {A.~H.}\ \bibnamefont {MacDonald}},\ }\bibfield  {title}
  {\bibinfo {title} {Universal intrinsic spin hall effect},\ }\href
  {https://doi.org/10.1103/PhysRevLett.92.126603} {\bibfield  {journal}
  {\bibinfo  {journal} {Phys. Rev. Lett.}\ }\textbf {\bibinfo {volume} {92}},\
  \bibinfo {pages} {126603} (\bibinfo {year} {2004})}\BibitemShut {NoStop}%
\bibitem [{\citenamefont {Bauer}\ \emph {et~al.}(2004)\citenamefont {Bauer},
  \citenamefont {Hilscher}, \citenamefont {Michor}, \citenamefont {Paul},
  \citenamefont {Scheidt}, \citenamefont {Gribanov}, \citenamefont {Seropegin},
  \citenamefont {No\"el}, \citenamefont {Sigrist},\ and\ \citenamefont
  {Rogl}}]{Bauer_PRL}%
  \BibitemOpen
  \bibfield  {author} {\bibinfo {author} {\bibfnamefont {E.}~\bibnamefont
  {Bauer}}, \bibinfo {author} {\bibfnamefont {G.}~\bibnamefont {Hilscher}},
  \bibinfo {author} {\bibfnamefont {H.}~\bibnamefont {Michor}}, \bibinfo
  {author} {\bibfnamefont {C.}~\bibnamefont {Paul}}, \bibinfo {author}
  {\bibfnamefont {E.~W.}\ \bibnamefont {Scheidt}}, \bibinfo {author}
  {\bibfnamefont {A.}~\bibnamefont {Gribanov}}, \bibinfo {author}
  {\bibfnamefont {Y.}~\bibnamefont {Seropegin}}, \bibinfo {author}
  {\bibfnamefont {H.}~\bibnamefont {No\"el}}, \bibinfo {author} {\bibfnamefont
  {M.}~\bibnamefont {Sigrist}},\ and\ \bibinfo {author} {\bibfnamefont
  {P.}~\bibnamefont {Rogl}},\ }\bibfield  {title} {\bibinfo {title} {Heavy
  fermion superconductivity and magnetic order in noncentrosymmetric
  ${\mathrm{c}\mathrm{e}\mathrm{p}\mathrm{t}}_{3}\mathrm{S}\mathrm{i}$},\
  }\href {https://doi.org/10.1103/PhysRevLett.92.027003} {\bibfield  {journal}
  {\bibinfo  {journal} {Phys. Rev. Lett.}\ }\textbf {\bibinfo {volume} {92}},\
  \bibinfo {pages} {027003} (\bibinfo {year} {2004})}\BibitemShut {NoStop}%
\bibitem [{\citenamefont {Sato}\ and\ \citenamefont
  {Fujimoto}(2009)}]{Sato_PRB}%
  \BibitemOpen
  \bibfield  {author} {\bibinfo {author} {\bibfnamefont {M.}~\bibnamefont
  {Sato}}\ and\ \bibinfo {author} {\bibfnamefont {S.}~\bibnamefont
  {Fujimoto}},\ }\bibfield  {title} {\bibinfo {title} {Topological phases of
  noncentrosymmetric superconductors: Edge states, majorana fermions, and
  non-abelian statistics},\ }\href {https://doi.org/10.1103/PhysRevB.79.094504}
  {\bibfield  {journal} {\bibinfo  {journal} {Phys. Rev. B}\ }\textbf {\bibinfo
  {volume} {79}},\ \bibinfo {pages} {094504} (\bibinfo {year}
  {2009})}\BibitemShut {NoStop}%
\bibitem [{\citenamefont {Soriano}\ and\ \citenamefont
  {Lado}(2021)}]{Soriano_2021}%
  \BibitemOpen
  \bibfield  {author} {\bibinfo {author} {\bibfnamefont {D.}~\bibnamefont
  {Soriano}}\ and\ \bibinfo {author} {\bibfnamefont {J.~L.}\ \bibnamefont
  {Lado}},\ }\bibfield  {title} {\bibinfo {title} {Spin{\textendash}orbit
  correlations and exchange-bias control in twisted janus dichalcogenide
  multilayers},\ }\href {https://doi.org/10.1088/1367-2630/ac12fb} {\bibfield
  {journal} {\bibinfo  {journal} {New Journal of Physics}\ }\textbf {\bibinfo
  {volume} {23}},\ \bibinfo {pages} {073038} (\bibinfo {year}
  {2021})}\BibitemShut {NoStop}%
\bibitem [{\citenamefont {Topp}\ \emph {et~al.}(2021)\citenamefont {Topp},
  \citenamefont {Eckhardt}, \citenamefont {Kennes}, \citenamefont {Sentef},\
  and\ \citenamefont {T\"orm\"a}}]{PRB_Topp}%
  \BibitemOpen
  \bibfield  {author} {\bibinfo {author} {\bibfnamefont {G.~E.}\ \bibnamefont
  {Topp}}, \bibinfo {author} {\bibfnamefont {C.~J.}\ \bibnamefont {Eckhardt}},
  \bibinfo {author} {\bibfnamefont {D.~M.}\ \bibnamefont {Kennes}}, \bibinfo
  {author} {\bibfnamefont {M.~A.}\ \bibnamefont {Sentef}},\ and\ \bibinfo
  {author} {\bibfnamefont {P.}~\bibnamefont {T\"orm\"a}},\ }\bibfield  {title}
  {\bibinfo {title} {Light-matter coupling and quantum geometry in moir\'e
  materials},\ }\href {https://doi.org/10.1103/PhysRevB.104.064306} {\bibfield
  {journal} {\bibinfo  {journal} {Phys. Rev. B}\ }\textbf {\bibinfo {volume}
  {104}},\ \bibinfo {pages} {064306} (\bibinfo {year} {2021})}\BibitemShut
  {NoStop}%
\bibitem [{\citenamefont {Topp}\ \emph {et~al.}(2019)\citenamefont {Topp},
  \citenamefont {Jotzu}, \citenamefont {McIver}, \citenamefont {Xian},
  \citenamefont {Rubio},\ and\ \citenamefont {Sentef}}]{PRR_Topp}%
  \BibitemOpen
  \bibfield  {author} {\bibinfo {author} {\bibfnamefont {G.~E.}\ \bibnamefont
  {Topp}}, \bibinfo {author} {\bibfnamefont {G.}~\bibnamefont {Jotzu}},
  \bibinfo {author} {\bibfnamefont {J.~W.}\ \bibnamefont {McIver}}, \bibinfo
  {author} {\bibfnamefont {L.}~\bibnamefont {Xian}}, \bibinfo {author}
  {\bibfnamefont {A.}~\bibnamefont {Rubio}},\ and\ \bibinfo {author}
  {\bibfnamefont {M.~A.}\ \bibnamefont {Sentef}},\ }\bibfield  {title}
  {\bibinfo {title} {Topological floquet engineering of twisted bilayer
  graphene},\ }\href {https://doi.org/10.1103/PhysRevResearch.1.023031}
  {\bibfield  {journal} {\bibinfo  {journal} {Phys. Rev. Research}\ }\textbf
  {\bibinfo {volume} {1}},\ \bibinfo {pages} {023031} (\bibinfo {year}
  {2019})}\BibitemShut {NoStop}%
\bibitem [{\citenamefont {Latini}\ \emph {et~al.}(2019)\citenamefont {Latini},
  \citenamefont {Ronca}, \citenamefont {De~Giovannini}, \citenamefont
  {H{\"u}bener},\ and\ \citenamefont {Rubio}}]{Latini2019}%
  \BibitemOpen
  \bibfield  {author} {\bibinfo {author} {\bibfnamefont {S.}~\bibnamefont
  {Latini}}, \bibinfo {author} {\bibfnamefont {E.}~\bibnamefont {Ronca}},
  \bibinfo {author} {\bibfnamefont {U.}~\bibnamefont {De~Giovannini}}, \bibinfo
  {author} {\bibfnamefont {H.}~\bibnamefont {H{\"u}bener}},\ and\ \bibinfo
  {author} {\bibfnamefont {A.}~\bibnamefont {Rubio}},\ }\bibfield  {title}
  {\bibinfo {title} {Cavity control of excitons in two-dimensional materials},\
  }\href {https://doi.org/10.1021/acs.nanolett.9b00183} {\bibfield  {journal}
  {\bibinfo  {journal} {Nano Letters}\ }\textbf {\bibinfo {volume} {19}},\
  \bibinfo {pages} {3473} (\bibinfo {year} {2019})}\BibitemShut {NoStop}%
\bibitem [{\citenamefont {Li}\ \emph {et~al.}(2020)\citenamefont {Li},
  \citenamefont {Fertig},\ and\ \citenamefont {Seradjeh}}]{PRR_Babak}%
  \BibitemOpen
  \bibfield  {author} {\bibinfo {author} {\bibfnamefont {Y.}~\bibnamefont
  {Li}}, \bibinfo {author} {\bibfnamefont {H.~A.}\ \bibnamefont {Fertig}},\
  and\ \bibinfo {author} {\bibfnamefont {B.}~\bibnamefont {Seradjeh}},\
  }\bibfield  {title} {\bibinfo {title} {Floquet-engineered topological flat
  bands in irradiated twisted bilayer graphene},\ }\href
  {https://doi.org/10.1103/PhysRevResearch.2.043275} {\bibfield  {journal}
  {\bibinfo  {journal} {Phys. Rev. Research}\ }\textbf {\bibinfo {volume}
  {2}},\ \bibinfo {pages} {043275} (\bibinfo {year} {2020})}\BibitemShut
  {NoStop}%
\bibitem [{\citenamefont {Schleder}\ \emph {et~al.}(2021)\citenamefont
  {Schleder}, \citenamefont {Focassio},\ and\ \citenamefont {Fazzio}}]{ML2021}%
  \BibitemOpen
  \bibfield  {author} {\bibinfo {author} {\bibfnamefont {G.~R.}\ \bibnamefont
  {Schleder}}, \bibinfo {author} {\bibfnamefont {B.}~\bibnamefont {Focassio}},\
  and\ \bibinfo {author} {\bibfnamefont {A.}~\bibnamefont {Fazzio}},\
  }\bibfield  {title} {\bibinfo {title} {Machine learning for materials
  discovery: Two-dimensional topological insulators},\ }\href
  {https://doi.org/10.1063/5.0055035} {\bibfield  {journal} {\bibinfo
  {journal} {Applied Physics Reviews}\ }\textbf {\bibinfo {volume} {8}},\
  \bibinfo {pages} {031409} (\bibinfo {year} {2021})}\BibitemShut {NoStop}%
\end{thebibliography}
\end{document}